\documentclass{aa}
\usepackage{graphicx,natbib}
\usepackage[varg]{txfonts}
\usepackage{afterpage}
\usepackage{color}
\usepackage{upgreek}
\usepackage{soul}
\usepackage{hyperref}

\newcommand{\Prot}{\ensuremath{P_{\mathrm{rot}}}}
\newcommand{\Bm}{\ensuremath{\langle B\rangle}}
\newcommand{\Bz}{\ensuremath{\langle B_z\rangle}}
\newcommand{\Bzrms}{\ensuremath{\langle B_z\rangle}_\mathrm{rms}}
\newcommand{\Bq}{\ensuremath{\langle B_\mathrm{q}\rangle}}
\newcommand{\vsi}{\ensuremath{v\,\sin i}}
\newcommand{\Teff}{\ensuremath{T_{\mathrm{eff}}}}
\newcommand{\kms}{km\,s$^{-1}$}
\newcommand{\Feline}{Fe~\textsc{ii}~$\lambda\,6149$\,\AA}
\newcommand{\Zeeman}{\Delta\lambda_{\rm Z}}
\newcommand{\ew}{W_\lambda}
\newcommand{\RI}{R_I^{(2)}(\lambda_I)}

\defcitealias{2020A&A...639A..31M}{Paper~I}
\defcitealias{2022A&A...660A..70M}{Paper~II}
\defcitealias{2024A&A...683A.227M}{Paper~III}
\bibpunct[ ]{(}{)}{;}{a}{}{,}

\begin{document}

\title{Super-slowly rotating Ap (ssrAp) stars: \\ Spectroscopic study}

\author{G.~Mathys\inst{1}
  \and D.~L.~Holdsworth\inst{2,3}
  \and M.~Giarrusso\inst{4}  
  \and D.~W.~Kurtz\inst{5,2}
  \and G.~Catanzaro\inst{4}
  \and F.~Leone\inst{6,4}}

\institute{European Southern Observatory,
  Alonso de Cordova 3107, Vitacura, Santiago, Chile\\\email{gmathys@eso.org}
\and
Jeremiah Horrocks Institute, University of Central Lancashire, Preston
PR1 2HE, UK
\and
South African Astronomical Observatory, PO Box 9, Observatory 7935,
Cape Town, South Africa
\and
INAF–Osservatorio Astrofisico di Catania, via S. Sofia 78, 95123 Catania, Italy
\and
Centre for Space Research, North-West University, Mahikeng 2745, South
Africa
\and
Dipartimento di Fisica e Astronomia, Sezione Astrofisica, Universit\`a
di Catania, Via S. Sofia 78, I-95123 Catania, Italy}

\date{Received $\ldots$ / Accepted $\ldots$}

\titlerunning{Spectroscopy of ssrAp stars}

\abstract
{The fact that the rotation periods of Ap stars span five to six
  orders of magnitude and that the longest ones reach several hundred
  years represents one of the main unsolved challenges of stellar
  physics.}
{Our goal is to gain better understanding of the occurrence and properties
  of the longest period Ap stars.}
{We obtained high resolution spectra of a sample of super-slowly
  rotating Ap (ssrAp) star candidates identified by a TESS photometric
  survey
  to confirm that they are indeed Ap stars, to check that their
  projected equatorial velocities are compatible with super-slow
  rotation, and to obtain a first estimate of their magnetic field
  strengths. For the confirmed Ap stars, we determined whenever possible their
  mean magnetic field modulus, their mean quadratic magnetic field,
  and an upper limit of their projected equatorial velocities.}
{Eighteen of the 27 stars studied are typical Ap stars; most of the
  other nine appear to be misclassified. One of the Ap stars is not a
  slow rotator; it must be seen nearly pole-on. The properties of the
  remaining 17 are compatible with moderately to extremely long
  rotation periods. Eight new stars with resolved magnetically split
  lines in the visible range were discovered; their mean magnetic field
  modulus and their mean quadratic magnetic field were measured. The
  mean quadratic field could also be determined in five more stars. Five
  spectroscopic binaries containing an Ap star, which were not
  previously known, were identified. Among the misclassified stars,
  one double-lined spectroscopic binary with two similar, sharp-lined
  Am components was also discovered.}
{The technique that we used to carry out a search for ssrAp star
  candidates using TESS data is validated. Its main limitation appears
  to arise from uncertainties in the spectral classification of Ap
  stars. The new magnetic field measurements obtained as part of this
  study lend further support to the tentative conclusions of our
  previous studies: the absence of periods $\Prot\gtrsim150$\,d
    in stars with 
  $B_0\gtrsim7.5$\,kG, the lower rate of occurrence of super-slow
  rotation for field strengths $B_0\lesssim2$\,kG than in the range
  $3\,{\rm kG}\lesssim B_0\lesssim7.5$\,kG, and the deficiency of
slowly rotating Ap stars with (phase-averaged) field strengths between
$\sim$2 and $\sim$3\,kG.}

\keywords{stars: chemically peculiar --
  stars: magnetic field --
  stars: rotation --
  stars: oscillations}

\maketitle

\section{Introduction}
\label{sec:intro}
The rotation periods of the Ap stars range from about half a day to
several centuries, possibly even $\sim$1000 years or more
\citep{2017A&A...601A..14M}. The loss of angular momentum undergone by
these stars during their main sequence lifetime is at most marginal
\citep{2006A&A...450..763K,2007AN....328..475H}. Hence, period
differentiation spanning five to six orders of magnitude must have been
achieved at the pre-main sequence stage. Understanding how this can be
done represents a major challenge for the theories of stellar
formation and evolution. For further progress in this area, additional
observational constraints are needed to guide the theoretical
developments. The project described in this paper represents part of
the current effort made to provide this kind of information.

The existence of a considerable population of Ap stars that have very
long rotation periods has become increasingly well established in recent
years. \citet{2017A&A...601A..14M} concluded that several percent of
all Ap stars must have rotation periods longer than 1\,yr. This
subsequently led him to define a class of super-slowly rotating Ap
(ssrAp) stars \citep{2020pase.conf...35M}, consisting of the Ap stars
that have rotation periods $\Prot>50$\,d. While arbitrary, this lower
limit makes physical sense in a number of respects
\citep{2024A&A...683A.227M}. As highlighted by these authors, for
  an Ap star observed equator-on, $\Prot=50$\,d is a value for which
  the rotational broadening of iron spectral lines is comparable to
  their thermal broadening in Ap stars, and close to, or slighhtly
  below, the instrumental broadening of most current high-resolution
  spectrographs. Longer rotation periods cannot be constrained from
  line broadening alone.

Studying the most extreme representatives of a special class of
objects often has the potential to provide insight into the origin of
the physical properties shared by all the members of the class. That
is why it appears particularly valuable to undertake systematic work
to characterise the Ap stars that have the longest rotation
periods. The ssrAp stars known until recently, which were mostly the
results of serendipitous discoveries, did not lend themselves well to
such a systematic effort. Accordingly, we undertook a coordinated
effort to carry out an exhaustive search for ssrAp stars in an
objective sample. 

To this effect, \citet{2020A&A...639A..31M} (hereafter Paper~I)
introduced an automated method to identify ssrAp star candidates from
the observations performed during Cycle~1 of the TESS (Transiting
Exoplanet Survey Satellite) mission. In a second step, they
performed similar searches based on TESS Cycle~2 data
(\citeauthor{2022A&A...660A..70M} \citeyear{2022A&A...660A..70M},
hereafter Paper~II) and Cycles~3 and 4 data 
(\citeauthor{2024A&A...683A.227M} \citeyear{2024A&A...683A.227M},
hereafter Paper~III). Their approach is based on the consideration
that Ap stars present brightness inhomogeneities over their surface,
which are stable over (very) long timescales and are, to first order, 
symmetrically distributed about the magnetic axis. The latter makes an
angle $\beta$ with the rotation axis, which is inclined to the line of
sight by an angle $i$. Accordingly,
photometric variations are observed with the rotation period of the
star \citep[e.g.][]{1971PASP...83..571P}. This is called
$\alpha^2$~CVn variability, from the name of the 
star in which it was first observed, which represents a
prototype of the class. The underlying basic principle of the
above-mentioned method is that those Ap stars that do not show
$\alpha^2$~CVn photometric variations over the duration of a TESS
sector (27\,d) are ssrAp star candidates. 

In practice, application of this simple principle unsurprisingly
involves some complications. For instance, the lack of observable
photometric variations may be due to a very low obliquity $\beta$ of
the magnetic axis or a very low inclination $i$ of the rotation axis
to the line of sight. These, and other possible complications, have
been discussed in detail in the references listed above. Because of
them, it is necessary to confirm spectroscopically the slow rotation
of the ssrAp star candidates. Here we report on the
first part of a follow-up project carried out to this effect. 

In Sect.~\ref{sec:spectrobs}, we present the spectroscopic
observations that were performed and their
outcome. Section~\ref{sec:Bvsi} describes the way in which the
obtained spectra were analysed to determine the magnetic fields of the
studied stars and to constrain their projected equatorial velocities,
and it gives the results of this analysis. The properties of each star
are discussed on an individual basis in
Appendix~\ref{sec:individual}. As part of this discussion, we also
revisit the 
available TESS data; many stars were reobserved in more TESS sectors
since we completed our original search. Section~\ref{sec:disc} is
devoted to a discussion of the chemical peculiarity, rotation,
binarity and magnetic field of the stars of our sample, on a
statistical basis. Finally, in Sect.~\ref{sec:conc}, we draw the
conclusions of the present work and sketch our plans for follow-up
studies. 

\section{Spectroscopic observations}
\label{sec:spectrobs}
\subsection{Observation details}
\label{sec:obs}
For spectroscopic confirmation of the ssrAp star candidates identified
in our TESS photometric survey, we started a dedicated observational
project. Our main objectives are (1) to confirm that the ssrAp star
candidates are indeed Ap stars; (2) to check that their projected
equatorial velocities are compatible with super-slow rotation; and (3)
to obtain a first estimate of their magnetic field strengths. The
observations were performed with HARPS-N (the High Accuracy Radial
velocity Planet Searcher for the Northern hemisphere;
\citealt{2012SPIE.8446E..1VC}) at the TNG 
(Telescopio Nazionale {\it Galileo\/}), SALT-HRS (the Southern African Large
Telescope High Resolution \'echelle Spectrograph;
\citealt{2010SPIE.7735E..4FB}), and CAOS (the Catania 
Astrophysical Observatory Spectropolarimeter; \citealt{2016AJ....151..116L})
at the 0.9-m telescope of OAC (Catania  
Astrophysical Observatory). In addition, we analysed two
spectra obtained with FEROS (the Fiber-fed Extended Range Optical
Spectrograph; \citealt{1999Msngr..95....8K}) fed by the 2.2-m
telescope at ESO (European Southern 
Observatory). Finally, for two stars, we
used spectropolarimetric observations from the CFHT
(Canada-France-Hawaii Telescope) ESPaDOnS (Echelle SPectropolarimetric
Device for the Observation of Stars; \citealt{2006ASPC..358..362D})
archive.     

HARPS-N at the TNG is an echelle spectrograph covering the wavelength
range 3830--6930\,\AA, with a resolving power 
$R=115,000$. According to the magnitudes of the targets, the S/N varies
from 150 to 200.

The SALT-HRS is a fibre-fed, dual-beam, echelle spectrograph with
wavelength coverage of 3700--5500\,\AA\ and 5500--8900\,\AA\ in the
blue and red arms, respectively
\citep{2010SPIE.7735E..4FB,crause2014}. Of the three available
observing modes, we used the High Resolution mode that achieves a
resolving power of $R\sim45,000$ with the current reduction pipeline. The
observations were 
automatically reduced using this pipeline, which is based
on the ESO's {\sc{midas}} pipeline \citep{pyhrs2,pyhrs3}. The
resultant spectra achieved a S/N in the range 200-280 in the spectral
range of interest. The
pipeline reduced spectra were normalised to unity in the continuum
using the SUPPN{\sc{et}} package \citep{2022A&A...659A.199R}. 

The CAOS spectra cover the spectral range 3900--6800\,\AA, with a
resolving power $R=40,000$, as measured from the ThAr lines of the
wavelength calibration arc spectrum. The S/N ranges from 50 to 150
according to the magnitudes of the targets.

The collection of FEROS observations from which the spectra analysed
here were obtained was described by \citet{2008MNRAS.389..441F} and
\citet{2012MNRAS.420.2727E}; we used the data as reduced by these
authors. These spectra, which cover the wavelength 
range 3530--9220\,\AA, were recorded at a
resolving power $R\sim48,000$, with S/N$\sim$150--200.

The ESPaDOnS archive spectra analysed in this study are similar to
those described by \citet{2017MNRAS.471..926K}. These Stokes $IV$
spectra, which range from 3700 to 10,000\,\AA, were reduced by the
CFHT team using the software package 
Libre-ESPriT \citep{1997MNRAS.291..658D}. The achieved resolving power
is $R\sim65,000$, at a S/N$\sim$500.

\subsection{Spectra}
\label{sec:spectra}
Figures~\ref{fig:spec6150_1} to \ref{fig:spec6150_5} show a
30\,\AA-long portion of
the spectrum of 25 of the 26 sharpest-lined stars that were observed
until now as part of this project. Near its centre is the \Feline\
line, whose high magnetic sensitivity and doublet Zeeman pattern have
made it the most used diagnostic line for determination of the mean
magnetic field modulus $\Bm$ (the line-intensity weighted average over
the visible stellar disk of the modulus of the magnetic field). The
surrounding wavelength interval includes lines of various chemical
elements, some of them in two different ionisation states. Among them
are elements that are often overabundant in Ap stars, such as Si,
Cr, Nd, and Pr, as well as elements that are generally observed in
normal A stars (O, Ca, Fe, and Ba). Using the Vienna Atomic Line Database
\citep[VALD;][]{1999A&AS..138..119K}, complemented by the National
Institute of Standards and Technology (NIST) Atomic
Spectra Database\footnote{https://physics.nist.gov/asd}
\citep{NIST-ASD}, we identified the main contributors 
of a number of 
lines that are present in different stars. These identifications are
only provided for illustrative purposes; they are not meant to be
exhaustive. In particular, some of the lines in some of the stars are
definitely blends resulting from the wavelength coincidence of
transitions of different elements. One of the most prominent examples
is that of the line located close to 6145\,\AA: in many Ap stars, it
is predominantly due to the Nd~{\sc iii} transition that is identified
in the figures, but the Si~{\sc i}~$\lambda\,6145.015$\,\AA\ line also
contributes to it. 

 The spectra are presented in Figs.~\ref{fig:spec6150_1} to
\ref{fig:spec6150_5} in order of increasing effective temperature
\Teff. The values of the latter are as listed in
\citetalias{2020A&A...639A..31M} and
\citetalias{2022A&A...660A..70M}. One can see in the figures a general
trend for many of the lines to become stronger or weaker following this
sequence. For instance, the Ca~{\sc i}~$\lambda\,6162.2$\,\AA\ line is
prominent in the coolest stars, but vanishes completely above
$\Teff\sim9700$\,K. Overall, the trend is similar for the Fe~{\sc i}
and Si~{\sc i} lines. The former remain visible up to
$\Teff\sim10,700$\,K (with the exception of TIC~444094235 -- see below),
while the latter, which also show more irregular star-to-star
differences, are no longer seen at $\Teff\gtrsim8700$\,K. Conversely,
the O~{\sc i} lines start to appear at temperatures $\Teff\sim9700$\,K
and above. The Fe~{\sc ii} lines are present throughout the whole
temperature range of consideration, a behaviour mostly shared by the
Cr~{\sc ii} lines, which however are conspicuously absent from the
spectra of a number of stars. The Nd~{\sc iii}~$\lambda\,6145.1$\,\AA\
line tends to be prominent in most stars up to $\Teff\sim10,700$\,K
(again, TIC~444094235 represents an exception), and the Pr~{\sc
  iii}~$\lambda\,6160.2$\,\AA\ line is also rather strong in some of
them.

However, throughout the whole temperature range of interest,
there are stars that do not show lines of any of the elements Si, Cr,
Nd or Pr: TIC~334505323 (HD~106322, $\Teff=7683$\,K), TIC~207468665
(HD~148330, $\Teff=9700$\,K), TIC~286965228 (HD~127304,
$\Teff=9950$\,K), TIC~301918605 (HD~17330, $\Teff=10,250$\,K),
TIC~80486647 (HD~67658, $\Teff=12,018$\,K), and TIC~124998213
(HD~44979, $\Teff=12,593$\,K).  These stars are definitely not 
typical Ap stars. Misclassification as Ap cannot be ruled out. In
fact, \citet{2020MNRAS.499.2701M} assigned the spectral type
A4IV to TIC~80486647. The difference between the typical effective
temperature of an A4 star and the much higher value reported here is
striking. The strength of the Ba~{\sc ii}~$\lambda\,6141.7$\,\AA\ line
appears much more consistent with the former than with the
latter. This lends further support to the suspicion that the
classification underlying the inclusion of this star in the present
sample was 
wrong. An additional marginal case is that of TIC~291561579
(HD~171420, $\Teff=6793$\,K). Its spectrum is very different from 
that of the other stars with effective temperatures below 7000\,K. The
Fe lines are weak, there are no visible lines of Cr or of rare earths,
but the Si~{\sc i} lines are present, and not exceedingly weak,
especially when compared with Fe. This may be an Ap star with milder
peculiarities than most other cool Ap stars.

\begin{figure}[t]
  \centering
  \includegraphics[scale=0.85]{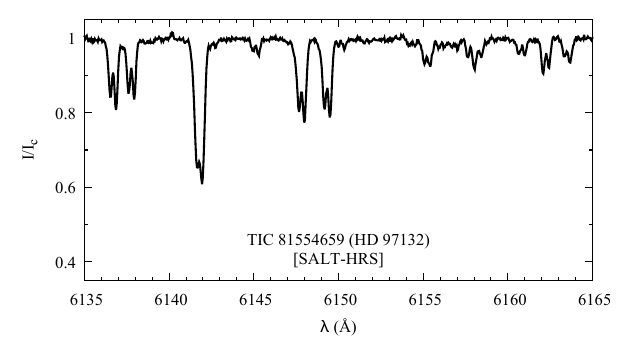}
  \caption{Portion of the spectrum of the SB2 star HD~97132, obtained
    with SALT-HRS on HJD~2,459,968.455, covering
    the same range as in Figs.~\ref{fig:spec6150_1} to
    \ref{fig:spec6150_5}. The wavelengths are in the heliocentric reference frame.}
  \label{fig:hd97132}
\end{figure}

\begin{figure*}
  \centering
  \includegraphics[scale=0.81]{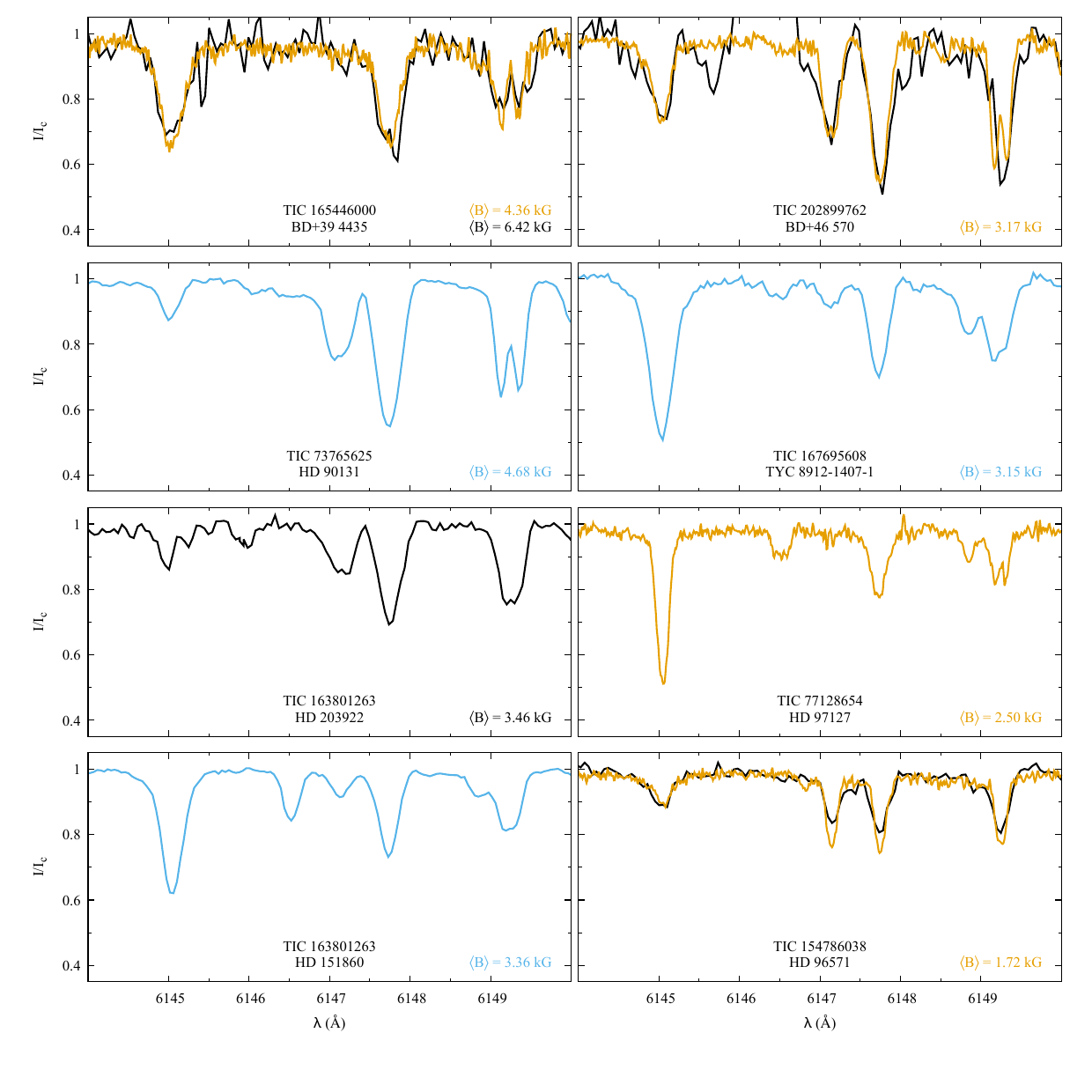}
  \caption{Blown-up portion of the spectra of the eight Ap stars
    in which the presence of resolved, or marginally resolved,
    spectral lines is reported for the first time here, including the
    mean magnetic field modulus diagnostic line \Feline. The CAOS spectra
    are plotted in black; the SALT-HRS spectra in sky blue; and the
    HARPS-N spectra in orange. For three stars, spectra were obtained
    both with HARPS-N and with CAOS. The two spectra of each pair are
    shown together, to illustrate the value of achieving the highest
    possible resolution to diagnose the magnetic field. The derived
    values of $\Bm$ appear in the bottom right corner of each panel;
    the CAOS spectra of BD+46~570 and of HD~96571 have insufficient
    resolution to determine $\Bm$ in these stars. The wavelengths are
    in the laboratory reference frame. The broad emission-like
    feature in the CAOS spectrum of BD+46~570 is due to an instrumental
    glitch. The Zeeman patterns of the lines
    seen in the considered spectral range are illustrated in
    Fig.~\ref{fig:zp6149}.}
    \label{fig:spec6149}
  \end{figure*}

On the other hand, the value of the effective temperature of
TIC~444094235 (HD~85284), $\Teff=13,640$\,K, appears to be wrong. This is
definitely a typical Ap star, and in the sequence shown in the figures,
the aspect of its spectrum would rather locate it close to
$\Teff=10,700$\,K.

\begin{figure}
  \includegraphics[width=\hsize]{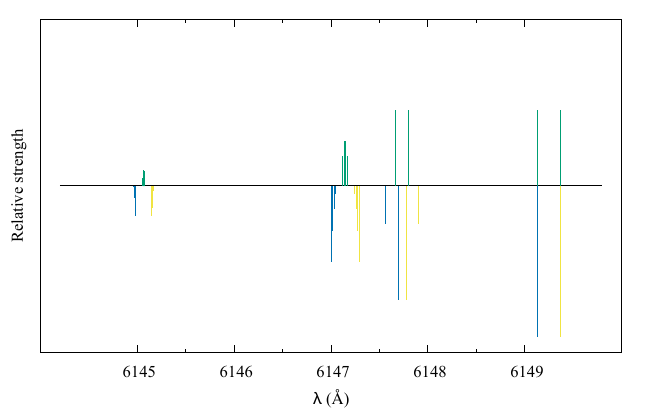}
  \caption{Zeeman patterns of the main lines observed in the spectral
    range covered in Fig.~\ref{fig:spec6149}: Nd~{\sc
      iii}~$\lambda\,6145.1$\,\AA, Cr~{\sc ii}~$\lambda\,6147.1$\,\AA,
  Fe~{\sc ii}~$\lambda\,6147.7$\,\AA, and Fe~{\sc
    ii}~$\lambda\,6149.2$\,\AA. The amplitude of 
  the splitting corresponds to a magnetic field strength
  $B=5$\,kG. The length of each vertical bar is proportional to the
  relative strength of the corresponding line component. The $\pi$
  components appear above the horizontal line (in green), the
  $\sigma_+$ and $\sigma_-$ components below it (in blue and yellow
  respectively).}
  \label{fig:zp6149}
\end{figure}  

One more star belongs to the group of the sharpest-lined observed
stars that is illustrated in Figs.~\ref{fig:spec6150_1} to
\ref{fig:spec6150_5}: TIC~84554659 (HD~97132). The same portion of its
spectrum as considered for the stars discussed above is shown in
Fig.~\ref{fig:hd97132}. This is obviously a double-lined spectroscopic
binary (SB2). To the best of our knowledge, the binarity is reported
here for the first time. The two components have very sharp spectral
lines and appear strikingly similar
to each other; they probably have nearly the same spectral type, and
the same rotational velocity. However, while the lines are very sharp,
neither of the two components seems to be a typical Ap star. It seems
more likely that they may be Am stars that were mistakenly classified
as Ap. As we do not know
the radial velocity of the barycentre of the system, we have no
meaningful reference to plot the spectrum in the laboratory reference
system. Thus, contrary to
Figs.~\ref{fig:spec6150_1}--\ref{fig:spec6150_5}, for
Fig.~\ref{fig:hd97132}, the wavelengths have been left in the
heliocentric reference system.

Besides the temperature and chemical composition effects that are
revealed by the intensity differences between the lines of different
ions,  the spectra shown in Figs.~\ref{fig:spec6150_1} to
\ref{fig:spec6150_5} also differ from each other in terms of line
profiles. Not only do they show a range of line widths, but also more
complex features, which in particular reflect the presence of more or
less strong magnetic fields. Most prominently, there are eight stars in
which the magnetically split components of the \Feline\ line are
fully or marginally resolved. For these eight stars, a blown-up
portion of the spectrum including the \Feline\ line is shown in
Fig.~\ref{fig:spec6149} to allow the magnetic resolution to be
visualised better than at the more compressed scale of
Figs.~\ref{fig:spec6150_1} to \ref{fig:spec6150_5}. The spectra are
presented in the order of decreasing magnetic field strength (see
Table~\ref{tab:meas}), from top to bottom in the left column, then
from top to bottom in the right column. The Zeeman
patterns of the main lines present in the displayed wavelengh range
are shown in Fig.~\ref{fig:zp6149}.

The most magnetically sensitive among them is the \Feline\ line. In
the visible range, it is the most used diagnostic line for
determination of the mean magnetic field modulus $\Bm$. Its
characteristics have been discussed in detail by
\citet{1990A&A...232..151M}. Its Zeeman pattern is a simple doublet,
resulting from a transition between an unsplit level and a split
level, in which each of the two $\pi$ components is shifted from the
line centre by the same amount as each of the single $\sigma_+$ and
$\sigma_-$ components. Among the stars shown in
Fig.~\ref{fig:spec6149}, the doublet is well resolved in the HARPS-N
spectra of BD+39~4435, BD+46~570 and HD~97127, in the SALT-HRS
spectrum of HD~90131, and in the CAOS spectrum of BD+39~4435, which
however is noisier, hence more difficult to exploit. The
resolution is more marginal, but still sufficient to untangle the two
components in the SALT-HRS spectra of TYC~8912-1407-1
and of HD~151860, in the CAOS spectrum of HD~203922, and in the HARPS
spectrum of HD~96571. The presence of (marginally) resolved
magnetically split lines in the eight stars of Fig.~\ref{fig:spec6149}
is reported here for the first time.

\begin{figure*}
  \centering
  \includegraphics[scale=0.81]{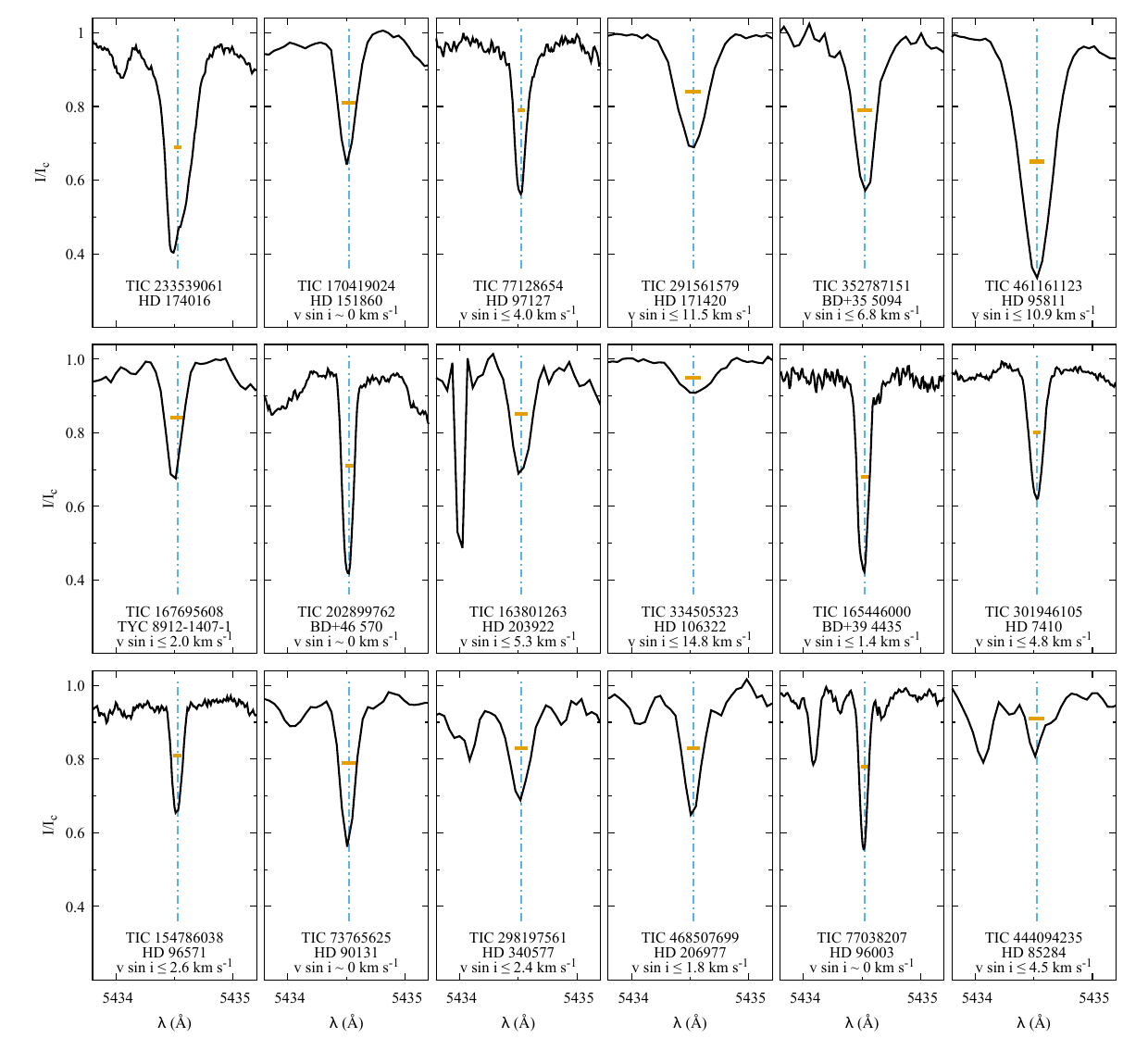}
  \caption{Profile of the magnetically insensitive Fe~{\sc
      i}~$\lambda\,5434.5$\,\AA\ line as 
    observed in the 18 stars of the present sample in which it is
    visible.  The spectra are shown in order of increasing effective
    temperature, from left to right, then from top to bottom. The
    orange horizontal line at approximate mid-depth of each profile
    represents the estimated contribution of the instrumental profile
    and of thermal Doppler broadening to the line full width. An
    upper limit of \vsi, as given in Table~\ref{tab:meas}, is indicated at the
    bottom of each panel; $\vsi\sim0$\,\kms\ means that the rotational
    broadening is below the detection limit at the achieved spectral
    resolution. The upper limit of \vsi\ was not determined for
    HD~174016 since it is a SB2 system for which the lines of the two
    components cannot be separated. For those stars for which both
    HARPS-N and CAOS 
    spectra were obtained, the \vsi\ constraint is the one determined 
    from the higher resolution HARPS-N spectrum. The wavelengths are
    in the laboratory reference frame. The deep, narrow feature at
    $\lambda\sim5434$\,\AA\ in the CAOS spectrum of TIC~163801263 is
    due to an instrumental glitch.}
\label{fig:spec5434}
\end{figure*}

\begin{figure*}
  \centering
  \includegraphics[scale=0.75]{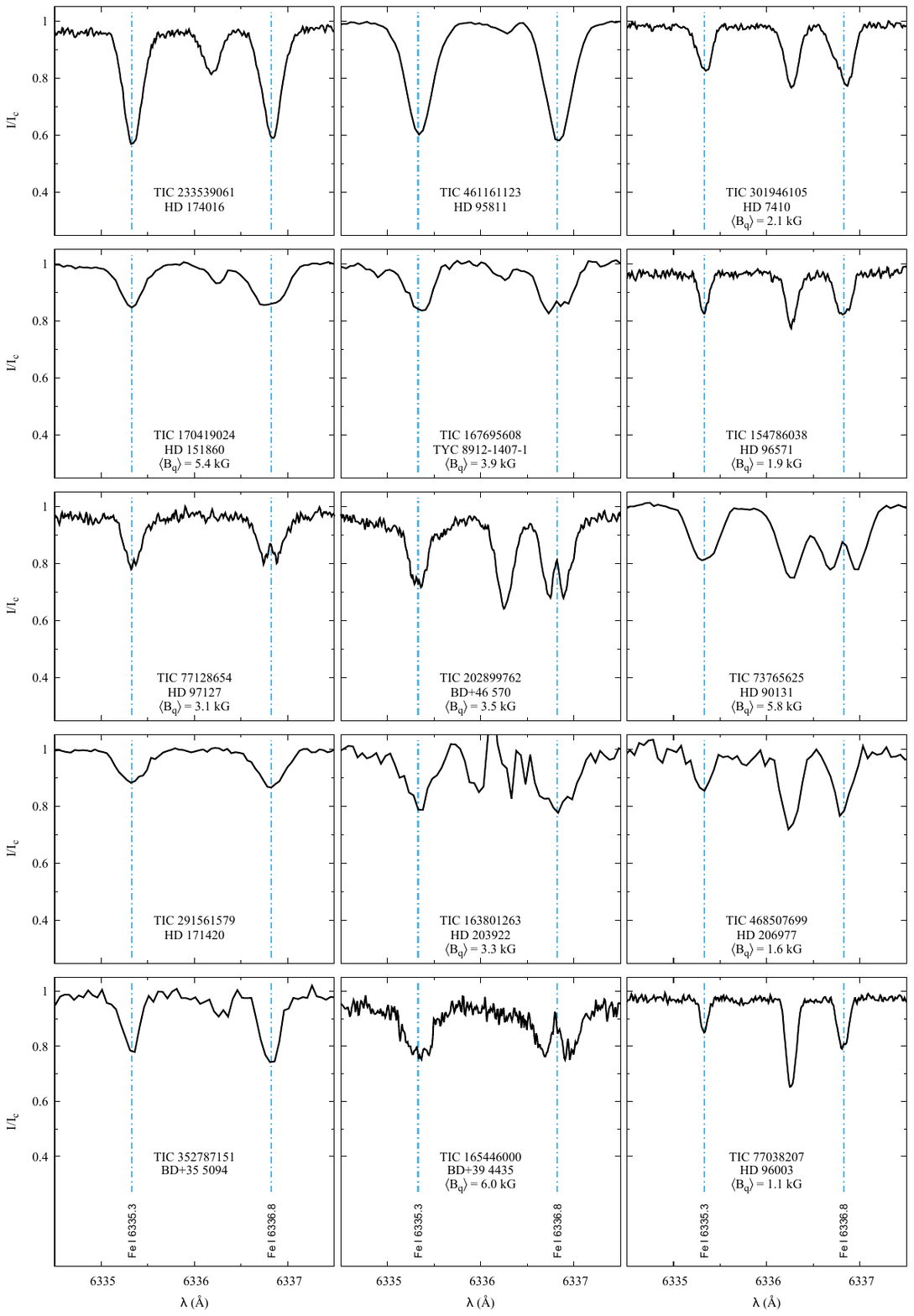}
  \caption{Comparison of the profiles of the Fe~{\sc
      i}~$\lambda\,6335.3$\,\AA\ and $\lambda\,6336.8$\,\AA\ lines as 
    observed in the 15 stars of the present sample in which both are
    well visible.  The spectra are shown in order of increasing effective
    temperature, from top to bottom, then from left to right. The
    value of the mean quadratic magnetic field $\Bq$ is indicated at the
    bottom of each panel, for all stars in which its line broadening
    effect was above the limit of detection at the achieved spectral
    resolution. For those stars for which both HARPS-N and CAOS
    spectra were obtained, the $\Bq$ value is the one determined 
    from the higher resolution HARPS-N spectrum. The wavelengths are
    in the laboratory reference frame. The Zeeman patterns of the lines
    seen in the considered spectral range are illustrated in
    Fig.~\ref{fig:zp6336}. The emission feature at
    $\lambda\sim6336$\,\AA\ in the CAOS spectrum of HD~203922 is
    due to an instrumental glitch.}
\label{fig:spec6336}
\end{figure*}

 In Figs.~\ref{fig:spec6150_1} to \ref{fig:spec6150_5}, one can also
appreciate star-to-star line width differences. The latter result from
a number of factors. The main ones are  the spectrograph
resolution, the Doppler effect due to stellar rotation, and the
stellar magnetic field strength. However, in the 18 stars of
Figs.~\ref{fig:spec6150_1} to \ref{fig:spec6150_4} that have effective
temperatures $\Teff\lesssim9500$~K, one magnetic null line, Fe~{\sc
  i}~$\lambda\,5434.5$\,\AA, is clearly visible and mostly
free from blends. This line, which has no magnetic sensitivity, 
also seems weakly present, albeit with a significant red blend, in the
spectrum of TIC~444094235 (HD~85284), whose 
effective temperature is certainly overestimated (see
above). 

\begin{figure}
  \includegraphics[width=\hsize]{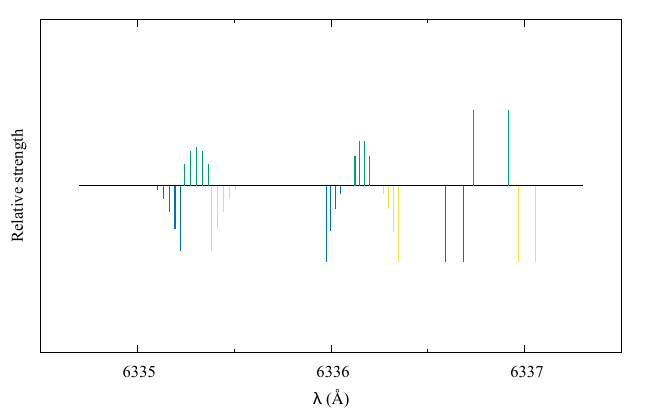}
  \caption{Zeeman patterns of the main lines observed in the spectral
    range covered in Fig.~\ref{fig:spec6336}: Fe~{\sc
      i}~$\lambda\,6335.3$\,\AA, Fe~{\sc ii}~$\lambda\,6336.2$\,\AA,
    and Fe~{\sc i}~$\lambda\,6336.8$\,\AA. The amplitude of
  the splitting corresponds to a magnetic field strength
  $B=5$\,kG. See the caption of Fig.~\ref{fig:zp6149} for more details.}
  \label{fig:zp6336}
\end{figure}  

Thus, the width of the Fe~{\sc i}~$\lambda\,5434.5$\,\AA\ line does not
depend on the magnetic field. Accordingly, it is well suited to
constraining the projected equatorial velocity \vsi\ in those stars in
which it is observable. Its profile in 18 of the 19 stars of the present
sample in which this is the case is shown in
Fig.~\ref{fig:spec5434}. (We omitted TIC~206461701, whose lines are
visibly the broadest ones among the 19 stars of interest -- see
Fig.~\ref{fig:spec6150_2}.) An orange horizontal line was drawn in
each 
panel to represent the approximate minimum value of the Full Width at
Half Depth (FWHD) that is expected to be observed in the considered
spectrum. There are two significant contributions to the latter that
can be readily estimated: the instrumental profile and thermal
broadening. The contribution of thermal broadening to the FWHD is
given by:
\begin{equation}
  \Delta\lambda_{\rm th}={\lambda_0\over c}\,\sqrt{2k\,\Teff\over
    m_{\rm ion}},
\end{equation}
where $\lambda_0$ is the wavelength of the observed line, $c$ is the
velocity of light, $k$ is Boltzmann's constant, and $m_{\rm ion}$ is
the mass of the ion responsible for the considered transition. The
instrumental contribution to the FWHD is $\Delta\lambda_{\rm
  inst}=\lambda_0/R$, where $R$ is the resolving power of the
spectrograph with which the observation was performed. The minimum
FWHD that can be observed results from the quadratic combination of
the instrumental and thermal components:
\begin{equation}
  \Delta\lambda_{\rm min} = (\Delta\lambda^2_{\rm
    instr}+\Delta\lambda^2_{\rm th})^{1/2}.
\end{equation}
Numerically, for the observations of the Fe~{\sc
  i}~$\lambda\,5434.5$\,\AA\ line with the spectrographs used in the
present study, this corresponds to
\begin{equation}
  \Delta\lambda_{\rm min}(\AA)=\sqrt{(3.103\,10^{-7}\,\Teff({\rm K})+w)}.
\end{equation}
The value of $w$ depends on the spectrograph:
$1.46\,10^{-2}$\,\AA$^2$ for SALT-HRS, $9.76\,10^{-3}$\,\AA$^2$
for CAOS, and $2.23\,10^{-3}$\,\AA$^2$ for HARPS-N.

Not unexpectedly, the width of all line profiles illustrated in
Fig.~\ref{fig:spec5434} is at least equal to, and often greater than,
the 
length of the orange bar representing the estimated minimum FWHD
value. Small differences between the two may partly originate from
unaccounted for line broadening factors such as 
microturbulence (which is unknown), as well as to low enough projected
equatorial velocity. They are consistent with the identification of
the stars showing them as ssrAp star candidates. The stars
TIC~202899762 (BD+46 570), TIC~165446000 (BD+39 4435), TIC~154786038
(HD~96571), TIC73765625 (HD~90131), and TIC~77038207 (HD~96003) show
the smallest differences between the FWHD of the 
Fe~{\sc i}~$\lambda\,5443.5$\,\AA\ line profile and the length of the
orange bar representing the minimum line width. They all have typical
Ap spectra in Figs.~\ref{fig:spec6150_1} to \ref{fig:spec6150_5}, and
four of them show resolved, or marginally resolved, magnetically split
lines: the probability 
that they are ssrAp stars is very high. The observed profile FWHD
seems to exceed slightly more the estimated minimum line width in
TIC~170419024 (HD~151860), TIC~77128654 (HD~97127), TIC~352787151
(BD+35~5094), TIC~167695608 (TYC~8912-1407-1),
TIC~301946105 (HD~7410), TIC~468507699 (HD~206977), and TIC~444094235
(HD~85284), but the difference
between them is still small enough for these stars to be super-slow
rotators, or at least to have moderately long rotation periods
($20\,{\rm d}\lesssim\Prot\lesssim50$\,d). Again, the spectra of all of
them are typical of Ap stars, and four of them show (marginally)
resolved magnetically split lines. Very likely, most of them are ssrAp
stars.

Of the four remaining stars shown in Fig.~\ref{fig:spec5434},
two do not have a typical Ap star spectrum: TIC~291561579 (HD~171420) and
TIC~334505323 (HD~106322). We suspect that the latter has been mistakenly
classified as Ap; the former may either be misclassified or be a mild
Ap star (see above and Appendix~\ref{sec:hd_171420}). By contrast, the
spectrum of TIC~461161123 (HD~95811) is 
similar to that of Ap stars, albeit with weak or absent Cr lines. The
rotational broadening of its lines is too large for it to be rotating
extremely slowly. It may have a moderately long rotation period, or be
an example of a near-alignment of the magnetic and rotation axes.

The case of TIC~233539061 (HD~174016) is different. The Fe~{\sc
  i}~$\lambda\,5443.5$\,\AA\ line is much wider than the estimated
minimum line width, but it also shows definite asymmetry. This
asymmetry reflects the presence of this line in the two components of
what is actually a double-lined spectroscopic binary, composed of an
Ap star and of a G giant. More details are given in
Appendix~\ref{sec:hd_174016}, but at the epoch of observation
(HJD~2,460,214.383), the 
spectral lines of the Ap component were blueshifted  with respect to
those of the giant. The Ap component is responsible for the blue dip
seen in Fig.~\ref{fig:spec5434}. This dip may be due to a line narrow
enough to belong to a ssrAp star. The very low projected equatorial
velocity was also confirmed by \cite{1999A&AS..140..279G}, who noted
that the CORAVEL (CORrelation-RAdial-VELocities) correlation peak for
this component is 
too narrow to derive a significant \vsi\ estimate. This represents
a very strict constraint, taking into account that magnetic broadening
must also contribute to the width of this peak. 

Unfortunately, for the hotter stars ($\Teff\gtrsim9500$\,K), we could
not identify any suitable magnetic null line. Rotation will be further
discussed, on a more quantitative basis, in Appendix~\ref{sec:individual}.

\section{Magnetic field and equatorial velocity}
\label{sec:Bvsi}

\subsection{Mean magnetic field modulus}
\label{sec:Bm}
For the stars shown in Fig.~\ref{fig:spec6149}, in which the \Feline\
line is resolved  into its two magnetically split
components, the wavelength separation of these two components is
proportional to the mean magnetic field modulus:
\begin{equation}
  \lambda_{\rm r}-\lambda_{\rm b}=g\,\Delta\lambda_{\rm Z}\,\Bm,
\label{eq:Bm}
\end{equation}
where $\lambda_{\rm r}$ and $\lambda_{\rm b}$ are the wavelengths of
the red and blue line components, respectively; $g=2.70$ is the Land\'e
factor of the split level of the transition; $\Delta\lambda_{\rm
  Z}=k\,\lambda_0^2$, with $k=4.67\,10^{-13}$\,\AA$^{-1}$\,G$^{-1}$;
and $\lambda_0=6149.258$\,\AA\ is the nominal wavelength of the
transition. The wavelengths are expressed in angstr\"oms and the
magnetic field in gauss.

As explained by \citet{1997A&AS..123..353M}, thanks to the doublet
structure of the \Feline\ line, the relation given in
Eq.~(\ref{eq:Bm}) is almost approximation-free. Therefore, its
application allows one to determine a physically meaningful value of
the mean magnetic field modulus. We also followed the method described
by \citeauthor{1997A&AS..123..353M} for measurement of the wavelengths
$\lambda_{\rm r}$ and 
$\lambda_{\rm b}$ of the split line components. Since in all stars that we
analysed, these components show at least some overlap, we preferred
Gaussian fitting to direct integration for these measurements. Each
component was fitted with a Gaussian; a third Gaussian was added to
the fit for the blending line affecting the blue wing of the \Feline\
line when its contribution could significantly impact the
derived values of $\lambda_{\rm b}$ and $\lambda_{\rm r}$.  Such a
three-Gaussian fit was used for the stars HD~151860, TYC~8912-1407-1,
and HD~97127, as well as for the HARPS-N spectrum of BD+39~4435.
The Gaussian fitting technique is applicable even to
  observations in which the \Feline\ line is almost flat-bottomed
  rather than showing clear splitting, such as the SALT-HRS spectrum
  of HD~151860 and the HARPS spectrum of HD~96571 (the resolution of
  the CAOS spectrum of this star is too low to use this approach). In
  these cases, it proves possible to fit two Gaussians of similar
  depth and width to the observed \Feline\ line profile without major
  ambiguity. The validity of this approach is also demonstrated by
  past experience, such as the consistency of the lowest $\Bm$ values
  derived for HD~9996 and HD~18078 with the variation curve defined
  from consideration of the phase range at which the magnetic field of
  these stars is stronger \citep{2017A&A...601A..14M}.

The difficulty of determining the uncertainties of the derived $\Bm$
values for stars for which we have at most a handful of measurements,
and more often only one, has been discussed by
\citet{1997A&AS..123..353M}. Like these authors, we estimated the
uncertainties by comparing 
the analysed spectra with those of stars well observed
over a full rotation cycle (or at least over a wide enough range of
phases), for which the uncertainties are given by the scatter of the
individual measurements around a smooth variation curve. The factors
taken into account in this comparison include the resolution and S/N
of the spectra, the separation of the blue and red components of the
\Feline\ line, and the amount of blending affecting this
line. Suitable reference stars abound in studies such as those of
\citet{2017A&A...601A..14M} and of \citet{2022MNRAS.514.3485G}. 
While this procedure inevitably involves some degree of subjectivity,
our experience suggests that the resulting uncertainty estimates
should be correct to within $\pm30$\%.

The $\Bm$ values that we derived for the stars of
Fig.~\ref{fig:spec6149} and their estimated uncertainties are given in
Cols.~8 and 9 of Table~\ref{tab:meas}. They are further discussed in
Appendix~\ref{sec:individual}. 

\begin{figure*}
  \centering
  \includegraphics[width=\hsize]{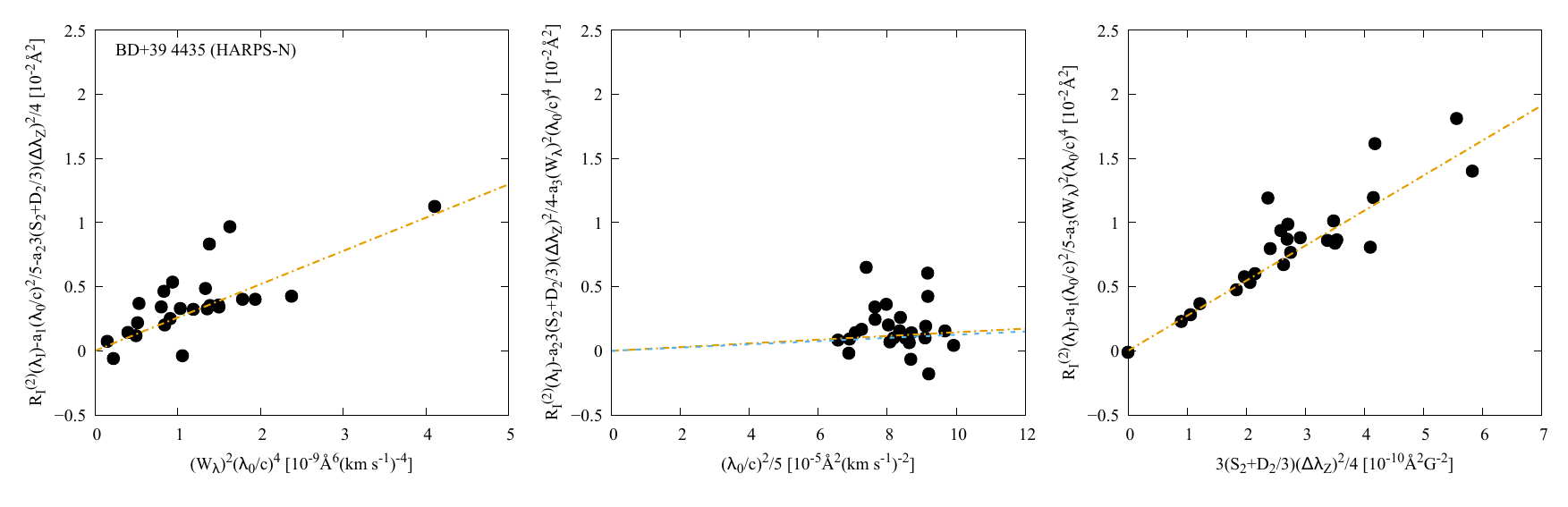}
  \caption{Contribution of the intrinsic ({\em left\/}), Doppler ({\em
    centre\/}) and magnetic ({\em right\/}) terms to the
  second-order moment about their centre of the profiles of the
  Fe~{\sc i} lines analysed in the HARPS-N spectrum of BD+39~4435. The
slopes of the dashed-dotted orange lines are, respectively, the
coefficients $a_3$, $a_1$ and $a_2$ of the least-squares fit of the
measured values of $\RI$ by a function of the form given in
Eq.~(\ref{eq:Bq}). The sky blue dashed line in the centre panel
represents the minimum contribution of the instrumental profile and of
thermal Doppler broadening to the observed line width (see text for
details). In BD+39~4435, because the rotational broadening is
negligibly small, this line coincides almost exactly with the $a_1$
slope line, from which it cannot be readily distinguished. The two
lines are clearly separated from each other in stars that show
significant rotation, as illustrated in Figs.~\ref{fig:bq46570} and
\ref{fig:bq90131}.} 
\label{fig:bq394435}
\end{figure*}

\subsection{Mean quadratic magnetic field and projected equatorial velocity}
\label{sec:Bq}
\subsubsection{Method}
\label{sec:Bqmethod}
For stars that do not show resolved magnetically split lines, 
the mean quadratic field $\Bq$ is a magnetic moment that represents a
valuable alternative to the mean magnetic field modulus for the
characterisation of the intrinsic strength of the stellar magnetic
field, especially in the context of statistical studies of its
distribution in star samples of interest. The mean quadratic magnetic
field is the square root of the sum of the mean square magnetic field
modulus and of the mean square longitudinal magnetic field
\citep{1995A&A...293..746M}. The latter are line-intensity weighted
averages over the visible stellar disk of the square of the modulus of
the magnetic vector and of the square of its component along the line
of sight. In practice, the sensitivity of the mean quadratic magnetic
field to the geometry of the observation is moderate, in
contrast with that of the mean longitudinal magnetic field $\Bz$ (the
line-intensity weighted average over the visible stellar disk of the
component of the magnetic vector along the line of sight). Thus, the
value of $\Bq$ is much more representative of the intrinsic stellar
magnetic field strength than the value of $\Bz$.

The main observable manifestation of the mean quadratic magnetic field
is the differential broadening of lines of different magnetic
sensitivities. This is illustrated in Fig.~\ref{fig:spec6336}, in
which the profiles of the Fe~{\sc i}~$\lambda\,6335.3$\,\AA\ and Fe~{\sc
  i}~$\lambda\,6336.8$\,\AA\ lines are compared to each other. These
lines are seen in stars with effective temperatures
$\Teff\lesssim9500$\,G, except in HD~106322, HD~340577, and HD~85284
(whose effective temperature is uncertain -- see above), in which the
Fe~{\sc i}~$\lambda\,6335.3$\,\AA\ line is too weak to provide useful
information at the achieved S/N. The Zeeman patterns of the two
Fe~{\sc i} lines of interest are shown in Fig.~\ref{fig:zp6336}. The
pattern spread is 
considerably wider for Fe~{\sc i}~$\lambda\,6336.8$\,\AA\ than for
Fe~{\sc i}~$\lambda\,6335.3$, especially since the strongest Zeeman
components of the former are the outermost ones, while in the latter,
the innermost components are stronger. The Fe~{\sc
  i}~$\lambda\,6336.8$\,\AA\ line is one of the
most magnetically sensitive lines of the red spectrum of the Ap stars. In
the eight stars of this study in which the \Feline\ line is
magnetically resolved, Fe~{\sc i}~$\lambda\,6336.8$\,\AA\ shows at least
a broadened, roughly flat bottom (HD~96571, HD~151860,
TYC~8912-1407-1), possibly with some additional structure (HD~203922),
or it is clearly split into two `components' (HD~97127, BD+46~570,
HD~90131, BD+39~4435). Because of the complex Zeeman pattern of the
Fe~{\sc i}~$\lambda\,6336.8$\,\AA\ transition, these `components'
are in fact 
the result of the combination of several individual components with
different wavelength shifts and relative strengths. Therefore, the
value of $\Bm$ cannot be derived in a straightforward manner from
their separation, contrary to the case of the \Feline\ line. Proper
determination of the mean magnetic field modulus from the splitting of
the Fe~{\sc i}~$\lambda\,6336.8$\,\AA\ line would require a numerical analysis
taking into account the radiative transfer effects.

Besides the stars listed above, HD~96003 also shows a hint of
broadening of Fe~{\sc i}~$\lambda\,6336.8$\,\AA\ relative to Fe~{\sc
  i}~$\lambda\,6335.3$\,\AA, or even some incipient splitting of the
former. The two lines also appear to have different widths in HD~7410
but, in addition, they seem to be somewhat asymmetric, with a more
extended blue wing, for an undetermined reason. As to the SB2
HD~174016, the two components cannot be distinguished from each other
here. The observed wavelengths of the two Fe~{\sc i} lines suggests
that they originate predominantly from the G giant primary. The other
stars shown in Fig.~\ref{fig:spec5434}, with the possible exception of
BD+35~5094, do not show any conspicuous additional broadening of the
Fe~{\sc i}~$\lambda\,6336.8$\,\AA\ line.

For determination of the mean quadratic magnetic field from
the observed differential magnetic broadening of spectral
lines, we used the moment technique
\citep{1995A&A...293..746M}. In practice, to derive the value of $\Bq$,
we applied the following formula \citep{2006A&A...453..699M}:
\begin{equation}
\RI=a_1\,{1\over5}\,{\lambda_0^2\over
  c^2}+a_2\,{3S_2+D_2\over4}\,\Zeeman^2+a_3\,\ew^2\,{\lambda_0^4\over c^4}\,,
\label{eq:Bq}
\end{equation}
where $a_2=\Bq^2$. 
In this equation, $\RI$ is the second-order moment of
the line profile observed in natural light (Stokes $I$) about its centre of gravity
$\lambda_I$, $\ew$ is the equivalent width of the line, and $S_2$ and
$D_2$ are atomic parameters characterising the Zeeman pattern of the
considered transition. For details about these parameters as well as
about the definition and actual measurement of
$\RI$, the reader is referred to
\citet{2017A&A...601A..14M}. 

Following this author, in practice, $\RI$ is measured for a sample of
lines and a linear least-squares fit to these data of the form given
by Eq.~(\ref{eq:Bq}) is computed. This fit is weighted by the
uncertainty of the $\RI$ values measured for the individual lines,
following the procedure originally introduced by
\citet{1994A&AS..108..547M} for determination of the mean longitudinal
magnetic field.

The difficulty in the derivation of the mean quadratic magnetic field
is to disentangle the contributions of the three terms appearing on the
right-hand side of Eq.~(\ref{eq:Bq}) to the
second-order moments of the Stokes $I$ line profiles. These contributions are
illustrated in Fig.~\ref{fig:bq394435} , in the case of the analysis of the
HARPS-N spectrum of   BD+39~4435, which yields the highest value of
$\Bq$ derived in this study.

To build this figure, we computed a weighted fit of the values of
$\RI$ measured in 25 Fe~{\sc i} diagnostic lines to a function of the
form given by Eq.~(\ref{eq:Bq}). We isolated the contribution of each
of the terms appearing on the right-hand side of this equation by
subtracting from the measured values of $\RI$ the fitted contributions
of the other two terms. We then plotted the residuals against the relevant
parameter, as follows. In the left panel of Fig.~\ref{fig:bq394435},
$[\RI-a_1\,\lambda_0^2/(5c^2)-a_2\,(3S_2+D_2)\,\Zeeman^2/4]$
is plotted against $(\ew^2\lambda_0^2/c^4)$. The middle panel shows
$\RI-a_2\,(3S_2+D_2)\,\Zeeman^2/4-a_3\,\ew^2\,\lambda_0^4/c^4]$
against $[\lambda_0^2/(5c^2)]$. In the right panel, one sees
$[\RI-a_1\,\lambda_0^2/(5c^2)-a_3\,\ew^2\,\lambda_0^4/c^4]$ against
$[(3S_2+D_2)\Zeeman^2/4]$.  The least-squares fits of the residuals
against the corresponding relevant parameter are illustrated by the
dashed-dotted orange straight lines. The slopes of the latter, from
left to right, are respectively the fit coefficients $a_3$, $a_1$, and
$a_2$.

The main contribution to the second-order moments of the line profiles
in the HARPS-N spectrum of BD+39~4435 is that of the second term of
the right-hand side of Eq.~(\ref{eq:Bq}). This term corresponds to the
magnetic broadening. The value of the mean quadratic magnetic field
can be derived from that of the fit coefficient $a_2$:
$\Bq=\sqrt{a_2}$. Its uncertainty is computed from the standard
deviation $\sigma(a_2)$: this formal value corresponds to the
line-to-line scatter about the best fit regression; it does not
consider other possible sources of error. One can see from
Fig.~\ref{fig:bq394435} that the linear dependence that determines the
value of $\Bq$ is very well defined.

The second significant contribution to the moments $\RI$ in the
HARPS-N spectrum of BD+39~4435 comes from the third term of the
right-hand side of Eq.~(\ref{eq:Bq}). Here too, the linear dependence
that constrains the value of the fit coefficient ($a_3$) is well
defined (see Fig.~\ref{fig:bq394435}). This term corresponds to what
\citet{2006A&A...453..699M} 
refer to as the intrinsic part of the line profile. The physical
meaning of this broadening contribution, which depends, in particular,
on the line equivalent widths, has no relevance in the present
context, but it is important to take this term duly into account to
avoid overestimating the other two.

\begin{figure*}
  \centering
  \includegraphics[width=\hsize]{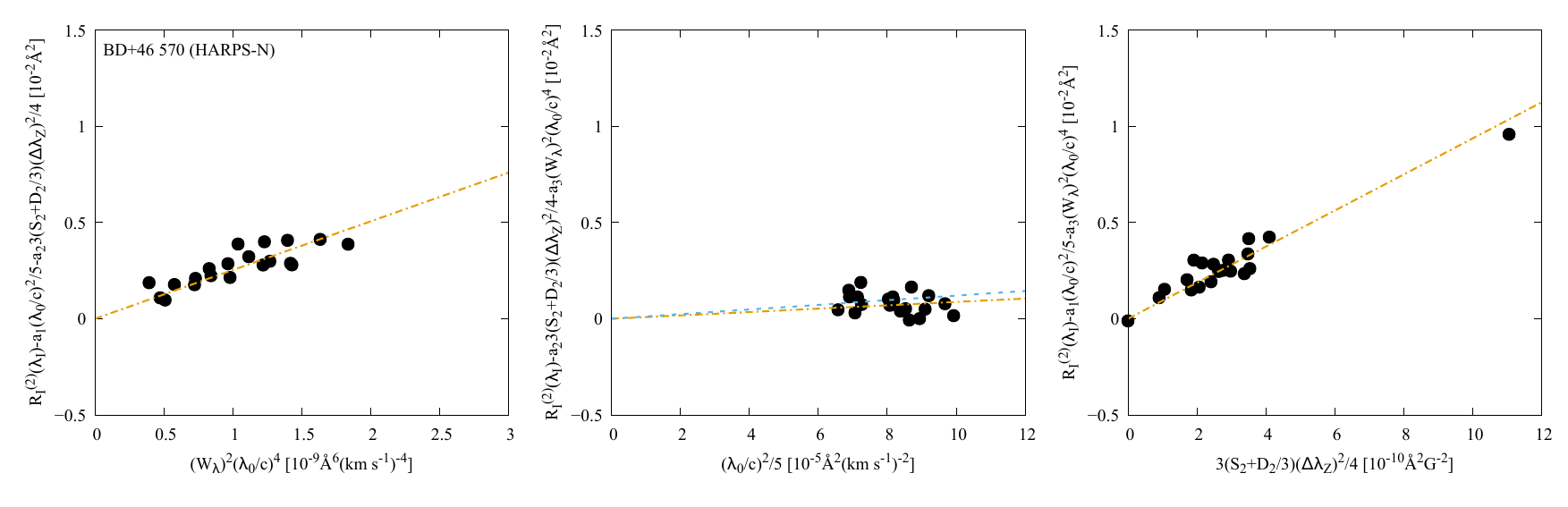}
  \includegraphics[width=\hsize]{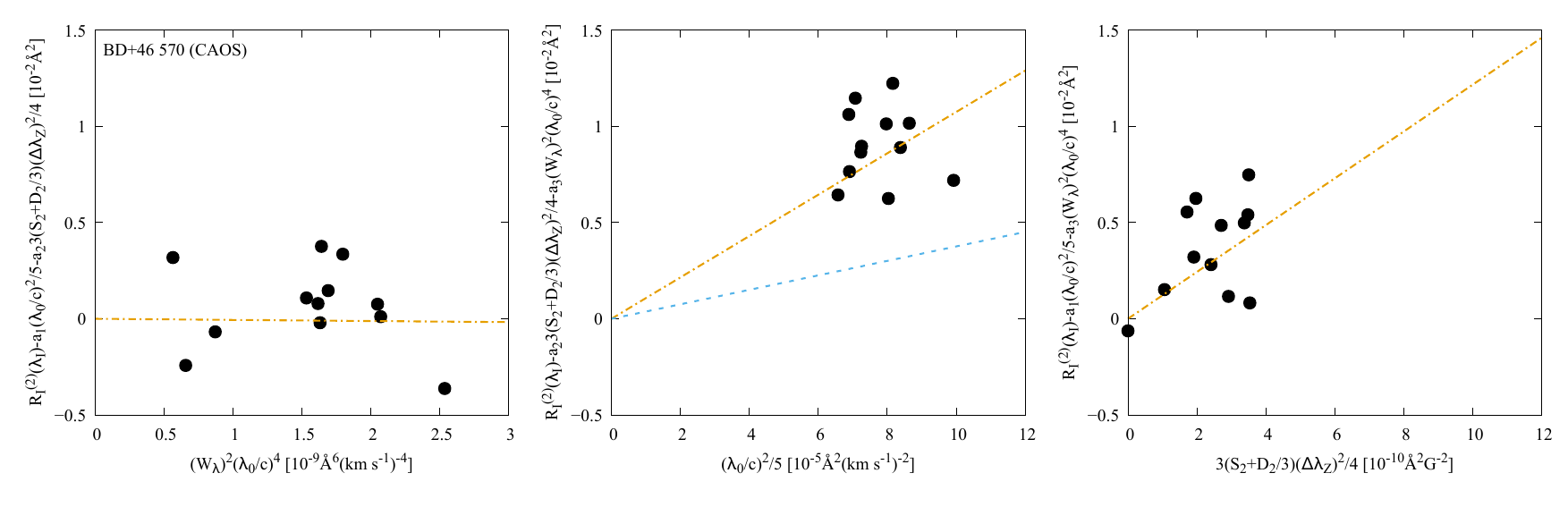}
  \caption{Same as Fig.~\ref{fig:bq394435}, for the HARPS-N ({\em
      top\/}) and CAOS ({\em bottom\/}) spectra of BD+46~570. While
    the magnetic broadening is unambiguously seen in the 
      right panels for both spectra, the scatter of the individual
    points about the respective best fit lines reflects the better
    precision achievable in the determination of the value of $\Bq$ at
    the higher resolution of HARPS-N. A much stricter upper limit 
    on \vsi\ can also be set from the latter, but the picture is
    confused by crosstalk between the intrinsic and Doppler
    contributions to the line width in the analysis of the CAOS
    spectrum (see text for details).} 
\label{fig:bq46570}
\end{figure*}

 The first term of the right-hand side of Eq.~(\ref{eq:Bq}) makes, at
most, a marginal contribution to the second-order moments of the
Stokes $I$ line profiles of BD+39~4435. This term corresponds to the
Doppler-like line broadening agents, which are proportional to the
wavelength. (Line width proportionality to the wavelength translates
into quadratic proportionality of the corresponding terms in the
second-order moment of the line profile.) The $a_1$ fit coefficient
can be expanded as follows \citep{2006A&A...453..699M}:
\begin{equation}
  a_1=v^2\,\sin^2i+5F\,c^2+10k\,\Teff/m_{\rm ion}+a^{\prime}_1\,,
  \label{eq:a1}
\end{equation}
with $F=1/(8\ln2\,R^2)$.
On the right-hand side of this equation, the terms, in order, account
respectively for rotational Doppler broadening, instrumental
broadening, thermal broadening, and all the other contributions to the
line width with a Doppler-like wavelength dependence (such as
microturbulence or pulsation). In general, we cannot readily determine
the value of $a^{\prime}_1$, but we can calculate the instrumental and
thermal terms (provided that we have an estimate of the effective
temperature). By subtracting these two terms from $a_1$, we can derive
an upper limit of the projected equatorial velocity. For the Fe
diagnostic lines used in this study, the numerical expression of this
upper limit is:
\begin{equation}
  \vsi\leq(a_1-a_{\rm inst}-1.474\times10^{-3}\,\Teff)^{1/2}\,,
\label{eq:vsimax}
\end{equation}
where $a_{\rm
  inst}=1.6205\times10^{10}\,R^{-2}$\,km$^2$\,s$^{-2}$. For SALT-HRS, 
$a_{\rm inst}=40.02$\,km$^2$\,s$^{-2}$; for FEROS, $a_{\rm
  inst}=35.17$\,km$^2$\,s$^{-2}$; for CAOS, $a_{\rm  
  inst}=26.79$\,km$^2$\,s$^{-2}$; for ESPaDOnS, $a_{\rm 
  inst}=19.18$\,km$^2$\,s$^{-2}$; and for HARPS-N, $a_{\rm
  inst}=1.23$\,km$^2$\,s$^{-2}$. The upper limit of \vsi\ that is
derived by application 
of Eq.~(\ref{eq:vsimax}) is meaningful because typically
$a^{\prime}_1$ is small.

\begin{figure*}
  \centering
  \includegraphics[width=\hsize]{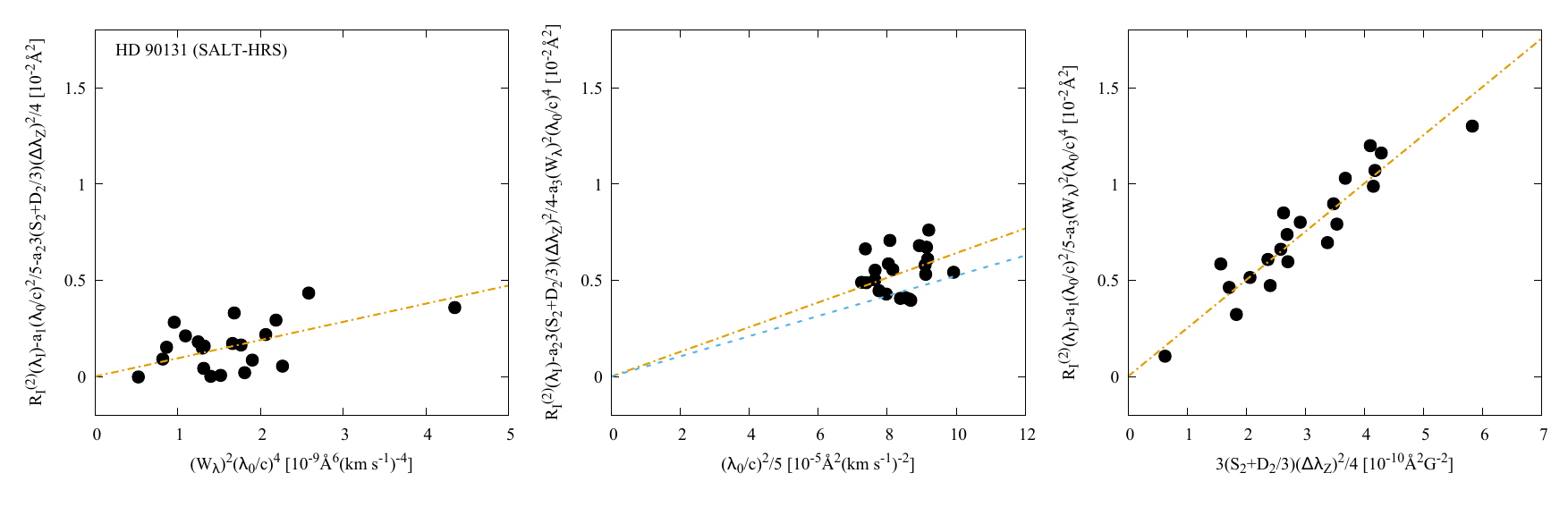}
  \includegraphics[width=\hsize]{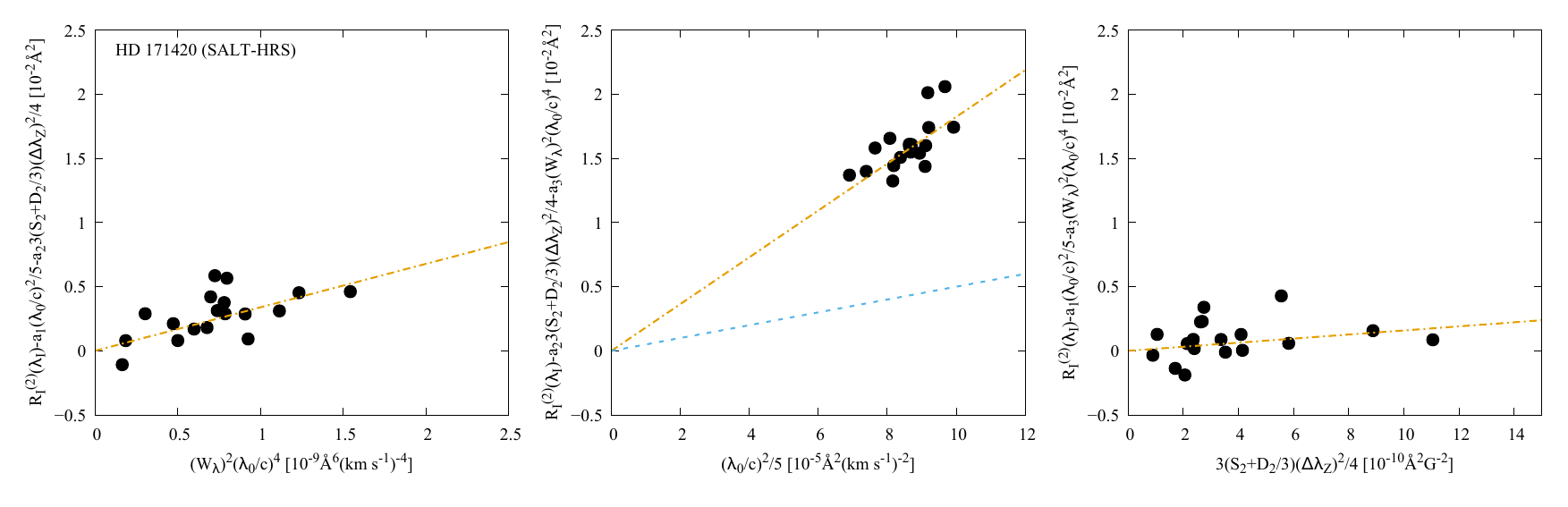}
  \caption{Same as Fig.~\ref{fig:bq394435}, for SALT-HRS spectra 
    of HD~90131 (at HJD 2460127.224) and HD~171420. The magnetic
    broadening is 
    unambiguously seen in the former but its contribution is at most
    marginally significant in the latter. Conversely, HD~171420 shows
    definite rotational broadening, but the Doppler-like term in
    HD~90131 hardly exceeds the combined contributions of the
    intrinsic profile width and the thermal broadening.} 
\label{fig:bq90131}
\end{figure*}

For the HARPS-N spectrum of BD+39~4435, the upper limit of \vsi\ is
below the detection limit (that is, it is negative, which strictly
speaking makes no physical sense, but compatible with a null or very
small value of the projected equatorial velocity within the
uncertainties). This is illustrated in the middle panel of
Fig.~\ref{fig:bq394435} by the sky blue dashed line. In this case, the
slope of the latter, $(a_{\rm inst}+1.474\times10^{-3}\,\Teff)$, is
almost equal to the slope $a_1$ of the best fit line, so that both
lines are almost exactly superimposed onto each other. 

As one may expect, our ability to detect the magnetic and/or
rotational broadening, the precision that can be achieved in the
determination of the value of the mean quadratic magnetic field, and
the significance with which an upper limit of the projected equatorial
velocity can be set depend considerably on the spectral resolution of
the observation. This is illustrated in Fig.~\ref{fig:bq46570}, in
which we compare the analysis of the HARPS-N spectrum and the CAOS
spectrum of BD+46~570. In both, the contribution of magnetic
broadening to the second-order moments of the Stokes $I$ line profiles
is unambiguously seen in the right panels. The greater scatter of the
individual line measurement points about the straight line fit in the
CAOS observations reflects the greater uncertainty of the mean
quadratic field determinations achievable with this instrument
compared with the HARPS-N measurement: with HARPS-N, we derive
$\Bq=(3537\pm148)$\,G; with CAOS, $\Bq=(4029\pm1003)$\,G. Both values
are significant, and they are mutually consistent (although they do
not have to be -- the magnetic field strength could plausibly have
varied between the two epochs of observation). The higher uncertainty
of the value derived with CAOS not only results from the lower S/N of
the spectrum (to which the uncertainty is to first order inversely
proportional) and to the fact that the
line profiles are less well defined at lower resolution. Because of
the limited resolution of CAOS, the number of lines (12) that are
sufficiently free from blending to be used to diagnose the magnetic
field is lower than in the HARPS-N case (20). Previous experience
\citep[for instance,][]{2017A&A...601A..14M} has shown that in
general, using a greater number of diagnostic lines allows one to
achieve a higher precision in magnetic field determinations. In cases
such as BD+46~570, the size of the diagnostic line samples also
contributes to the differences in the uncertainties of the $\Bq$ values
determined with different spectrographs.

 For the HARPS-N spectrum of BD+46~570, one can also see in the left
panel of Fig.~\ref{fig:bq46570} the tight dependence of the intrinsic
part of $\RI$ on the line intensity variable,
$\ew^2\,\lambda_0^4/c^4$. By contrast, the fit of the Doppler-like
term (orange dash-dotted line) is marginally below the line showing
the combination of the instrumental and thermal line broadening (sky
blue dashed line), indicating that, like for BD+39~4435, the
rotational Doppler effect is below the detection limit. This
conclusion is consistent with the fact that the FWHD of the profile of
the Fe~{\sc i}~$\lambda\,5434.5$\,\AA\ line does not significantly
exceed the length of the horizontal bar representing the contributions
to the line width of the  instrumental profile and of the thermal
broadening (see Fig.~\ref{fig:spec5434}).

The conclusion that rotational broadening is below the detection limit
in the HARPS-N spectrum of BD+46~570 seems to
be in contradiction with the result of the analysis of its CAOS
spectrum. Indeed, in the middle panel of the lower row of
Fig.~\ref{fig:bq46570}, the representative points of the inferred
contribution of the first term of the rigt-hand side of
Eq.~(\ref{eq:Bq}) are located well above the intrinsic and thermal
broadening line. Thus rotation appears to make
a significant, though poorly constrained (note the large scatter),
contribution to the line width. This results from the
occurrence of crosstalk between the $a_1$ and $a_3$ terms of the
regression analysis of the CAOS spectrum of BD+46~570. This is
indicated by the lack of dependence of $\RI$ on the equivalent width,
which does not make physical sense.

The possible
occurrence of crosstalk between the $a_1$ and $a_2$ terms in mean
magnetic quadratic field determinations has been discussed in some
detail by \citet{2006A&A...453..699M}, who showed some examples of
instances in which it probably played a role. It results from the fact
that some correlation may exist between the independent variables of
the two terms of interest of the right-hand side of
Eq.~(\ref{eq:Bq}). While this type of crosstalk is the most likely to
present itself in the $\Bq$ determination, the fact that 
\citet{2006A&A...453..699M} did not find any correlation between
$(\ew^2\,\lambda_0^4/c^4)$ and the other two independent variables in the
examples that they considered does not rule out the possibility of
such correlations in other cases. In particular, given the wavelength
depence of the intrinsic contribution to the second-order moment of
the Stokes $I$ line profiles, crosstalk between the
$a_1$ and $a_3$ terms is more likely to occur in the case of moderate
resolution spectra recorded at comparatively low S/N in which the line
equivalent widths span a narrow range of values. The CAOS spectrum of
BD+46~570 fulfils all these conditions. 

This example serves as a warning. Critical evaluation of the results
of the analysis carried out to determine the mean quadratic magnetic
field is required to validate them, as far as possible. This can be done
through consideration of the respective contributions of the three
terms of the right-hand side of Eq.~(\ref{eq:Bq}) using graphs such as
those shown in Figs.~\ref{fig:bq394435} and \ref{fig:bq46570} and
visual inspection of the analysed spectra. In the case of BD+46~570,
one can confirm from the appearance of lines with high magnetic
sensitivity, such as shown in Figs.~\ref{fig:spec6149} and
\ref{fig:spec6336}, that the value of $\Bq$ is almost certainly
determined correctly, despite the crosstalk between the $a_1$ and
$a_3$ terms. 

In Fig.~\ref{fig:bq90131}, we compare the results of the analysis of
the SALT-HRS spectra of HD~90131 and HD~171420. The former is one of
the most strongly magnetic stars of this study, as confirmed by the
well resolved split components of the \Feline\ line. This is reflected
in the top right panel of Fig.~\ref{fig:bq90131} by the slope of the
tight linear dependence of the magnetic part of $\RI$ on the magnetic
sensitivity $(3S_2+D_2)\,\Zeeman^2/4$. By contrast, the representative
points of the Doppler-like contribution to $\RI$ (top centre panel)
hardly lie above the representative line of the intrinsic and thermal
line broadening: rotational broadening is below the detection limit,
consistently with the FWHD of the Fe~{\sc i}~$\lambda\,5434.5$\,\AA\
line as seen in Fig.~\ref{fig:spec5434}. 

The opposite situation prevails for HD~171420: significant rotational
line broadening is observed in the bottom centre panel of
Fig.~\ref{fig:bq90131}, corresponding to an upper limit
$\vsi\lesssim11.3$\,\kms, but the mean quadratic magnetic field is
below the detection limit achievable with SALT-HRS. Here too, the
results of  the analysis are consistent with the observed profiles of
the Fe~{\sc i}~$\lambda\,5434.5$\,\AA\ (Fig.~\ref{fig:spec5434}) and
$\lambda\,6336.8$\,\AA\ (Fig.~\ref{fig:spec6336}) lines. 

\subsubsection{Measurements}
\label{sec:Bqmeas}
In practice, for determination of the mean quadratic magnetic field, a
suitable set of Fe diagnostic lines was built. This set consists of
lines that appear to be (almost) free from blends. Iron lines are
preferred because they are present in the spectra of all Ap stars
within the whole effective temperature range spanned by the ssrAp star
candidates. Furthermore, experience indicates that Fe lines in Ap stars tend to
undergo, at most, low-amplitude intensity variations: in most stars, the
Fe distribution over the stellar surface is either uniform or shows
only moderate inhomogeneitites. This ensures that the value of the
magnetic field moment that is derived from analysis of a 
Fe line sample is as representative as possible of the actual strength
and structure of the field rather than a convolution of the latter
with an unknown elemental abundance distribution. 

In most cases, the diagnostic lines were selected in the wavelength
range $\sim$5400--6700\,\AA. This choice is based on various
considerations. Blueward of 5400\,\AA, in most stars, the increase of
the line density drastically reduces the number of lines that are
sufficiently free from blends, especially in the SALT-HRS, CAOS and FEROS
spectra, which have lower resolution than the HARPS-N spectra.
This wavelength also approximately
coincides with the dividing wavelength between the spectral ranges
covered by the blue and red arms of SALT-HRS. For this instrument, we
prefer to use a sample of diagnostic lines from a single arm, out of
concern that different instrumental effects between one arm and the
other could introduce systematic errors in the $\Bq$ determinations if we
combined diagnostic lines from both arms. On the other end, above
$\sim$6700\,\AA, the stellar line density drops quickly, and telluric
lines abound in some wavelength intervals, so that it becomes difficult
to find suitable diagnostic lines. With such limited prospects, it may
actually be counterproductive to try to select diagnostic lines over a
broader wavelength interval, as the magnetic field may be differently
sampled at the blue and red ends of too long a range \citep[see for
instance][]{2020MNRAS.499.2734J}, thereby increasing the error
of the $\Bq$ determination.

Most stars have low enough temperatures so that Fe~{\sc i} lines can
be best used for determination of their mean quadratic magnetic
field. However, at the hot end of the studied star sample, not enough
lines of this ion are present in the spectra of HD~148330, HD~127304,
HD~17330, HD~236298, HD~11187, HD~67658, and HD~44979. We used Fe~{\sc
  ii} lines instead to diagnose 
$\Bq$ in these stars. Furthermore, in the spectra of the two hottest
ones (HD~67658 and HD~44979), and in the CAOS spectrum of HD~148330,
even the Fe~{\sc ii} line density is too 
low in the $\sim$5400--6700\,\AA\ wavelength interval, so that we
selected the diagnostic lines in the $\sim$4100--5500\,\AA\
range. Lines of Fe~{\sc i} from this range also had to be used for the
analysis of HD~106322, given the weakness of the Fe~{\sc i} lines in
the red part of its spectrum. 

For historical reasons, SALT-HRS spectra of TYC~8912-1407-1 that can
be used for determination of the mean quadratic magnetic field were
recorded at five different epochs. Following
\citet{2017A&A...601A..14M}, we took advantage of these
multi-epoch observations to untangle more efficiently
the magnetic contribution ($a_2$) to the second-order moments of the
unpolarised line profiles from the instrinsic ($a_3$) and Doppler
broadening ($a_1$) terms. Indeed, to first order, contrary to the
former, the latter two can be expected to show little, if any,
rotational modulation. Hence, the values of the fit parameters $a_1$
and $a_3$ can be determined at a higher S/N from the average over the
observed phases of the second-order moments $\RI$ of the diagnostic
lines. By subtracting from the 
$\RI$ values observed at each epoch the fitted intrinsic and Doppler-like
contributions to the multi-epoch $\RI$ average, one can then isolate
the magnetic contribution to the line width at each phase from those
of the other broadening agents. In practice, we applied this method in
the analysis of the SALT-HRS spectra of TYC~8912-1407-1. The reader is
referred to \citet{2017A&A...601A..14M} for the details of the
procedure.

This approach proves particularly effective if the mean quadratic
magnetic field of the studied star indeed shows significant
variability between the various epochs of observations, or at least,
if the magnetic broadening is large enough compared to the other line
broadening terms. The former condition is not fulfilled in our series
of spectra of TYC~8912-1407-1: no $\Bq$ variations are detected over
the time range spanned by these observations. However, the mean
quadratic magnetic field is rather strong, so that both it and the
Doppler-like broadening term can be better constrained by application
of the above-described two-step procedure than through separate
analysis of each of the individual spectra.

By contrast, the application of this two-step method to the ESPaDOnS
archive spectra of HD~127304 and HD~148330 recorded at different
epochs yields physically meaningless results. Indeed, these two stars,
which are not typical Ap stars (see Appendices~\ref{sec:hd_148330} and
\ref{sec:hd_127304}), do not show any 
magnetic broadening of their spectral lines. Since they also have low
projected equatorial velocities, the results of the analysis of the
averages over the different epochs of the $\RI$ moments of the
diagnostic lines suffer from crosstalk, which leads to badly
overestimated magnetic broadening. In this case, the only sensible
option is to analyse each individual spectrum independently from the
others. Doing so for the ESPaDOnS spectra of HD~127304 and HD~148330,
we found that the mean quadratic magnetic field is below the detection
threshold in all of them. The derived upper limits of the projected
equatorial velocity are $\vsi\lesssim7.34$\,\kms\ for HD~127304 and
$\vsi\lesssim10.23$\,\kms\ for HD~148330.

Further consideration of
these observations would be outside the scope
of this paper, since the probable non-Ap nature of HD~127304 and
HD~148330 makes them irrelevant for the study of ssAp stars. Therefore, we
restrict ourselves to presenting in Table~\ref{tab:meas} the results
of our analysis of the CAOS spectra of these two stars, which were
obtained for the specific purpose of this project, but we omit the
detailed data obtained from the ESPaDOnS spectra at individual epochs,
which are without 
interest in the present context. The non-zero but formally
insignificant value of $\Bq$ derived from the CAOS spectrum of
HD~148330 reflects primarily 
the occurrence of crosstalk between the $a_2$ and $a_3$ terms of the
fit of the observed second-order moments of the unpolarised line
profiles by a function of the form given in Eq.~(\ref{eq:Bq}). Indeed,
the derived value of $a_3$ is negative, which is physically
meaningless, as it would indicate an anti-correlation between the
intrinsic width of the spectral lines and their intensity. 

\subsubsection{Results}
\label{sec:results}
We analysed the spectra of 24 of the 25 stars shown in
Figs.~\ref{fig:spec6150_1} to \ref{fig:spec6150_5}. The exception is
the SB2 system TIC~233529061 (HD~174016-7), for which we cannot
untangle the lines of the two components. We also analysed a spectrum
of TIC~80486647 (HD~67658), whose spectral lines are visibly broader than
those of the stars appearing in the figures, but considerably narrower
than those of all other ssrAp star candidates that we observed until
now as part of this project. The latter are fast rotators. It will be
important to understand why they were unduly identified as ssrAp star
candidates in our TESS-based systematic search, but this discussion is
better left to a future study.

For each of the 25 stars studied here, we determined the radial
velocity $v_{\rm r}$, the mean quadratic magnetic field $\Bq$, and an
upper limit of the projected equatorial velocity $(\vsi)_{\rm
  max}$. For the eight stars among them that show (marginal) resolution of
the magnetically split components of the \Feline\ line, we also
derived the value of the mean magnetic field modulus $\Bm$. The
results are presented in Table~\ref{tab:meas}. They are discussed on a
star-to-star basis in Appendix~\ref{sec:individual}.

\begin{figure}
\resizebox{\hsize}{!}{\includegraphics{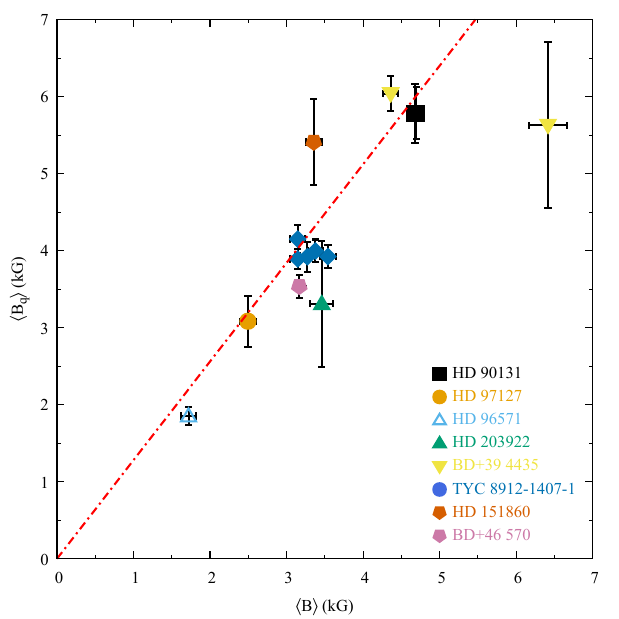}}
\caption{Comparison of the values of the mean magnetic
  field modulus $\Bm$ and the mean quadratic magnetic field $\Bq$
  derived at individual epochs for the eight stars of this study in
  which the magnetically split components of the \Feline\ line are
  resolved. Each of these stars is identified by a different
  symbol/colour combination, as indicated in the figure legend. The
  dashed-dotted red line, whose slope is 1.28, represents the relation
  between $\Bq$ and $\Bm$, as derived by \citet{2017A&A...601A..14M}
  (see text for details).}
\label{fig:Bq_vs_B}
\end{figure}

For the eight stars in which the \Feline\ line is magnetically
resolved, Fig.~\ref{fig:Bq_vs_B} illustrates the relation between the
derived values of $\Bm$ and $\Bq$. This figure should be compared with
Fig.~12 of \citet{2017A&A...601A..14M}, from which this author
established the existence of a linear relation between the two
magnetic field moments, with a slope of 1.28 determined through
least-squares fitting. This is the slope of the blue dashed line of
Fig.~12 of \citet{2017A&A...601A..14M} and of the red dashed-dotted
line plotted in Fig.~\ref{fig:Bq_vs_B}. Each data point appearing in
Fig.~12 of \citet{2017A&A...601A..14M} corresponds to the average of
the values of $\Bm$ and $\Bq$ derived from spectra recorded at
different sets of epochs. Thus, the two field moments do not sample
their respective phase variation curves in the same way. By contrast,
each data point in Fig.~\ref{fig:Bq_vs_B} corresponds to the
determination of both $\Bm$ and $\Bq$ from a spectrum recorded at a
single epoch. This figure demonstrates that the same correlation that
was found from consideration of epoch-averaged data is also valid on
an epoch-by-epoch basis. This result is established here for the first
time. Its validity is not questioned by the only point in the figure
that departs significantly from the linear trend: the low S/N of the
CAOS spectrum of BD+39~4435 to which it corresponds implies that the
uncertainties affecting the derived $\Bm$ and $\Bq$ values are
exceptionally large, possibly even more than the formal error bars
represent.  

\section{Discussion}
\label{sec:disc}

\subsection{Rotation and chemical peculiarity}
\label{sec:rotation}
In this study, we analysed spectra of 27 ssrAp star candidates that
had been identified in \citetalias{2020A&A...639A..31M} and
\citetalias{2022A&A...660A..70M}, focussing on their rotation, their
magnetic fields, and their binarity. The results are summarised in
Table~\ref{tab:properties}. Eighteen of these 27 stars are typical Ap
stars with lines sharp enough to be consistent with moderately to very
long rotation periods. The upper limit of the projected equatorial
velocity and TESS photometric variability constraints that we derived
for eleven of them indicate that they are 
almost certainly ssrAp stars; three more could either have
$\Prot>50$\,d (the defining criterion of super-slow rotation) or have
moderately long periods ($20\,{\rm d}\lesssim\Prot\lesssim50$\,d). The
somewhat higher upper limits of \vsi\ and/or the detection of TESS
variations for the 
remaining four seem to rule out super-slow rotation but are
suggestive of moderately long periods. In fact, for three stars, the
actual value of the period was determined: the ssrAp star HD~340577
($\Prot=116\fd7$), and two stars with moderately long periods:
HD~236298 ($\Prot\sim24\fd3$) and HD~7410 ($\Prot=37\fd08$). For the
latter, the rotation period was known prior to this study, but this 
had been overlooked in \citetalias{2022A&A...660A..70M}.

\begin{table}
  \scriptsize
  \caption{Properties of the studied stars.}
  \label{tab:properties}
  \centering
  \begin{tabular}{ccccc}
    \hline\hline\\[-4pt]
    TIC&Other ID&Resolved&Magnetic&Binarity\\
       &        &lines   &field   &\\[4pt]
    \hline\\[-4pt]
    \multicolumn{5}{c}{ssrAp stars}\\[4pt]
    \hline\\[-4pt]
 73765625&HD~90131       &x&x&   \\
 77038207&HD~96003       & &x&SB1\\
154786038&HD~96571       &x&x&SB1\\ 
165446000&BD+39 4435     &x&x&   \\
167695608&TYC 8912-1407-1&x&x&   \\
170419024&HD~151860      &x&x&SB1\\
202899762&BD+46 570      &x&x&SB1\\
233539061&HD~174017      & & &SB2\\ 
298197561&HD~340577      & &x &   \\ 
301918605&HD~17330       & &x&SB1\\
444094235&HD~85284       & &x&SB1\\[4pt]
    \hline\\[-4pt]
    \multicolumn{5}{c}{Long period Ap stars}\\[4pt]
    \hline\\[-4pt]
 77128654&HD 97127       &x&x&   \\
352787151&BD+35~5094     & & &   \\
468507699&HD~206977      & &x&   \\[4pt]
    \hline\\[-4pt]
    \multicolumn{5}{c}{Ap stars with moderately long periods}\\[4pt]
    \hline\\[-4pt]
163801263&HD~203922      &x&x&   \\
301946105&HD~7410        & &x&SB1\\
347202840&HD~236298      & &x&   \\ 
461161123&HD~95811       & & &   \\[4pt]
    \hline\\[-4pt]
    \multicolumn{5}{c}{Short period Ap star}\\[4pt]
    \hline\\[-4pt]
403625657&HD~11187      & &x&   \\[4pt]
    \hline\\[-4pt]
    \multicolumn{5}{c}{Stars that are not typical Ap stars}\\[4pt]
    \hline\\[-4pt]
 80486647&HD~67658       & & &   \\
 81554699&HD~97132       & & &SB2\\
124988213&HD~44979       & & &   \\
206461701&HD~209364      & & &   \\
207468665&HD~148330      & & &   \\
286965228&HD~127304      & & &SB(2?)\\
291561579&HD~171420      & & &   \\
334505323&HD~106322      & & &   \\[4pt]
    \hline
  \end{tabular}
    \tablefoot{The table is divided in
    five segments. The top one includes those stars for which the
    analysis presented in Appendix~\ref{sec:individual} indicates that
    they are in all probability ssrAp stars. The spectroscopic
    properties of the stars listed in the second segment are
    consistent with their having either moderately or extremely long
    rotation periods. The rotation periods of the stars from the third
  segment definitely or very probably are moderately long ($20\,{\rm
    d}\lesssim\Prot\lesssim50$\,d). The fourth segment contains the
  only magnetic Ap star of this study to have a rotation period of a
  few days. In the bottom segment, we list those stars whose spectrum
  does not look like those of typical Ap stars. In each segment, the
  stars are listed in order of increasing TIC number (Col.~1); an
  alternative ID is given is Col.~2. The visible range spectrum of
  those stars for which `x' appears in Col.~3 shows resolved
  magnetically split lines; the `x' flag in Col.~4 identifies those
  stars for which a significant detection of the magnetic field was
  achieved and at least one field moment could be determined, either
  in this study or in the literature (see text). Finally, Col.~5
  indicates which 
  stars are spectroscopic binaries, distinguishing to the extent possible
  the single-lined (SB1) and double-lined systems.}
  \end{table}

One more typical Ap star, HD~11187, which also shows rather sharp
spectral lines, undergoes variations with a period of the order of a
few days, whose exact value remains ambiguous. This was another
oversight of our original search for ssrAp star candidates. The
inclination of the rotation axis of this star to the line of sight
must be low. Such configurations are expected to occur for a small
fraction of the ssrAp star candidates \citepalias[see][for
details]{2020A&A...639A..31M}. The presence of one such star in our
sample is compatible with this expectation. 

None of the remaining stars of Table~\ref{tab:properties} has a typical
Ap spectrum. While their exact spectral classification remains to be
definitely established, their rate of occurrence among the ssrAp star
candidates from our TESS-based survey suggests that
misclassification is one of the main causes responsible for spurious
identification of such candidates. This strengthens the suspicion
expressed in \citetalias{2024A&A...683A.227M} that Ap star lists, such
as \citet{2009A&A...498..961R}, tend to be significantly contaminated
by spectral classification errors (even for stars that are not flagged
as having uncertain chemical peculiarities), and justifies the care
taken in 
this paper to carry out a critical evaluation of the information
available about each ssrAp star candidate to ensure as much as
possible that it was indeed a bona fide Ap star. However, this does
not imply that all TESS-based ssrAp stars candidates for which
follow-up spectroscopy indicates fast rotation are not Ap
stars. Consideration of the stars of this group is outside the scope
of the present study, but the confirmation of their spectral
classification will be an important part of a next step of this
project. 

\subsection{Binarity}
\label{sec:binarity}
Eight of the 18 slowly rotating Ap stars of Table~\ref{tab:properties}
appear to be spectroscopic binaries. This represents a fraction of
44\%, which within the limits of small number statistics, is
consistent with the 51\% rate of occurrence of binarity among the Ap
stars with resolved magnetically split lines studied by
\citet{2017A&A...601A..14M}. The binarity of three of these stars,
HD~17330, HD~96003 and HD~174017, had already been mentioned in the
literature. To the best of our knowledge, the variability of the
radial velocity of the other five is reported here for the first
time. On the other hand, the HD~174016-7 system represents one of the
few known cases of 
an Ap star in a double-lined spectroscopic binary; the seven other
slowly rotating Ap binaries listed here appear to be single-lined. It
will be interesting to study these systems in detail, to compare their
distribution in a diagramme showing the rotation period against the
orbital period with that of the other spectroscopic binaries
containing a slowly rotating Ap star shown in Fig.~16 of 
\citet{2017A&A...601A..14M}.

Among the nine stars of Table~\ref{tab:properties} that do not have
typical Ap star spectra, two are also binaries. One, HD~97132, is a
particularly remarkable SB2 system, whose components, which show very
sharp lines, are Am stars that appear almost identical. We could not
find any previous 
mention of this double-lined spectrum in the literature.
By contrast, definite differences exist
between the published values of the radial velocity of HD~127304,
whose variability 
is fully confirmed by our measurements. The existing measurements
were obtained randomly; no systematic attempt was made to characterise
the orbit. We could not confirm the suggestion by
\citet{1989A&A...209..233R} that HD~127304 is probably a SB2 system,
but we cannot rule out this possibility. 

\subsection{Magnetic field distribution}
\label{sec:Bdist}
In this study, we discovered eight new Ap stars in which the components of
the \Feline\ line are resolved or marginally resolved into their
magnetically split components and we determined their mean
magnetic field modulus. We were also able to derive the value of the
mean quadratic magnetic field for five additional slowly rotating
stars for which this field moment had never been
measured. Furthermore, we identified four stars that
have a low projected equatorial velocity compatible with long rotation
periods in which $\Bq$ is below the detection threshold with the
currently available spectra. For these, we can estimate upper limits of
the field strength. A lower limit can also be set for two of them,
based on mean longitudinal magnetic field measurements from the
literature. One last star, HD~174016-7, appears to show line widths
compatible with a very long rotation period, but it belongs to a SB2
system. At the epoch of observation, the lines of the two components were
overlapping, so that it was impossible to constrain the magnetic field
of the Ap star. 

Thus, we have obtained new magnetic field contraints for 17
stars. Ten of them must with a very high probability be ssrAp
stars; the other seven may have moderately long rotation periods. At
this stage of our study, this distinction has little importance. As
explained in \citetalias{2024A&A...683A.227M}, including stars with
$20\,\mathrm{d}\lesssim P\lesssim50$\,d in the sample of ssrAp stars
used to study the distribution of the magnetic field strengths 
should have very little impact on the statistical conclusions that can
be drawn. We can update the knowledge of this distribution that was
achieved in \citetalias{2024A&A...683A.227M} by including the new
magnetic field determinations for the 17 stars specified above. Of
these 17 stars, three were already part of the sample of
\citetalias{2024A&A...683A.227M}, but only on the basis of mean
longitudinal magnetic field measurements. We can characterise the
magnetic field strength of two of them in a much more meaningful way
by using instead 
the $\Bm$ and $\Bq$ values derived here. In this way, we add 14 stars
to the sample of ssrAp star candidates for which the magnetic field is
constrained that was considered in \citetalias{2024A&A...683A.227M}. The
latter was incorrectly stated to include 70 stars while its size
actually was 71 stars, as can be inferred from consideration of Fig.~6 
of \citetalias{2024A&A...683A.227M}. However, one of these 71 stars,
HD~11187, was reported here (Appendix~\ref{sec:hd_11187}) to be a fast
rotator, and we argued that HD~148330 is not a typical Ap star
(Appendix~\ref{sec:hd_148330}). Accordingly, these two stars were
excluded from the present sample, which now contains 83 stars.  

\begin{figure}
\resizebox{\hsize}{!}{\includegraphics{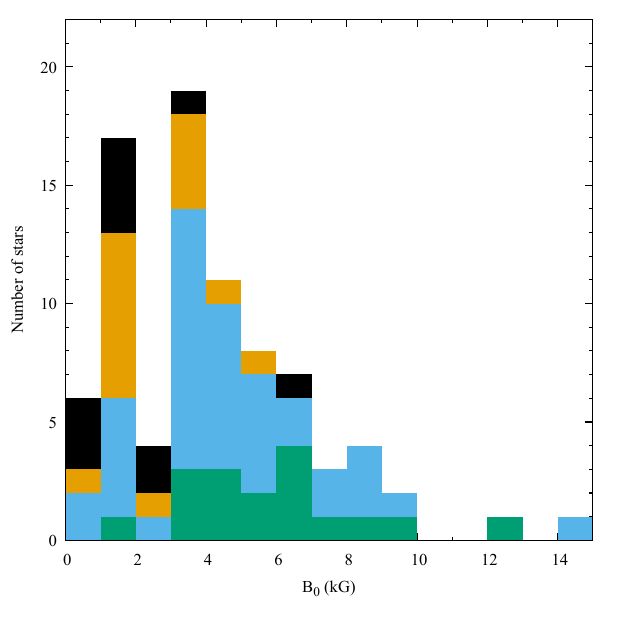}}
\caption{Distribution of the phase-averaged magnetic field strength
  $B_0$ for the long-period Ap stars 
  of Table~1 of \citetalias{2020A&A...639A..31M}, of Table~A.1 of
  \citetalias{2022A&A...660A..70M}, of Tables~1--3 of
  \citetalias{2024A&A...683A.227M}, and for which new $\Bm$ or $\Bq$
  determinations were obtained in this study. The green, sky blue and orange 
  parts of the histogram corresponds to the stars for which
  measurements of the mean magnetic field modulus or of the mean
  quadratic magnetic field are available; for the remaining stars
  (black part of the histogram), a
  lower limit of $B_0$ was inferred from the existing mean
  longitudinal magnetic field measurements. Orange identifies the
  stars for which the $\Bm$ and $\Bq$ measurements of this paper
  were used. Green distinguishes the
  known ssrAp stars that were not identified as ssrAp star candidates
  on the basis of our TESS-based photometric survey. Sky blue is
    used for the previously
  confirmed ssrAp star candidates identified with TESS with $\Bm$ and
  $\Bq$ values from earlier studies.} 
\label{fig:bavhist}
\end{figure}

Ideally, the statistical study of the distribution of the magnetic
field strengths in the slowly rotating Ap stars should be based on
field moment values averaged over a stellar rotation cycle. While this
was possible for a rather small fraction of the stars studied in
\citetalias{2024A&A...683A.227M}, here, there is only one star,
TYC~8912-1407-1, for which
more than a couple of magnetic measurements could be obtained, and
even then, the resulting values are unlikely to sample the variation
cycle adequately. Accordingly, the adopted field moment values on
which this analysis is based are in most cases snapshots of certain
rotation phases rather than rotation-averaged measurements. The reasons
why this is non-critical in the present context, given the typically
low amplitudes of variability of $\Bm$ and $\Bq$, were explained in
\citetalias{2024A&A...683A.227M}.

We denote by $B_0$ the reference values on which the statistical study
of the distribution of the magnetic field strengths of the ssrAp stars
and of the Ap stars with moderately long rotation periods is
based. The reader is referred to \citetalias{2024A&A...683A.227M} for
details about the definition of $B_0$ according to the different types of
magnetic measurements that were available in the previously studied
sample. As far as the new magnetic field determinations of this paper
are concerned, the following values were used. For those stars in
which the value of the mean magnetic field modulus could be derived,
$B_0$ was set to this value. When $\Bm$ measurements at several
epochs had been obtained, their average was adopted. In the case of
the stars that do not show resolved magnetically split lines for which
the mean quadratic magnetic field could be determined, we converted it
using the relation $B_0=\Bq/1.28$ (see Sect.~\ref{sec:results}). Only
the formally significant values of $\Bq$ were considered. In
three stars for 
which both a CAOS spectrum and a HARPS-N spectrum were available and
yielded formally significant $\Bq$ values, the adopted
$\Bq$ value was determined from the latter, given its better spectral
resolution. For two stars for which the $\Bq$ value was
below the detection threshold from a CAOS spectrum (HD~17330,
HD~340577), we used the lower limit of $B_0=3\,\Bzrms$ derived from mean
longitudinal field measurements from the literature. (The root-mean
square longitudinal magnetic field 
$\Bzrms$, defined by \citet{1993A&A...269..355B}, is the quadratic
mean over the observed phases of the mean longitudinal magnetic field
values $\Bz$.) Magnetic
broadening was also below the detection threshold in the analyses of a
CAOS spectrum of BD+35~5094 and of
a SALT-HRS spectrum of HD~95811. For these stars, we set
$B_0=1.4$\,kG. By comparison with the other stars of this
study, we estimate that this value represents a reasonable
approximation of the upper limit of $B_0$ in such cases.

Figure~\ref{fig:bavhist} shows the distribution of $B_0$ after
addition of the new magnetic field measurements 
presented in this paper to those considered in
\citetalias{2024A&A...683A.227M}. It should be compared to Fig.~6 of
the latter. It should be noted that none of the new $B_0$ values of
this study (identified by the orange colour) exceeds 6~kG. This
strengthens the views that previous 
studies tended to overlook ssrAp stars with weak to moderate magnetic
fields. While the longest rotation periods ($\Prot\gtrsim150$\,d)
  are not found among the most strongly magnetic Ap stars
  ($B_0\gtrsim7.5$\,kG; \citealt{2017A&A...601A..14M}), the previously
  observed trend for 
super-slow rotation to occur less frequently in the most weakly
magnetic Ap stars ($B_0\lesssim2$\,kG) than in the Ap stars with
moderate magnetic field strengths ($3\,{\rm kG}\lesssim
B_0\lesssim7.5$\,kG) 
remains visible in Fig.~\ref{fig:bavhist}. As argued in detail in
\citetalias{2022A&A...660A..70M}, this effect appears particularly
significant if one compares the $B_0$ distribution in slowly rotating
Ap stars with the distribution of the rms longitudinal magnetic field
in all Ap stars, which peaks abruptly well below 1\,kG
\citep{2009MNRAS.394.1338B}.

\begin{figure}
\resizebox{\hsize}{!}{\includegraphics{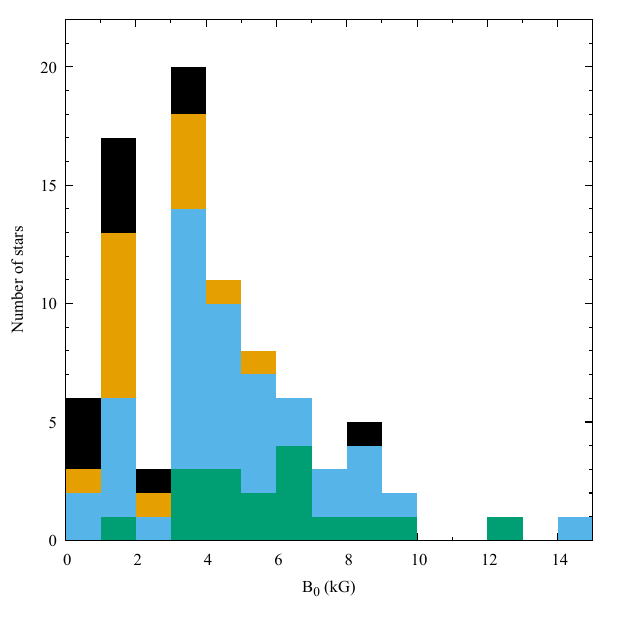}}
\caption{Same as Fig.~\ref{fig:bavhist} but using
  the relation $B_0=3.86\,\Bzrms$ to compute the values of $B_0$ for those
  stars for which the only magnetic field moment that has been
  determined until now is the mean longitudinal field. These stars are
  still represented by the black parts of the histogram; the meaning
  of the other colours is unchanged.}
\label{fig:bavhistok}
\end{figure}

Figure~\ref{fig:bavhist} also appears to confirm and strengthen the
reality of the existence of a gap, or deficiency, of slowly rotating
Ap stars with (phase-averaged) field strengths between $\sim$2 and
$\sim$3\,kG. By contrast, the distribution of
  \citet{2009MNRAS.394.1338B} shows a strictly monotonic decrease 
  from the lowest $\Bzrms$ values to the highest ones, without any
  hint of a gap or a discontinuity. While the gap in the distribution
  of the $B_0$ values in the ssrAp stars appears to be in the
  interval from $\sim$2.0 to  
  $\sim$3.0\,kG, this does not imply that it is an artefact of the chosen 
  binning. As the total number of ssrAp stars for which magnetic
  measurements become available increases, the number of such stars with a
  field strength between $\sim$2.0 and $\sim$3.0 kG hardly grows while the
  neighbouring bins become more and more populated. This is a genuine
  observational result, not the consequence of any selection
  effect. The targets that are observed are just the ssrAp stars that
  are identified; there is no other criterion applied in the
  definition of the sample. Admittedly, the statistics are limited,
  but the more they grow, the firmer are the conclusions that can be
  drawn from them. A more detailed comparison of the field strength
  distribution revealed by our study of the ssrAp stars with the
  distribution for all Ap stars as established by
  \citet{2009MNRAS.394.1338B} was given in
  \citetalias{2022A&A...660A..70M}. In this reference, further
  supporting evidence based on the study of a volume-limited sample
  \citep{2019MNRAS.483.3127S} was also presented. The new data of the
  present study confirm and strengthen the conclusions reached in
 \citetalias{2022A&A...660A..70M}.  

The origin of the gap in the $B_0$ value distribution in the ssrAp
stars remains unknown, but is plausibly
related with the existence of two channels of formation of Ap stars,
with different processes responsible for weakly and (more) strongly
magnetic stars. Or possibly, if all Ap stars form in the same way,
there may be two rotational braking processes allowing them to achieve
slow rotation, one below $\sim$2~kG and the other above $\sim$3~kG,
neither of which is efficient in the $\sim$2--3~kG field strength
interval. The observational evidence available until now is
insufficient to draw definitive conclusions. Further characterisation
of the magnetic field strength distribution as well as of possible
correlations between magnetic and other properties are needed to
provide the basis for further theoretical developments.

The magnetic field values $B_0$ used to build Fig.~\ref{fig:bavhist} were
defined in such a way as to be consistent with the data presented in
Fig.~6 of \citetalias{2024A&A...683A.227M}. In particular, for those
stars for which only the mean longitudinal magnetic field could be
determined until now, lower limits of $B_0$ were used. This lower
limit, $B_0=3\,\Bzrms$, is empirical. It is based on the consideration
of Fig.~8 of \citet{2017A&A...601A..14M}, which shows that with almost
no exception, for measurements of $\Bm$ and $\Bz$ distributed over a
stellar rotation cycle or obtained at a number of random epochs
(generally different for the two magnetic field moments), one has
$\Bzrms\geq\Bm_{\rm av}/3$ ($\Bm_{\rm av}$ is the average of the
individual $\Bm$ values over the set of measurements under
consideration).

Recently, \citet{2024A&A...686A.189K} carried out numerical
simulations in an attempt to establish, for stars having a dipolar
magnetic field, statistical relationships between the polar field
strength and the observable field moments $\Bz$ and $\Bm$. With
respect to the latter two, he derived the following conversion
equation:
\begin{equation}
  \Bm_{\rm av}=3.86^{+1.04}_{-0.48}\,\Bzrms\,.
    \label{eq:Bav_vs_Bzrms}
  \end{equation}
Taking this result into account, an alternative version of the
distribution of the magnetic field strength $B_0$ in the long-period
Ap stars was generated, adopting the value $B_0=3.86\,\Bzrms$ for
those stars for which the only available magnetic field measurements
are those of $\Bz$. This revised distribution is shown in
Fig.~\ref{fig:bavhistok}. The interest of this representation over
that of Fig.~\ref{fig:bavhist} is that, on a statistical basis, the
values of $B_0$ used for those stars for which only $\Bz$ measurements
are available should be more consistent with those derived from
$\Bm$ and $\Bq$ measurements. This difference concerns 11 stars out of
a total sample of 83, so that its impact on the overall aspect of the
figure is unavoidably limited. Nevertheless, the similarity between
both figures is noteworthy, especially with respect to the
absence of periods $\Prot\gtrsim150$\,d in stars with
  $B_0\gtrsim7.5$\,kG, the lower rate of occurrence of super-slow
  rotation for field strengths $B_0\lesssim2$\,kG than in the range
  $3\,{\rm kG}\lesssim B_0\lesssim7.5$\,kG, and the deficiency of
slowly rotating Ap stars with (phase-averaged) field strengths between
$\sim$2 and $\sim$3\,kG. The difference between the smooth
  behaviour of the $\Bzrms$ distribution from
  \cite{2009MNRAS.394.1338B} and the apparently bimodal distribution of
  Figs.~\ref{fig:bavhist} and \ref{fig:bavhistok} for the ssrAp stars
  does not depend in 
  any way on the value of the factor adopted to convert $\Bzrms$ to
  $B_0$.

One thing that we do not know for sure is whether the
observed gap at the low end of the magnetic field strength
distribution is also present in faster rotating stars. Observations
such as those presented here are ill-suited to address this question,
because resolved magnetically split lines can only be observed in the
stars that have the lowest projected equatorial velocity and the mean
quadratic magnetic field becomes increasingly difficult to untangle
from Doppler broadening as \vsi\ becomes increasingly
large. High-resolution infrared (IR) spectroscopy would lend itself better
to study the low end of the magnetic field strength distribution in
Ap stars that rotate faster than those considered here. Indeed, since
the Zeeman effect increases quadratically with wavelength, as opposed
to the linear dependence of the Doppler effect, at equal spectral
resolution, for a given field strength, the magnetic splitting can be
observed in shorter period Ap stars in the IR than in the
visible. The value of high resolution IR spectroscopy for
determination of Ap star magnetic fields has been illustrated by
\citet{2003A&A...409.1055L}. More recently,
\citet{2019ApJ...873L...5C} have observed resolved 
magnetically split lines in the H-band spectrum of more than 150 Ap
stars, most of which have projected equatorial velocities greater than those
of the Ap stars in which the \Feline\ line is resolved into its
components. But the lowest value of the mean magnetic field modulus
that they were able to measure is $\Bm=3.6$~kG, because of the
limited resolution of these IR observations. Higher resolution IR
spectroscopy is required to decide if the $\sim$2--3\,kG gap in the
magnetic field distribution is limited to the most slowly rotating
Ap stars or if it also exists for shorter rotation periods. 

\section{Conclusions}
\label{sec:conc}
Overall, the outcome of the present study validates the technique
proposed in \citetalias{2020A&A...639A..31M} and subsequently applied
in \citetalias{2022A&A...660A..70M} and
\citetalias{2024A&A...683A.227M} to carry out a systematic search for
ssrAp stars from the analysis of TESS photometric observations. It
also sheds light on its limitations.

As seen in Sect.~\ref{sec:rotation}, 18 of the 27 stars that 
were analysed in this work are definite Ap stars whose spectral lines
are sharp enough to be compatible with long rotation periods. An
additional Ap star, HD~11187, must have a rather short rotation
period, of the order of days, but also shows moderate broadening of
its spectral lines: its lack of photometric variability over the
duration of a TESS sector can plausibly be assigned to the low
inclination of its rotation axis to the line of sight. Given the size
of the studied sample, it is not surprising that at least one star may
show near-alignment of the rotation axis with the line of sight.

The lines of the remaining eight stars tend to be somewhat broader in
average. More importantly, none of these stars shows a typical Ap star
spectrum. We cannot rule out at this stage the occurrence of mild
peculiarities in some of them, but at first sight, the available
evidence rather suggests that most are stars that were misclassified
as Ap. Actually, one of them, HD~97132, is a SB2 system consisting of
two nearly identical, very sharp-lined Am components. Checking the
exact nature of all  
of these stars is better left to a future study including also those
TESS-based ssrAp star candidates whose spectra show broad to very
broad spectral lines. However, it appears inescapable that one of the
main reasons for misidentification of stars showing no TESS
photometric variability as ssrAp star candidates is that the published
Ap star lists are rather severely contaminated by misclassified stars
of other types, such as normal A stars, Am stars, HgMn stars, or
F\,str~$\lambda\,4077$ stars. This suspicion was already raised in
\citetalias{2024A&A...683A.227M}; the evidence presented here
strengthens it.

The other difficulty that has already been discussed in
\citetalias{2024A&A...683A.227M} is to distinguish the ssrAp stars
from the Ap stars that have moderately long rotation periods
($20\,{\rm d}\lesssim\Prot\lesssim50$\,d). This difficulty arises from
the fact that in the process of correcting the data for
instrumental effects, the TESS pipeline may often also remove actual
stellar variations occurring on timescales of the same order as the
length of a TESS sector. Spectroscopic observations obtained at one or
two epochs are in most cases insufficient to resolve the
ambiguity. Measurements of \vsi\ only yield lower limits of the
stellar rotation rate. Among the stars studied here, HD~95811 and
HD~203922 show line broadening that appears inconsistent with
super-slow rotation and rather seems indicative of a moderately long
rotation period, even though analysis of the TESS raw (SAP) data does
not show convincing evidence of variability. Analysis of the SAP data
also reveals the occurrence of photometric variations likely due to
rotation with moderately long periods for two more stars of our
sample, HD~7410 ($\Prot=37\fd08$) and HD~236298
($\Prot\sim24\fd3$). However, periodic variability typical of rotation
was not detected in either the SAP or PDCSAP (pipeline reduced) TESS
data for HD~11187, despite the definite rotational broadening of its
spectral lines and the observed variability of its mean longitudinal
magnetic field on timescales of days. 

The remaining 14 typical Ap stars of this study have lines sharp
enough to be compatible with super-slow rotation ($\Prot>50$\,d), but
some may have somewhat shorter periods, even though analysis of the
SAP data did not show convincing evidence of rotational variability
for any of them. The difficulties of this analysis are abundantly
illustrated in Appendix~\ref{sec:individual}. Even though for many of
these stars, TESS observations have become available for more sectors
than there were at the time when the initial systematic search for
ssrAp star 
candidates was carried out on the basis of the PDCSAP data
(\citetalias{2020A&A...639A..31M} and 
\citetalias{2022A&A...660A..70M}), low
amplitude stellar variations standing out of the instrumental effects
on timescales of a few weeks remain elusive. This does not imply
that none of these stars may have a moderately long rotation
period. However, for a random distribution of the inclination angles
of the rotation axes to the line of sight, one should expect the vast
majority of the Ap stars with lines sharp enough for consistency with
periods longer than 50 days to be indeed super-slow rotators.

In other words, the 14 sharp-lined Ap stars identified in this work
(listed under `ssrAp stars' or `Long period Ap stars' in
Table~\ref{tab:properties}) fufil the necesary condition to be ssrAp
stars. However, observation at more epochs are needed for final
confirmation that their rotation periods are indeed longer than
50\,d. Photometry is poorly suited to this effect, mostly because of
the difficulty of ensuring sufficient long-term stability in the
measurements so as to avoid confusion between variations of stellar
origin and instrumental drifts. Line-intensity variations may provide
useful constraints in some Ap stars, but their occurrence is not
ubiquitous, and in some cases, the variation amplitudes may be too low
to allow definite conclusions to be drawn. As a rule, consideration of
the magnetic field behaviour is the best-suited approach to study the
variability of the ssrAp stars \citep{2020pase.conf...35M}.

The mean quadratic magnetic field $\Bq$ and, in the most favourable
cases (stars with strong enough magnetic fields observed at
sufficiently high spectral resolution), the mean magnetic field
modulus $\Bm$ can be determined from the analysis of spectra recorded
in natural light. However, the relative amplitude of variation (that
is, the ratio of the maximum value to the minimum) of
these field moments only seldom exceeds 1.3 (for $\Bm$; see
\citealt{2017A&A...601A..14M}), so that high precision measurements
are required to characterise the variations in a meaningful
manner. This is more easily achieved for $\Bm$, whose values range
from $\sim$2\,kG up and can frequently be determined with
uncertainties in the $\sim$30\,--\,$\sim$50\,G range, and only very rarely
greater than $\sim$100\,G. The precisions with which the values of
$\Bq$ can be derived is less good. Values of the internal error as low
as 120\,G as reported here for the analysis of the HARPS-N spectrum
HD~96571 are exceptionally good; typical values are more often of the
order of a few hundred gauss. On the other hand, the orders of
magnitude of the mean quadratic magnetic field and of the mean
magnetic field modulus are similar to each other -- the former is, in
average, ~30\% greater than the latter (see Sect.~\ref{sec:results}
and Fig.~\ref{fig:Bq_vs_B}).

Thus, the relative precision of the $\Bm$
values is in general considerably better than for the $\Bq$ values:
consideration of the former, when possible, represents a better way to
constrain the magnetic variations of ssrAp stars. However, $\Bq$
presents the advantage that it can also be measured in stars that do
not have resolved magnetically split lines, so that in principle, it
gives access to weaker magnetic fields. This is illustrated by the fact
that we achieved a signficant detection of a mean quadratic magnetic
field as low as $\Bq=1110$\,G through analysis of a HARPS-N spectrum of
HD~96003. As argued in Sect.~\ref{sec:results}, this suggests that the
mean magnetic field of this star, whose lines are unresolved, should
be of the order of $\Bm\sim870$\,G. This is about half of the lowest
value of the mean magnetic field modulus that we actually determined,
$\Bm=1720$\,G for a star showing resolved lines. The latter, in turn,
matches well the estimated lower 
limit of the $\Bm$ values that can be derived from observations
performed in the visible range \citep{1997A&AS..123..353M}.

An aspect that the determinations of the mean magnetic field modulus
and of the mean quadratic magnetic field have in common is that both
strongly benefit from the usage of observations recorded at the
highest spectral resolution. Determining the lowest values of $\Bm$
and achieving the lowest uncertainties of $\Bq$ are only feasible
at resolutions such as that of HARPS-N. They are beyond reach for the
lower resolution of the spectra obtained with the other instruments
used in this study.

An even better way to constrain the timescales over which variability
occurs in (very) slowly rotating Ap stars is the consideration of
their mean longitudinal magnetic fields $\Bz$. Not only is the
relative amplitude of variation of the latter often considerably
greater than for $\Bm$ or $\Bq$ \citep[see Fig.~9
of][]{2017A&A...601A..14M}, but also the occurrence of sign
reversal in numerous stars makes unambiguous period determinations
more straightforward. However, the determination of
this field moment requires spectropolarimetric observations. Suitable
instruments are much less numerous than natural light spectrographs,
and they tend to be in higher demand. Securing observing time on them
for multi-epoch observations potentially to be spread over time
intervals of years, which do not lend themselves well to rapid
publications of significant results, may represent a challenge on
several levels. Convincing the evaluators of observing
proposals that the expected scientific return is sufficient to
place the project above the telescope oversubscription cutoff line in
the peer 
review process may not be enough. Observatory policy barriers that do
not affect short-term programmes may also need to be overcome.

An additional advantage of spectropolarimetric observations is their
ability to detect and constrain much weaker magnetic fields than those
that can be diagnosed from spectra recorded in 
natural light. For several stars in this study for which the mean
quadratic magnetic field was below the detection threshold, the
literature reports successful detections of the mean longitudinal
magnetic field: HD~17330, HD~340577, and HD~11187 (which is not a
ssrAp star). Even though $\Bz$ measurements are affected by
additional ambiguity given their high sensitivity to the geometry of
the observations, they represent the best approach for
characterisation of the lowest end of the magnetic field strength
distribution. 

Multi-epoch observations with a view to constraining rotation from
magnetic field measurements represent a genuine follow-up of the work
presented here. It will also be of interest to try to constrain the
strengths of the magnetic fields that were below the detection
threshold with the observations obtained until now, either by
recording spectra in natural light at higher resolution (and in some
cases, S/N) than those that could be analysed until now, or by
performing Stokes $I$ and $V$ spectropolarimetric observations with a
view to measuring the mean longitudinal magnetic field. More
generally, the ultimate goal must be to achieve the most detailed
characterisation of the period and magnetic field strength
distribution for the ssrAp stars, and of the possible correlations
between the rotational and magnetic properties of these stars. A
couple of specific questions to be answered have already been
identified. Is the rate of occurrence of super-slow rotation lower
among weakly magnetic stars than among strongly magnetic ones? Is
there a gap between $\sim$2\,kG and $\sim$3\,kG in the
distribution of the magnetic field strengths of the ssrAp stars? The
elements of information gathered until now support these views, but
further confirmation of their validity is needed to draw definitive
conclusions. On the other hand, as our knowledge of the ssrAp stars
improves, more questions will certainly arise.

The above-described follow-up represents one of our main objectives for
the next steps of this project. In parallel, we continue to acquire
spectroscopy for those ssrAp star candidates from
\citetalias{2020A&A...639A..31M} to \citetalias{2024A&A...683A.227M}
for which no information about the rotation period or the projected
equatorial velocity is available until now. We plan to analyse these
observations in the same way as the data considered in this
study. This will allow us to build the largest set 
  of confirmed ssrAp stars to date for which preliminary constraints on the
  magnetic field strength have been determined. This sample will lend
  itself to a statistical investigation of the other 
  properties that possibly distinguish the ssrAp stars from their
  faster rotating counterparts, with a view to gaining theoretical
  understanding of the physical processes responsible for the
  differentiation of the rotation rates among Ap stars. It would be
  premature to address this issue in more detail in the present study
  as we have the prospect of becoming able to do it on the basis of a
  more complete sample in a reasonably near future (within
  the next couple of years). That is why, in this paper, we focussed
  primarily on the description of the methods that we used to analyse
  the observations and on the presentation of the immediate results of
  the measurements. Once we have analysed the spectroscopic data whose
  acquisition is in progress for the TESS-based ssrAp star candidates
  that have not been studied here, we shall devote the next paper to
  the presentation of the resulting measurements and to the detailed
  discussion of the insight that the consideration of the full data
  set provides into the origin of extremely slow rotation in Ap stars.

As illustrated in this paper, the present work also yielded several
by-products, of which more should be delivered by the intended
follow-up. Obviously, the discovery of eight new stars showing
resolved magnetically split lines in the visible range is not
unexpected, in a study aimed 
at confirming the low projected equatorial velocity of magnetic Ap
stars. Nevertheless, it represents a significant addition to the
sample of known Ap stars having this property, which contains of the
order of 100 members.

We also reported the detection of radial velocity variations in five
Ap stars that were not previously known to be part of spectroscopic
binaries. The apparent existence of an intriguing connection between
rotation and binarity in ssrAp stars \citep[see Sect.~5.6
of][]{2017A&A...601A..14M} highlights the interest of determining the
orbital parameters of binaries containing a ssrAp star. Besides the
five recently discovered binaries mentioned above, the sample
considered in the present study contains two previously identified
ssrAp binaries whose orbits have not been characterised yet. The
follow-up observations to be obtained to 
study the magnetic variations of these seven stars will also
lend themselves to deriving constraints on their orbital
properties. Moreover, one 
should also expect the spectroscopic survey of unconfirmed TESS-based
ssrAp candidates to lead to the discovery of some new spectroscopic
binaries.

As an aside, we also identified a previously unknown spectroscopic
binary consisting of two very similar Am stars that have very sharp
spectral lines (HD~97132). While the
inclusion of this target in our sample resulted from misclassification
and HD~97132 is not directly relevant to our objectives, it is
nevertheless a rather infrequent and scientifically interesting system
that deserves further attention on its own. 

Finally, the analysis carried out in this work gave us unprecedented
insight into some aspects of the method of determination of the mean 
quadratic magnetic field. Its potential pitfalls were highlighted, the
origin of some of its limitations became clearer, and precautions to
be taken in its application were identified. The discussion in
Sect.~\ref{sec:Bqmethod} represents a valuable complement to 
the detailed considerations of the original presentation of this
magnetic field diagnosis approach \citep{2006A&A...453..699M}, which
should be taken into account in future applications.

\begin{acknowledgements}
DLH and DWK acknowledge support from the Funda\c c\~ao para a
Ci\^encia e a Tecnologia (FCT) through national funds
(2022.03993.PTDC). Based on observations made with the Italian
Telescopio Nazionale 
{\it Galileo\/} (TNG) operated on the island of La Palma by the Fundación
{\it Galileo Galilei\/} of the INAF (Istituto Nazionale di Astrofisica) at the
Spanish Observatorio del Roque de los Muchachos of the Instituto de
Astrofisica de Canarias; on observations obtained with the Southern
African Large Telescope under the proposal codes
2018-1-SCI-026, 2022-2-SCI-017, 2023-1-SCI-008 (PI: Holdsworth); on
ESPaDOnS data retrieved from the CFHT 
Science Archive; and on observations collected at the European
Southern Observatory under ESO programmes 081.D-2002 and
084.D-0067. We thank Martin Hall for making the reduced version of
these latter data available to us. This paper includes data collected
by the TESS mission. Funding for TESS is provided by
NASA's Science Mission Directorate. Resources used in this work were
provided by the NASA High End Computing (HEC) Program through the NASA
Advanced Supercomputing (NAS) Division at Ames Research Center for the
production of the SPOC data products. This work
has made use of the VALD database, operated at Uppsala 
University, the Institute of Astronomy RAS in Moscow, and the
University of Vienna. This research has also made use of the SIMBAD
database, operated at CDS, Strasbourg, France. The collocation of
Figs.~\ref{fig:spec6150_1} to \ref{fig:spec6150_5}, of
Table~\ref{tab:meas}, and of the notes on individual stars in 
an appendix was demanded by the A\&A Editorial Office. 
\end{acknowledgements}

\bibliographystyle{aa}
\bibliography{ssrAp1}

\appendix

\section{Spectra, measurement results and notes on individual stars}
\label{sec:individual}
The spectral properties of the analysed ssrAp star candidates are
described in Sect.~\ref{sec:spectra}. Figures~\ref{fig:spec6150_1} to
\ref{fig:spec6150_5} illustrate this description.

These spectra were measured as explained in Sect.~\ref{sec:Bvsi}. The
latter includes a discussion of the measurement results
(Sect.~\ref{sec:results}), which are presented in
Table~\ref{tab:meas}. The properties of the individual stars are
discussed below. 

\subsection{TIC~73765625 (HD~90131)}
\label{sec:hd_90131}
Two SALT-HRS spectra of HD~90131 have been obtained, at epochs
separated by 109 days. The two components of the \Feline\ line are
clearly resolved and the line is almost free from
blending (see Fig.~\ref{fig:spec6149}). Accordingly, the mean magnetic
field modulus was
determined with the highest precision that is achievable with
HRS. Good precision was also achieved in the determination of the
mean quadratic magnetic field, as more than 20 clean diagnostic
lines of Fe~{\sc i} were identified in the red arm spectral
range. Neither of the two field moments shows any significant
variability between the two epochs of observation. The upper limit of
the projected equatorial velocity derived from the first epoch
spectrum, $\vsi\lesssim3.43$\,\kms\ is small enough to be compatible
with super slow rotation. The second epoch spectrum does not show any
significant rotational broadening, consistently with the width of the
Fe~{\sc i}~$\lambda\,5434.5$\,\AA\ line seen in
Fig.~\ref{fig:spec5434}. The spectroscopic evidence
strongly supports the conclusion that HD~90131 must be a ssrAp
star. On the other hand, while its radial velocity is greater than
that of the majority of single Ap stars, there is no indication of
variability.

SAP data are available for TIC~73765625 for Sectors 22, 45, 46
  and 49. Sectors 22 and 49
  suffer strong instrumental excursions up to 8\,mmag in range. The
  S45-46 data are cleaner, and are used here for a constraint on
  possible rotational light variations. The two sectors comprise 34164
  data points after the removal of 96 problematic points (outliers or
  obvious instrumental excursions). Figure~\ref{fig:73765625_lc_ftd}
  shows the light curve and amplitude spectrum of the SAP data. The
  visible variations and both of the low-frequency peaks are plausibly
  instrumental in origin. The two low frequency peaks are significant
  and have periods of $25\fd5$ and $6\fd7$, but for rotational
  variations in $\alpha^2$\,CVn stars we expect only one rotational
  peak and that is usually at much higher amplitude than these two
  peaks, which have amplitudes of only 160\,$\upmu$mag. We also find
  the same light variations and two peaks for S45-46 data for the next
  star,  TIC~77038207, showing that these peaks are instrumental in
  origin. We therefore conclude that there is no rotational variation
  visible over the 52.4-d time span of S45-46, supporting the
  identification of TIC~73765625 as a ssrAp star.

\subsection{TIC~77038207 (HD~96003)}
\label{sec:hd_96003}
With more than 30 suitable Fe~{\sc i} diagnostic lines, the
high-resolution spectrum of HD~96003 recorded with HARPS-N lends
itself well to a very 
precise determination of the mean quadratic magnetic field. With
$\Bq=1.1$\,kG, this is the weakest field that we could definitely
detect in this study. It is well below the detection limit with
CAOS. The distortion of the Fe~{\sc 
  i}~$\lambda\,6336.8$\,\AA\ line by the magnetic field is clearly seen
in the HARPS-N spectrum (see Fig.~\ref{fig:spec6336}) but the \Feline\
line is not resolved into its magnetically split components so that 
the mean magnetic field modulus cannot be
precisely determined. The quadratic field value is consistent with the
weak mean longitudinal magnetic field measurements obtained at the
Special Astrophysical Observatory \citep[and references
therein]{2023AstBu..78..567R}, according to which $\Bz$ shows at most
low-amplitude variability about its rms value,
$\Bzrms=-170$\,G. 

\citeauthor{2023AstBu..78..567R} also report the occurrence of 
variations of the radial velocity, and argue that HD~96003 is a
spectroscopic binary. This conclusion is supported by the formally
significant difference between the 
values of the radial velocity that we 
determine from the CAOS and HARPS-N spectra, which were obtained only
10 days apart. These values are within the range covered by the
measurements of 
\citet{2023AstBu..78..567R}.

Our mean quadratic magnetic field determination
with HARPS-N did not yield a significant value of the projected equatorial
velocity. This suggests that rotational broadening of the spectral
lines must be substantially less than the HARPS-N resolution,
consistently with the profile of the Fe~{\sc
  i}~$\lambda\,5434.5$\,\AA\ line illustrated in
Fig.~\ref{fig:spec5434}. This
confirms that HD~96003 must be a ssrAp star. (The somewhat high value
of the upper limit, $\vsi\lesssim6.0$\,\kms, derived from the analysis
of the CAOS spectrum appears to result from crosstalk between the
$a_1$ and $a_3$ terms of the regression analysis.)

SAP data are available for TIC~77038207  for Sectors 22, 45, 46 and
49, as for the previous star, TIC~73765625. The light curve and the
amplitude spectrum for TIC~77038207 for the S45-46 data are
essentially identical to those for TIC~73765625, showing that the
variations are instrumental. We therefore conclude that there is no
rotational variation visible over the 52.4-d time span of S45-46,
supporting the identification of TIC~77038207 as a ssrAp star.

\afterpage{\clearpage}
\begin{figure*}[p]
  \centering
  \includegraphics[scale=0.81]{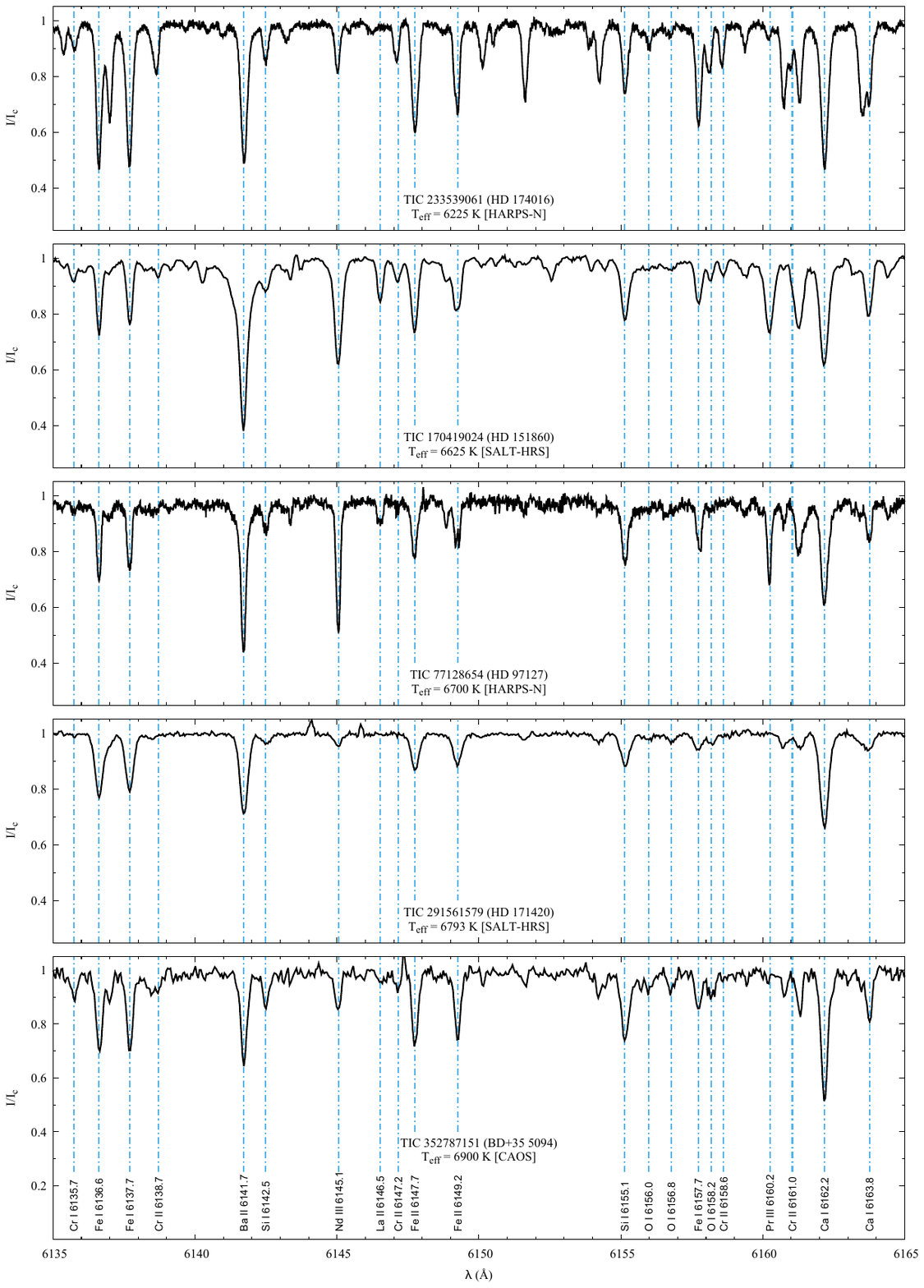}
  \caption{Portion of the spectrum of 5 sharp-lined ssrAp star
    candidates. The wavelengths are in the laboratory reference frame.}
  \label{fig:spec6150_1}
\end{figure*}

\afterpage{\clearpage}
\begin{figure*}[p]
  \centering
  \includegraphics[scale=0.81]{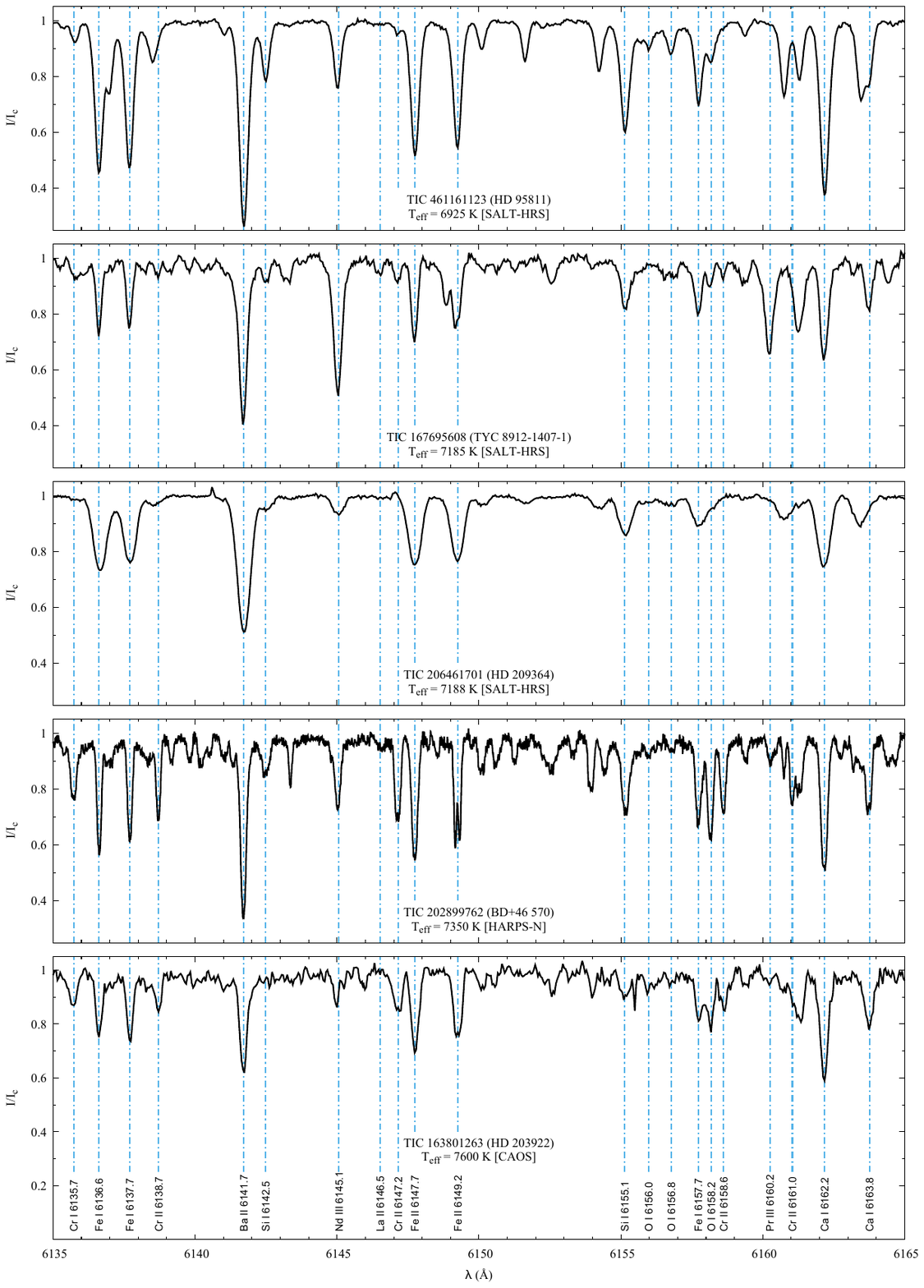}
  \caption{Portion of the spectrum of 5 sharp-lined ssrAp star
    candidates. The wavelengths are in the laboratory reference frame.}
  \label{fig:spec6150_2}
\end{figure*}

\afterpage{\clearpage}
\begin{figure*}[p]
  \centering
  \includegraphics[scale=0.81]{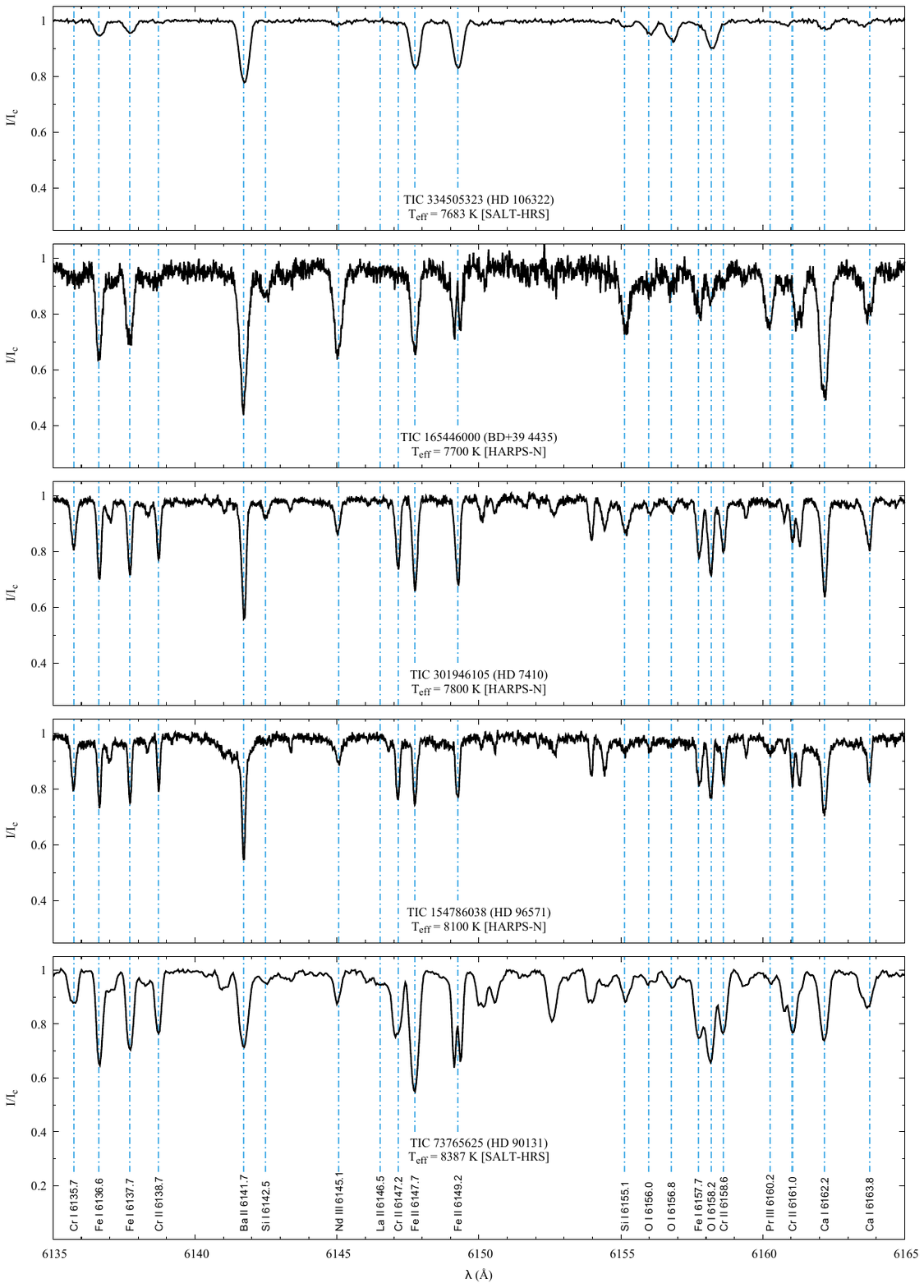}
  \caption{Portion of the spectrum of 5 sharp-lined ssrAp star
    candidates. The wavelengths are in the laboratory reference frame.}
  \label{fig:spec6150_3}
\end{figure*}

\afterpage{\clearpage}
\begin{figure*}[p]
  \centering
  \includegraphics[scale=0.81]{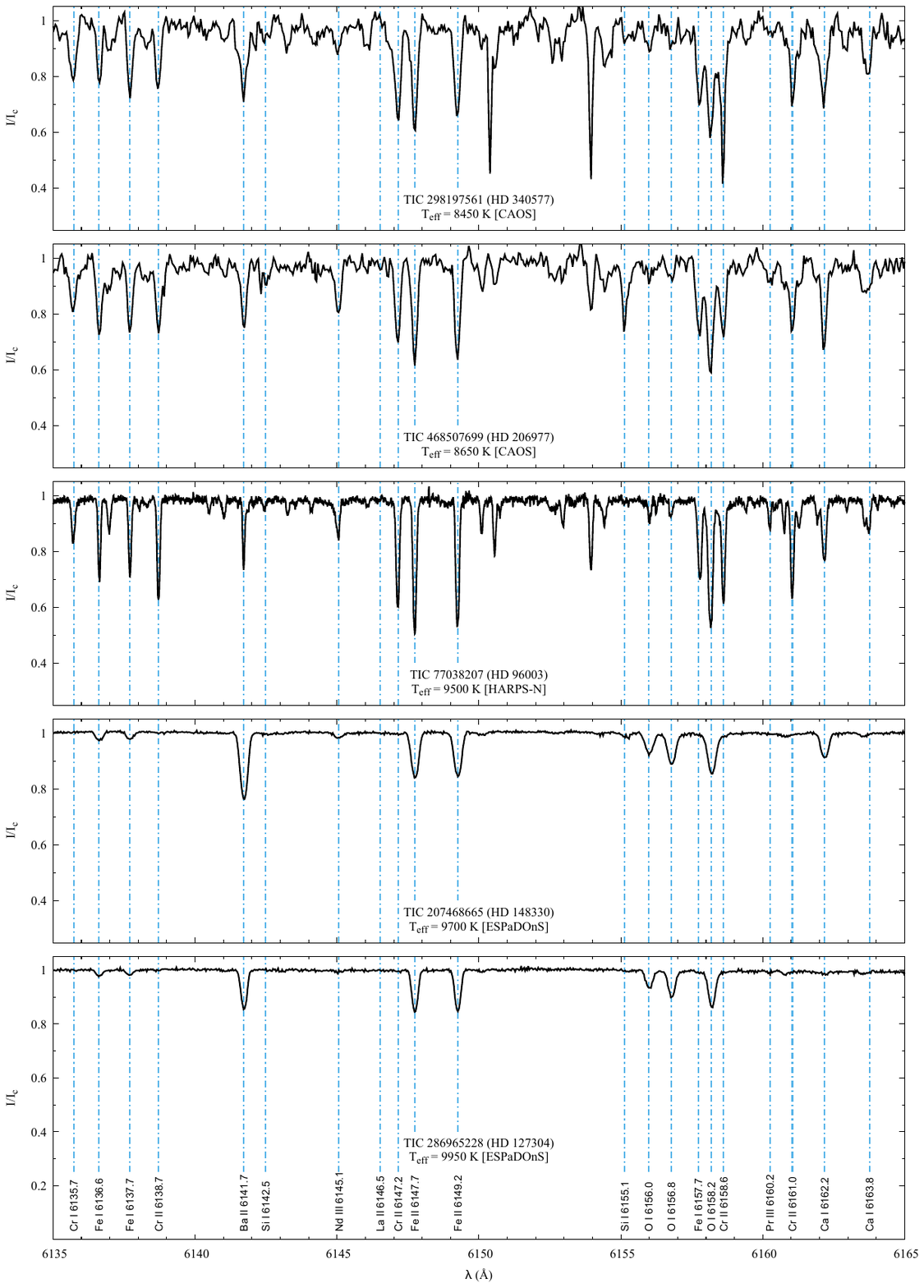}
  \caption{Portion of the spectrum of 5 sharp-lined ssrAp star
    candidates. The wavelengths are in the laboratory reference frame.}
  \label{fig:spec6150_4}
\end{figure*}

\afterpage{\clearpage}
\begin{figure*}[p]
  \centering
  \includegraphics[scale=0.81]{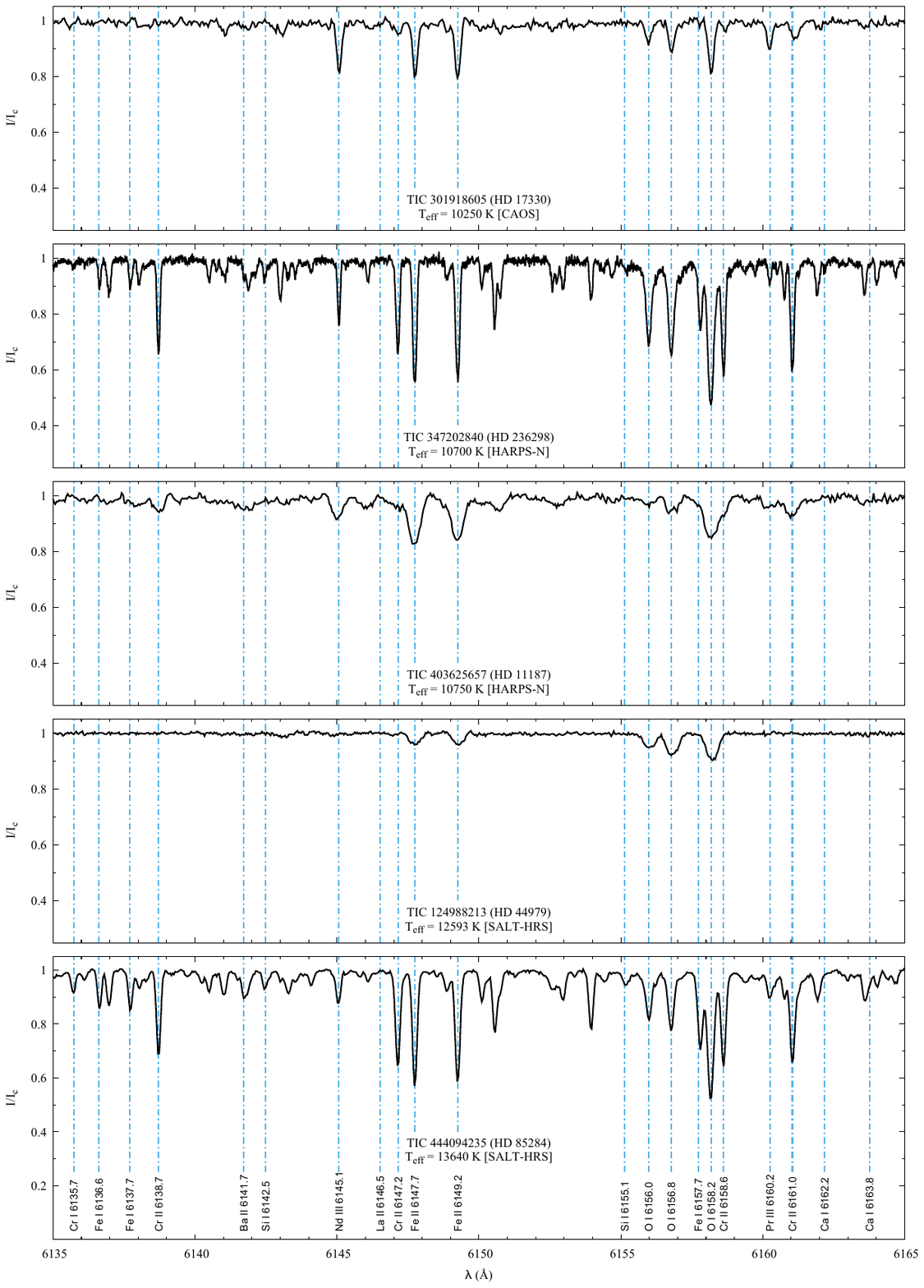}
  \caption{Portion of the spectrum of 5 sharp-lined ssrAp star
    candidates. The wavelengths are in the laboratory reference frame.}
  \label{fig:spec6150_5}
\end{figure*}

\afterpage{\clearpage}
\begin{table*}[p]
  \scriptsize
  \caption{Measurements of the spectra of the ssrAp star
    candidates.}
\label{tab:meas}
\setlength{\tabcolsep}{4pt}
\begin{tabular*}{\textwidth}[]{@{}@{\extracolsep{\fill}}rcrccrrrrrrcrrcrr}
\hline\hline\\[-4pt]
    \multicolumn{1}{c}{TIC}&Other ID&\multicolumn{1}{c}{V}&HJD&Instr&\multicolumn{1}{c}{S/N}&\multicolumn{1}{c}{\Teff}&\multicolumn{1}{c}{$v_{\rm r}$}&\multicolumn{1}{c}{$\sigma(v_{\rm r})$}&\multicolumn{1}{c}{$\Bm$}&\multicolumn{1}{c}{$\sigma(\Bm)$}&$N_{\rm g}$&\multicolumn{1}{c}{$\Bq$}&\multicolumn{1}{c}{$\sigma(\Bq)$}&$N_l$&\multicolumn{1}{c}{$(v\,\sin\,i)_{\rm max}$}&\multicolumn{1}{c}{$\sigma(v\,\sin\,i)$}\\
    &&&$(2,400,000.+)$&&&\multicolumn{1}{c}{(K)}&\multicolumn{1}{c}{(\kms)}&\multicolumn{1}{c}{(\kms)}&\multicolumn{1}{c}{(G)}&\multicolumn{1}{c}{(G)}&&\multicolumn{1}{c}{(G)}&\multicolumn{1}{c}{(G)}&&\multicolumn{1}{c}{(\kms)}&\multicolumn{1}{c}{(\kms)}\\[4pt]

\hline\\[-4pt]
 73765625&HD 90131       & 9.49&60018.524&S&320& 8387& 47.34&0.16&4698& 40&2&5785& 340&21& 3.4&0.6\\
         &               &     &60127.224&S&300&     & 47.49&0.24&4677& 40&2&5780& 380&21& 0.4&0.7\\[3pt]
 77038207&HD 96003       & 6.87&60071.390&C&140& 9500&-10.02&0.15&    &   & & 920& 920&29& 6.0&0.4\\
         &               &     &60081.356&H&130&     &-11.28&0.13&    &   & &1110& 210&32& 0.0&\\[3pt]
 77128654&HD 97127       & 9.43&60058.360&H& 80& 6700&  3.80&0.13&2496&100&3&3080& 330&22& 3.9&0.8\\[3pt]
 80486647&HD 67658       & 9.76&59916.466&S&315&12018& 19.54&0.30&    &   & &   0&    &13&28.8&0.5\\[3pt]
124988213&HD 44979       & 6.53&59891.466&S&350&12593& 26.78&0.16&    &   & &   0&    &19&16.3&0.2\\[3pt]
154786038&HD 96571       & 7.31&60071.424&C&100& 8100&  8.62&0.26&    &   & &2140& 620&14& 3.0&1.0\\
         &               &     &60282.773&H&130&     & -4.90&0.13&1720&100&2&1850& 120&27& 2.6&0.3\\[3pt]
163801263&HD 203922      & 8.50&59818.469&C& 60& 7600&-24.33&0.15&3461&150&2&3310& 820&20& 5.3&0.9\\[3pt]
165446000&BD+39 4435     & 9.3\phantom{0}&59819.454&C& 30& 7700&-26.03&0.34&6418&250&2&5410&1050&13& 3.5&1.7\\
         &               &     &60084.723&H& 50&     &-25.38&0.09&4362&100&3&6040& 230&25& 1.4&0.9\\[3pt]
167695608&TYC 8912-1407-1&11.51&58242.235&S& 85& 7185&  9.88&0.19&3541&100&3&3920& 150&20& 2.0&0.3\\
         &               &     &58388.584&S& 95&     &  9.58&0.20&3146&100&3&4150& 180&20& 2.0&0.3\\
         &               &     &58390.579&S&115&     &  9.50&0.16&3146&100&3&3890& 130&20& 3.0&0.3\\
         &               &     &58395.593&S& 70&     &  9.42&0.17&3272&100&3&3920& 200&20& 2.0&0.3\\
         &               &     &58400.609&S&100&     &  9.59&0.17&3377&100&3&4000& 150&20& 2.0&0.3\\[3pt]
170419024&HD 151860      & 9.01&54691.620&F&200& 6625& -0.82&0.20&    &   & &2510& 280&24& 0.0&\\ 
         &               &     &60005.596&S&205&     &  4.10&0.25&3356&100&3&5410& 560&19& 0.0&\\[3pt] 
202899762&BD+46 570      & 9.63&59887.477&C& 30& 7350&  8.29&0.27&    &   & &4030&1000&12& 8.4&1.7\\
         &               &     &60312.319&H&115&     & 10.42&0.18&3167& 30&2&3540& 150&20& 0.0&\\[3pt]
206461701&HD 209364      &10.03&60129.472&S&300& 7188&-18.78&0.23&    &   & &   0&    &20&17.4&0.5\\[3pt]
207468665&HD 148330      & 5.73&60165.425&C&150& 9700& -3.31&0.22&    &   & &3450&1300&17&14.0&0.4\\[3pt]
286965228&HD 127304      & 6.05&60071.467&C&150& 9950&-18.01&0.15&    &   & &   0&   0&15& 9.1&0.7\\[3pt]
291561579&HD 171420      &10.67&60084.437&S&205& 6793&  7.84&0.10&    &   & &1460& 660&18&11.5&0.5\\[3pt]
298197561&HD 340577      & 9.09&59880.313&C& 50& 8450& 19.48&0.35&    &   & &   0&    &15& 2.4&1.1\\[3pt]
301918605&HD 17330       & 7.11&59880.547&C& 80&10250&-13.10&0.12&    &   & &   0&    &20& 8.1&0.6\\[3pt]
301946105&HD 7410        & 9.07&59880.453&C& 50& 7800&  2.18&0.22&    &   & &   0&    &20& 7.2&0.9\\
         &               &     &60272.514&H&130&     & -0.22&0.12&    &   & &2110& 180&27& 4.8&0.3\\[3pt]
334505323&HD 106322      & 9.39&59977.466&S&240& 7683&-12.14&0.20&    &   & &   0&    &27&14.8&0.3\\[3pt]
347202840&HD 236298      & 9.45&59837.541&C& 50&10700& -7.91&0.24&    &   & &   0&    &16& 5.5&1.3\\
         &               &     &60272.328&H&120&     & -7.92&0.13&    &   & &2210& 290&24& 3.1&0.3\\[3pt]
352787151&BD+35 5094     & 9.08&59887.398&C& 70& 6900& -9.44&0.12&    &   & &   0&    &21& 6.8&0.6\\[3pt]
403625657&HD 11187       & 7.94&59880.495&C& 80&10750&  5.61&0.27&    &   & &   0&    &20&17.0&0.9\\[3pt]
444094235&HD 85284       & 9.82&55228.695&F&110&13640&  0.62&0.29&    &   & &1700& 340&19& 0.0&\\ 
         &               &     &59955.435&S&280&     &  9.47&0.33&    &   & &   0&    &13& 4.5&1.0\\[3pt] 
461161123&HD 95811       & 9.56&60017.553&S&375& 6925& -3.36&0.18&    &   & &   0&    &19&10.9&0.6\\[3pt]
468507699&HD 206977      & 8.98&59837.457&C& 80& 8650&  9.72&0.20&    &   & &1650& 520&20& 1.7&0.5\\[4pt]
\hline\\[-4pt]
\end{tabular*}
  \tablefoot{The stars are listed in order of increasing TIC (TESS
    Input Catalogue) number
  (Col.~1), with another ID (preferably the HD number) given in
  Col.~2 and the $V$ magnitude in Col.~3. The Heliocentric
  Julian Date (HJD) of mid-observation, 
  the instrument with which it was obtained (S\,=\,SALT-HRS; F\,=\,FEROS; 
  C\,=\,CAOS; H\,=\,HARPS-N), and the resulting S/N appear in
  Cols.~4 to 6. The values of \Teff\ in Col.~7 are as listed in
  \citetalias{2020A&A...639A..31M} and
  \citetalias{2022A&A...660A..70M}. The following columns contain the
  results of our measurements: the stellar radial velocity $v_{\rm
    r}$ and its uncertainty $\sigma(v_{\rm r})$ (Cols. 8 and 9); the
  mean magnetic field modulus $\Bm$, its uncertainty $\sigma(\Bm)$,
  and the number $N_{\rm g}$ of Gaussians fitted to the \Feline\ line
  for its determination (Cols.~10 to 12); the mean quadratic magnetic
  field $\Bq$, its 
  formal uncertainty $\sigma(\Bq)$, and the number $N_{\rm l}$ of
  diagnostic lines from which it was derived (Cols.~13 to 15); the
  upper limit $(\vsi)_{\rm max}$ of the projected equatorial velocity
  and the formal uncertainty $\sigma(\vsi)$ of the \vsi\
  determination (Cols.~16 and 17). The values of the radial
  velocity and of the projected equatorial velocity are derived from
  analysis of the same sample of $N_{\rm l}$ lines as used for the $\Bq$
  determination. For spectra in which the \Feline\ line is not
  resolved into its magnetically split components, the entries in
  Cols.~8 to 10 are left blank; `0' in Col.~11 identifies those
  spectra in which the mean quadratic magnetic field is below the
  detection threshold, and `0.00' in Col.~14, those spectra in which the
  rotational Doppler broadening is below the detection threshold.} 
\end{table*}

\subsection{TIC~77128654 (HD~97127)}
\label{sec:hd_97127}
The \Feline\ line is resolved into its magnetically split components
in our HARPS-N spectrum of the roAp star HD~97127
\citep{2014MNRAS.443.2049H}. The considerable blending of the blue
component (see Fig.~\ref{fig:spec6149})
induces significant uncertainty in the determination of the mean
magnetic field modulus. Its derived value, $\Bm=2.5$\,kG, is among the
lowest ever measured in an Ap star
\citep[see][]{2017A&A...601A..14M}. The mean quadratic magnetic field,
$\Bq=3.1$\,kG, can be determined with good precision. The value of the
upper limit of the projected equatorial velocity that can be derived
as a by-product, $\vsi\lesssim3.95$\,\kms, is not inconsistent with
super-slow rotation, but a moderately long rotation period ($20\,{\rm
  d}\lesssim\Prot\lesssim50$\,d) cannot be ruled out.

SAP data are available for TIC~77128654 for Sectors 22, 45, 46 and 49,
the same as for the previous two stars. Sectors 22, 46 and 49 suffer
strong instrumental excursions up to 60\,mmag in range. To illustrate
the problem of using the SAP data for the detection of long rotation
periods, as in  the ssrAp stars, the top panel of
Fig.\,\ref{fig:77128654_lc_ftd} shows the unedited light curve for the
S45-46 data. The large excursions for the S46 data occurred after the
field was recovered following a data transfer to ground. Immediately
after a field is recovered, the detectors are still temperature
stabilising resulting in sensitivity changes. These are modelled and
removed in PDCSAP data, but are still seen in the SAP data shown
here. Figure~\ref{fig:77128654_lc_ftd} shows the raw light curve (top),
the edited light curve (middle) and the amplitude spectrum of the SAP
data. All variations are plausibly instrumental in origin.  We
therefore conclude that there is no rotational variation evident over
the 52.4-d time span of S45-46. 
  
\subsection{TIC~80486647 (HD~67658)}
\label{sec:hd_67658}
According to a recent study \citep{2020MNRAS.499.2701M}, HD~67658 is a
normal A4IV star. Hence, it must have been misclassified as an Ap
star. Indeed, lines of the elements typically present in Ap stars do
not appear to be present in its HARPS-N spectrum. Visual inspection of
the latter also reveals that HD~67658 has the broadest spectral lines
among the stars of studied in this work. Of these stars, it is also
the one for which we derived the least stringent upper limit of the
projected equatorial velocity, $\vsi\lesssim28.43$\,\kms. With a mean
quadratic magnetic field below the detection threshold, HD~67658 is
definitely not a ssrAp star, and probably not even an Ap star at all. 
There is no plausible variation in the S7, 9 and 35 SAP data, hence no
$\alpha^2$\,CVn rotational variation, also suggesting that this is not
an Ap star. 

\begin{figure}
  \centering  
\includegraphics[width=1.0\linewidth,angle=0]{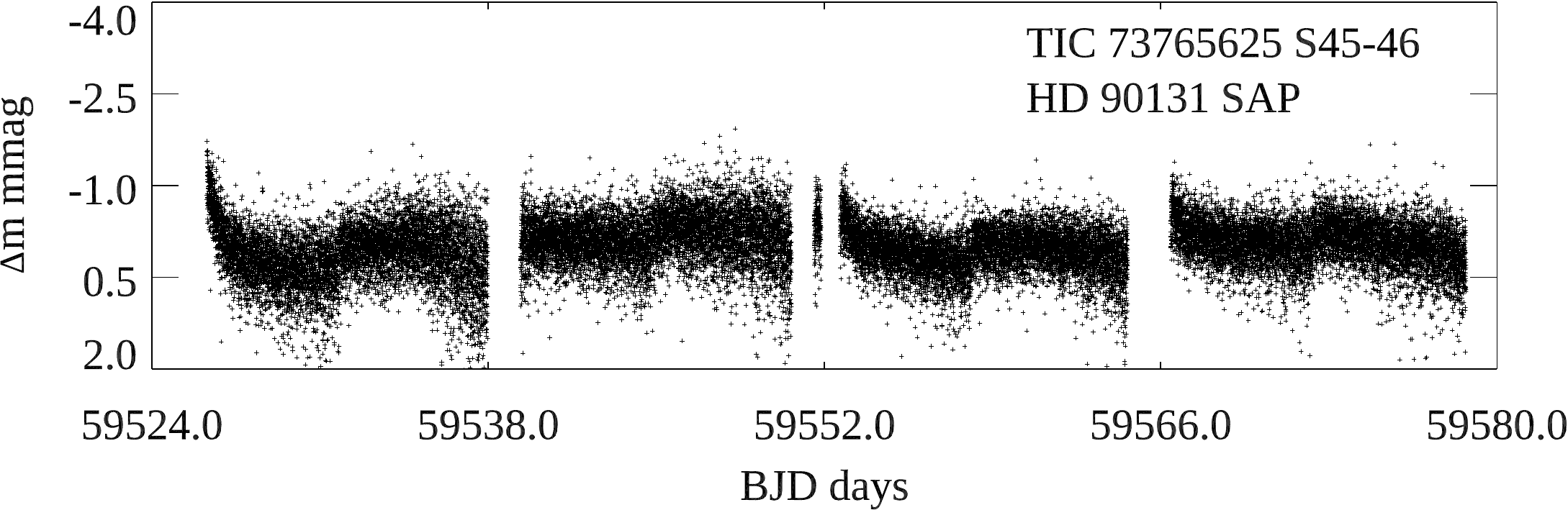}
\includegraphics[width=1.0\linewidth,angle=0]{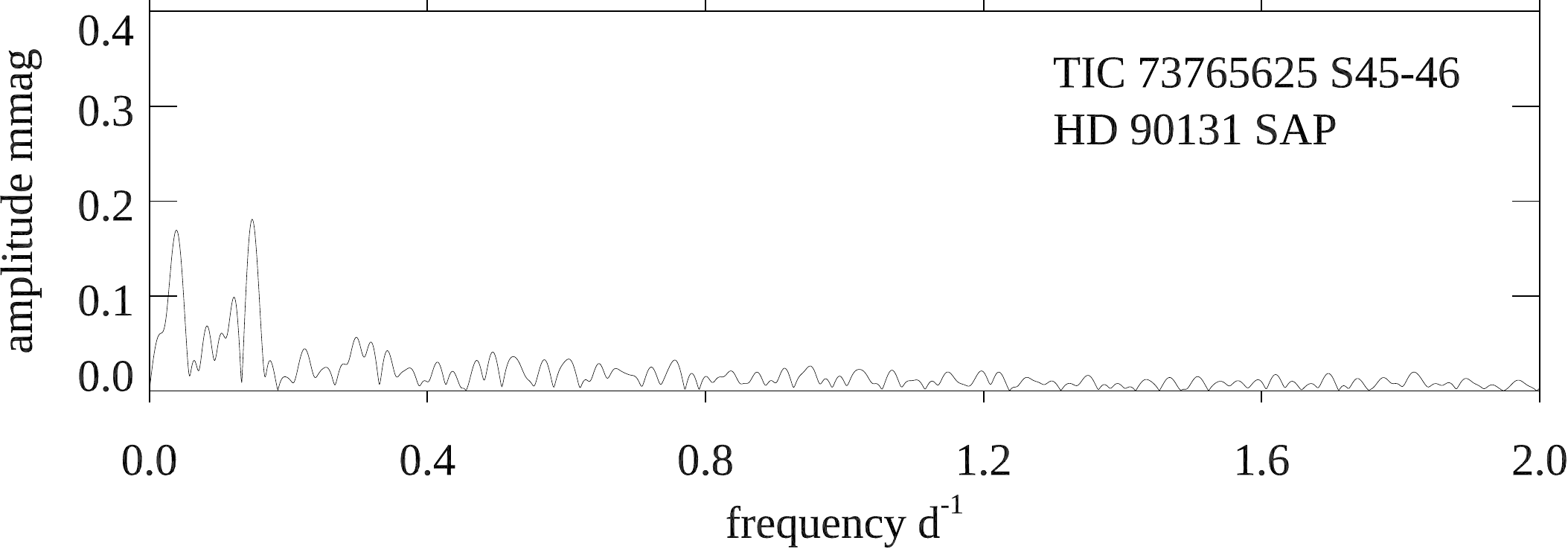}
  \caption{TIC~73765625 (HD~90131). Top: The light curve of the S45-46 SAP data. The visible variations are plausibly instrumental. Bottom: The amplitude spectrum of the S45-46 SAP data showing two low-frequency peaks, both of which are plausibly instrumental.  }
\label{fig:73765625_lc_ftd}
\end{figure}

\subsection{TIC~81554659 (HD~97132)}
\label{sec:hd_97132}
We observed HD~97132 with SALT-HRS and discovered that it is a
double-lined spectroscopic binary (SB2). To the best of our
knowledge, this has never been reported before. The two components
appear to be nearly identical to each other. Although the star has
been classified ApSrEuCr \citep{1978mcts.book.....H}, we could not
readily identify in its spectrum strong lines of these elements, or of
other rare earth elements. There is no plausible variation in the Sector~10,
and 36-37 SAP data, hence 
no $\alpha^2$\,CVn rotational variation, also suggesting that this is
not an Ap star.

 We carried out a preliminary abundance analysis through spectrum
synthesis, similar to \citet{2024A&A...685A.133C}. We adopted the same
fundamental parameters for both components, $\log g=4.0$ and
$\Teff=8250$\,K, which is lower than the value from the TIC
($\Teff=9257$\,K). This temperature matches better the intensity
ratio of the Fe~{\sc i} and 
Fe~{\sc ii} lines and the Balmer line profiles. Some portions of the
synthetic spectrum are shown in Fig.~\ref{fig:hd97132_synth}. This
analysis indicates that in both stars, the C and O abundances are
solar, Ca and Sc are underabundant, and Si, Fe, and Ba are
overabundant. Thus HD~97132 is a SB2 system consisting of two Am stars. 

\subsection{TIC~124988213 (HD~44979)}
\label{sec:hd_44979}
The line density in the spectrum of HD~44979 is low, and most lines
are weak. Accordingly, our attempt to determine the mean quadratic
magnetic field from a SALT-HRS spectrum was based on the analysis of
the blue arm part. The upper limit that we derived for the projected
equatorial velocity, $\vsi\lesssim16.26$\,\kms, is consistent with the
moderately broadened appearance of the spectral lines. This is not a
ssrAp star. Moreover, the mean quadratic magnetic field is below the
detection threshold. One may wonder if the ApSi classification
\citep{1982mcts.book.....H} of HD~44979 is mistaken, as the Si lines in its
spectrum do not appear particularly strong. 

SAP data are available for TIC~124988213 for Sectors
6-7. Fig.\,\ref{fig:124988213_lc_ftd} shows the light curve and
amplitude spectrum. All variations are plausibly instrumental. This
also suggests that the star is not an $\alpha^2$\,CVn star, hence not
an Ap star.  
  
\subsection{TIC~154786038 (HD~96571)}
\label{sec:hd_96571}
Two observations of HD~96571 were obtained, one with CAOS and the
other with HARPS-N, at an interval of 211 days. There is a highly
significant difference of radial velocity ($\Delta v_{\rm
  r}=13.5$\,\kms) between the two epochs. This represents a strong
indication that the star must be a spectroscopic binary, which has not
been previously identified. The radial velocity recorded with HARPS-N
does not significantly differ from the value published in the General
Catalogue of Stellar Radial Velocities \citep{1953GCRV..C......0W}.

  \begin{figure}
  \centering  
\includegraphics[width=1.0\linewidth,angle=0]{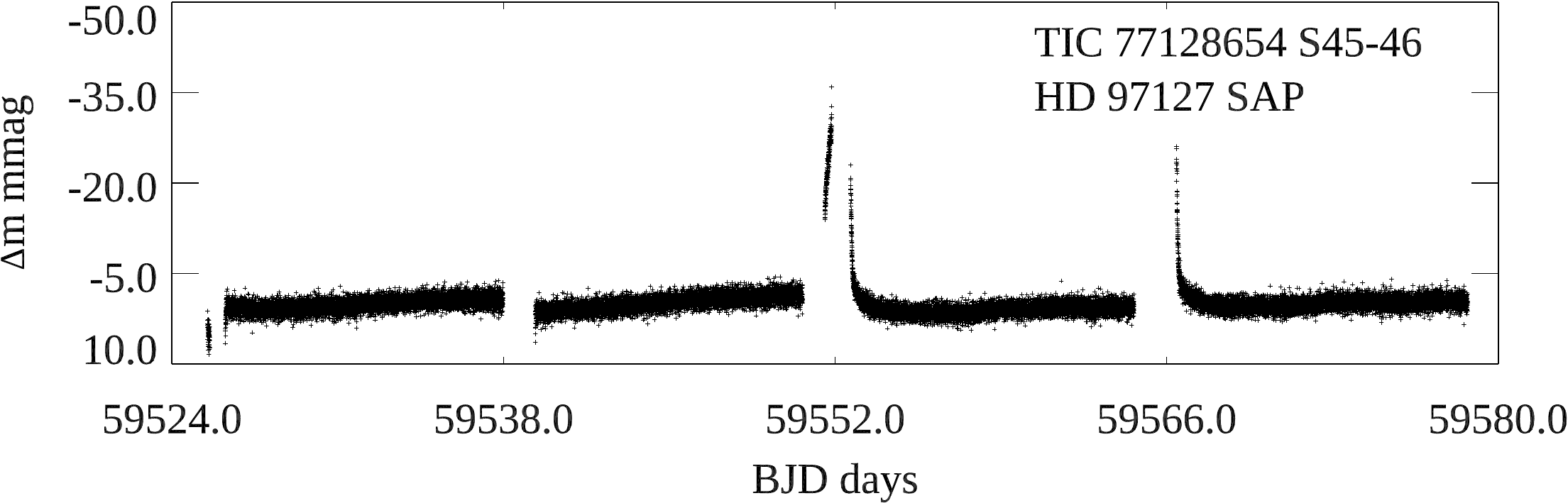}
\includegraphics[width=1.0\linewidth,angle=0]{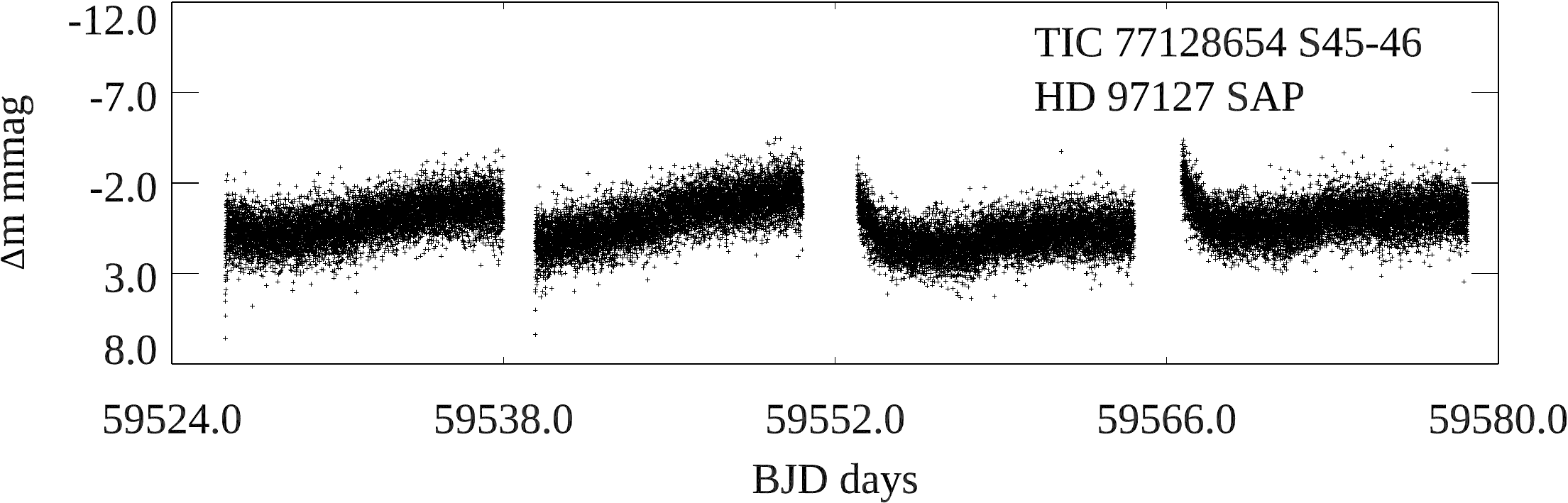}
\includegraphics[width=1.0\linewidth,angle=0]{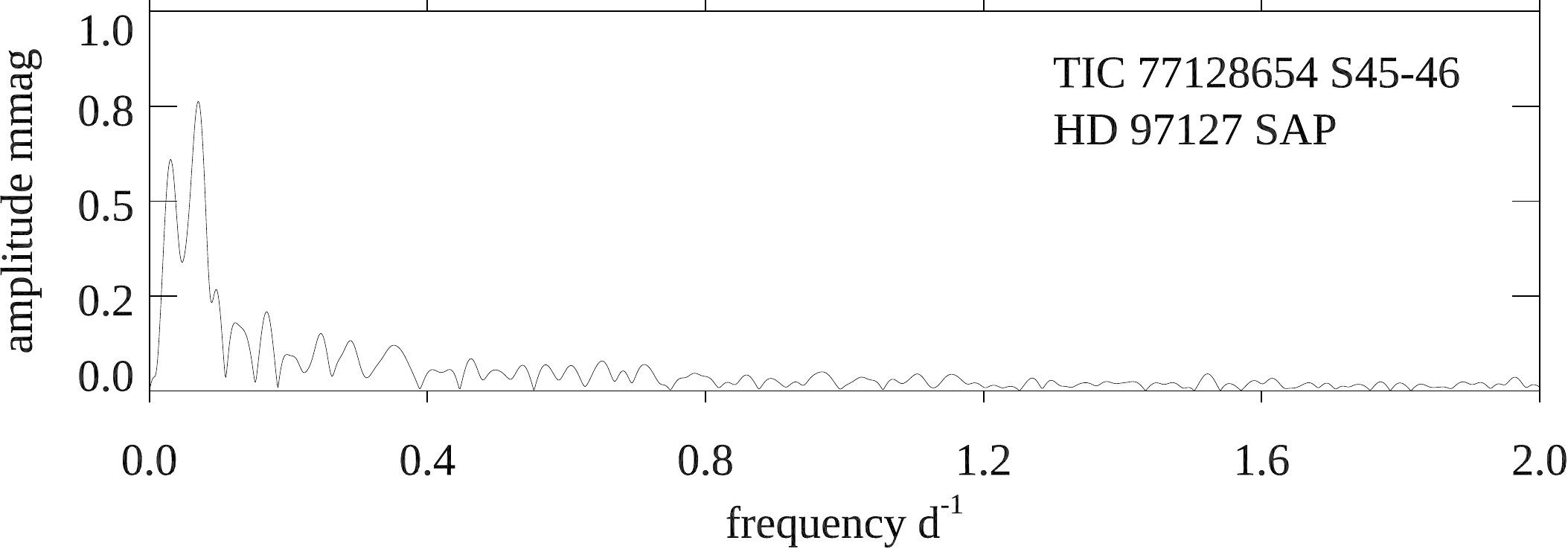}
 \caption{TIC~77128654 (HD~97127). Top: The light curve of the S45-46 SAP data without editing the data. The instrumental excursions in S46 are evident. S22 and S49 suffer similar excursions. This is not uncommon in the SAP data. Middle: The S45-46 SAP data after the larger instrumental excursions are filtered out.  Bottom: The amplitude spectrum of the edited S45-46 SAP data showing two low-frequency peaks, both of which are plausibly instrumental.}
\label{fig:77128654_lc_ftd}
\end{figure}

In the HARPS-N spectum, the \Feline\ line is only marginally resolved,
but it is free from blends (see Fig.~\ref{fig:spec6149}). This allowed
us to determine the mean 
magnetic field modulus, $\Bm=1.7$\,kG. This value is similar to the
lower limit predicted by \citet{1997A&AS..123..353M} for measurements
of the mean magnetic field modulus from spectra at a resolving power
$R\sim10^5$. Because of the marginal resolution of the \Feline\ line,
the estimated uncertainty of the $\Bm$ determination, 100\,G, is
greater than for higher mean field modulus values that can be derived
in the best cases from the analysis of spectra of similar resolution
(down to $\sim$25\,G, see Table~13 of \citealt{2017A&A...601A..14M}).

\begin{figure*}[t]
  \centering
  \includegraphics[scale=1.2]{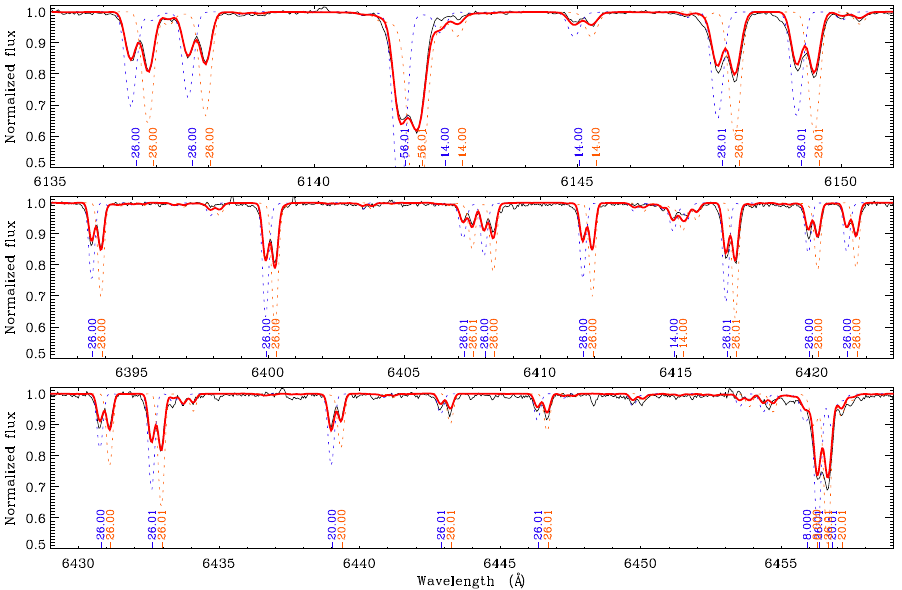}
  \caption{Synthetic spectrum of the SB2 star HD~97132, for selected
    wavelength ranges. The thin solid black line is the observed
    spectrum; the thick solid red line is the synthetic spectrum for
    the combination of the two components. The synthetic spectra of
    each individual component is represented by the thin dashed blue
    and red lines. The ions responsible for the fitted transitions are
  identified in the Kurucz notation \citep{1975SAOSR.362.....K}.}
      \label{fig:hd97132_synth}
\end{figure*}

  \begin{figure}
  \centering  
\includegraphics[width=1.0\linewidth,angle=0]{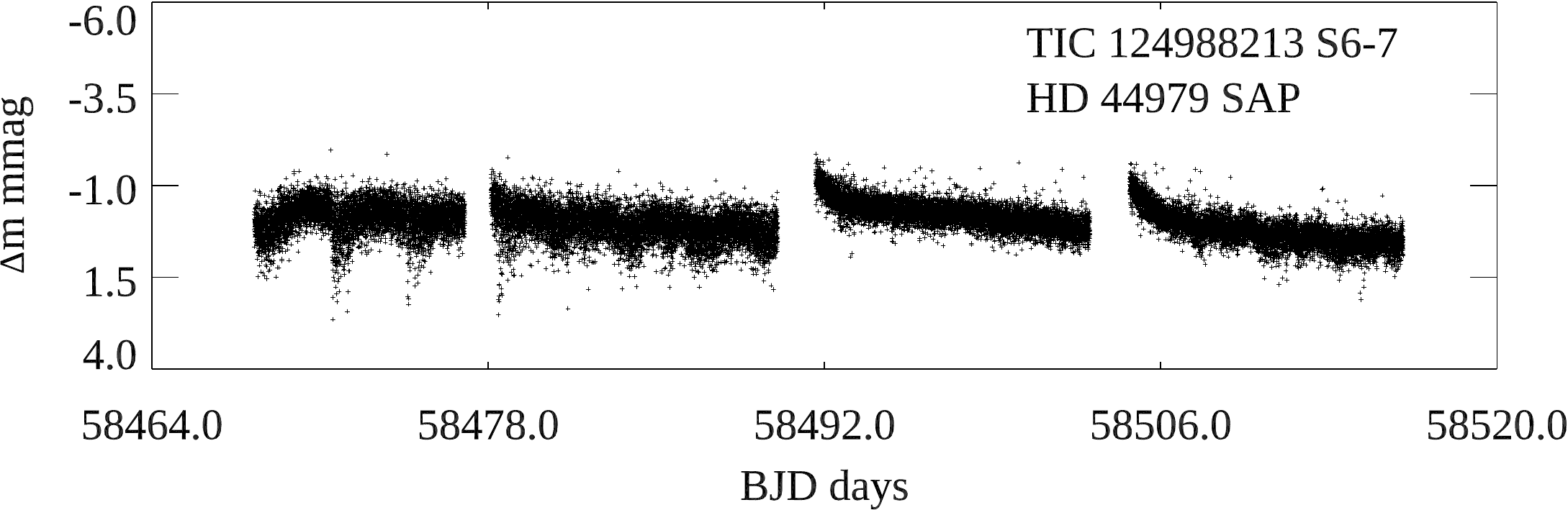}
\includegraphics[width=1.0\linewidth,angle=0]{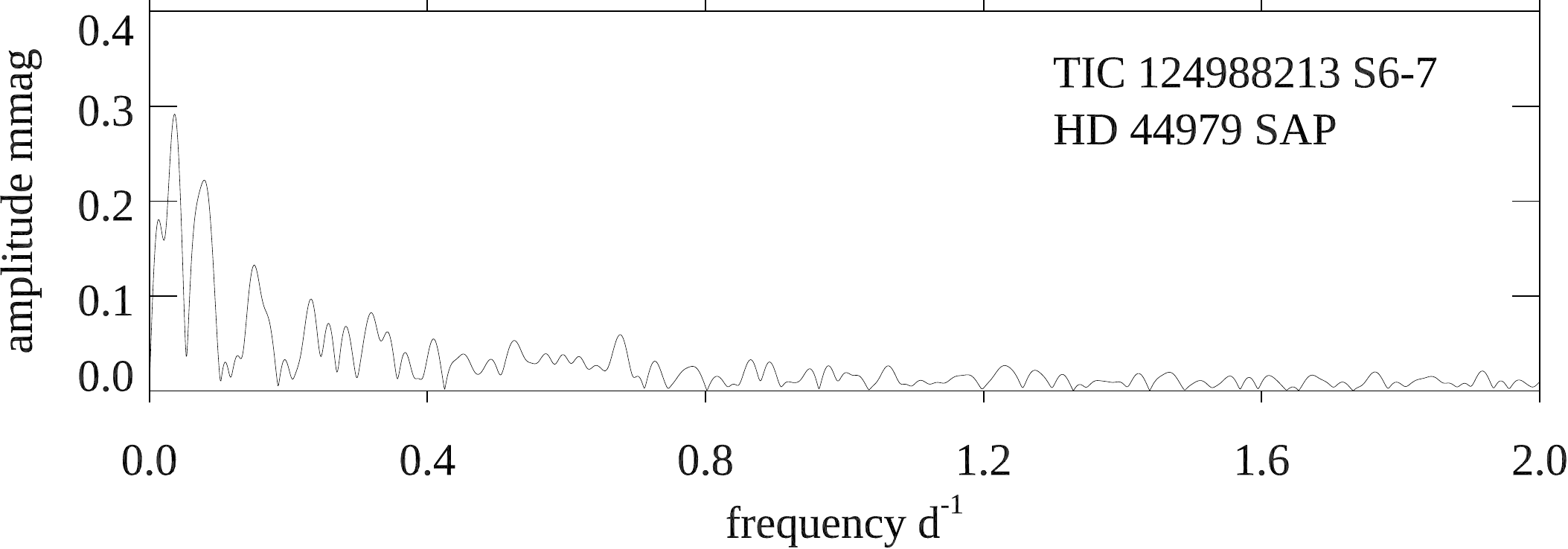}
 \caption{TIC~124988213 (HD~44979). Top: The light curve of the S6-7
   SAP data. Bottom: The amplitude spectrum of the S6-7 SAP data. All
   variations are plausibly instrumental.}  
\label{fig:124988213_lc_ftd}
\end{figure}

By contrast, the formal precision achieved in the determination of the
mean quadratic magnetic field from the HARPS-N spectrum is the best
one of this study. The distortion of the Fe~{\sc
  i}~$\lambda\,6336.8$\,\AA\ line by the magnetic field is clearly seen
in 
Fig.~\ref{fig:spec6336}. The value of $\Bq$ determined from analysis
of the CAOS spectra is also just above the threshold of formal
significance. Within the uncertainties, the mean quadratic field
modulus has not varied between the two epochs of
observations. However, this is a weak constraint, as for the CAOS
spectrum, the formal uncertainty of $\Bq$ is rather large.

The upper limit of the projected equatorial velocity that is derived
from analysis of the HARPS-N spectrum, $\vsi\lesssim2.56$\,\kms, is
consistent with super slow rotation, but a moderately long rotation
period ($20\,{\rm d}\lesssim\Prot\lesssim50$\,d) cannot be
definitely ruled out. Within the
uncertainty, the analysis of the CAOS spectrum yields a \vsi\ upper
limit that is not significantly different. In summary, all the
available information 
consistently indicates that HD~96571 must be a slowly rotating Ap star
in a binary system.

SAP data are available for Sectors 14, 19-20, 26, 40 and 53. There is a
probable instrumental variation with an amplitude of only 0.5\,mmag on
the timescale of 30\,d, which is close to the sector length. This
does not support any photometric variation in the $20\,{\rm
  d}\lesssim\Prot\lesssim50$\,d range. 

\subsection{TIC~163801263 (HD~203922)}
\label{sec:hd_203922}
The \Feline\ line is marginally resolved into its two components in
our CAOS spectrum of HD~203922.  It is not significantly affected by
any blend (see Fig.~\ref{fig:spec6149}). Accordingly, the value of the
mean magnetic field modulus, 
$\Bm=3.46$\,kG, could be determined with rather good precision. The
mean quadratic magnetic field, $\Bq=3.31$\,kG, could also be measured
at the $4.0\,\sigma$ level. Given their respective uncertainties, the
values of the two field moments are mutually consistent. The $\Bq$
analysis also yielded an estimate of the upper limit of the projected
equatorial velocity, $\vsi\lesssim5.27$\,\kms. Indeed, in
Fig.~\ref{fig:spec5434}, the
observed profile of the Fe~{\sc i}~$\lambda\,5434.5$\,\AA\ line
appears somewhat broader than the estimated contributions of
instrumental and thermal broadening. This difference is not
necessarily incompatible with super slow rotation, but it seems more
likely that HD~203922 has a moderately long rotation period. 

\begin{figure}
  \centering  
\includegraphics[width=1.0\linewidth,angle=0]{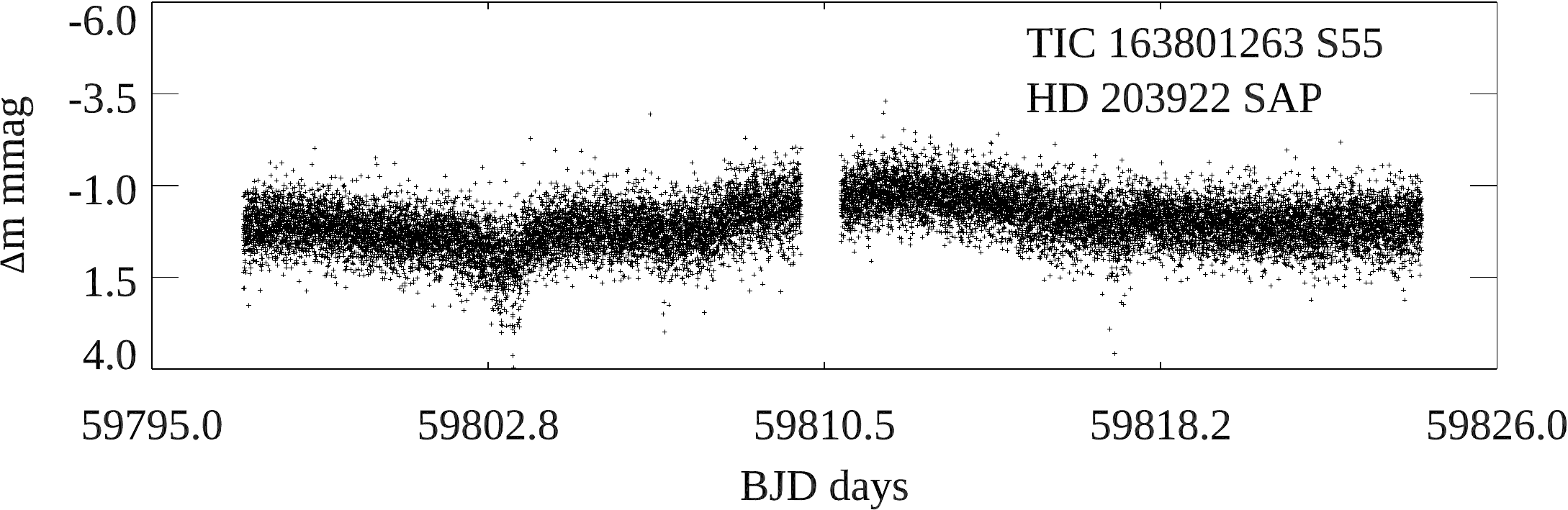}
\includegraphics[width=1.0\linewidth,angle=0]{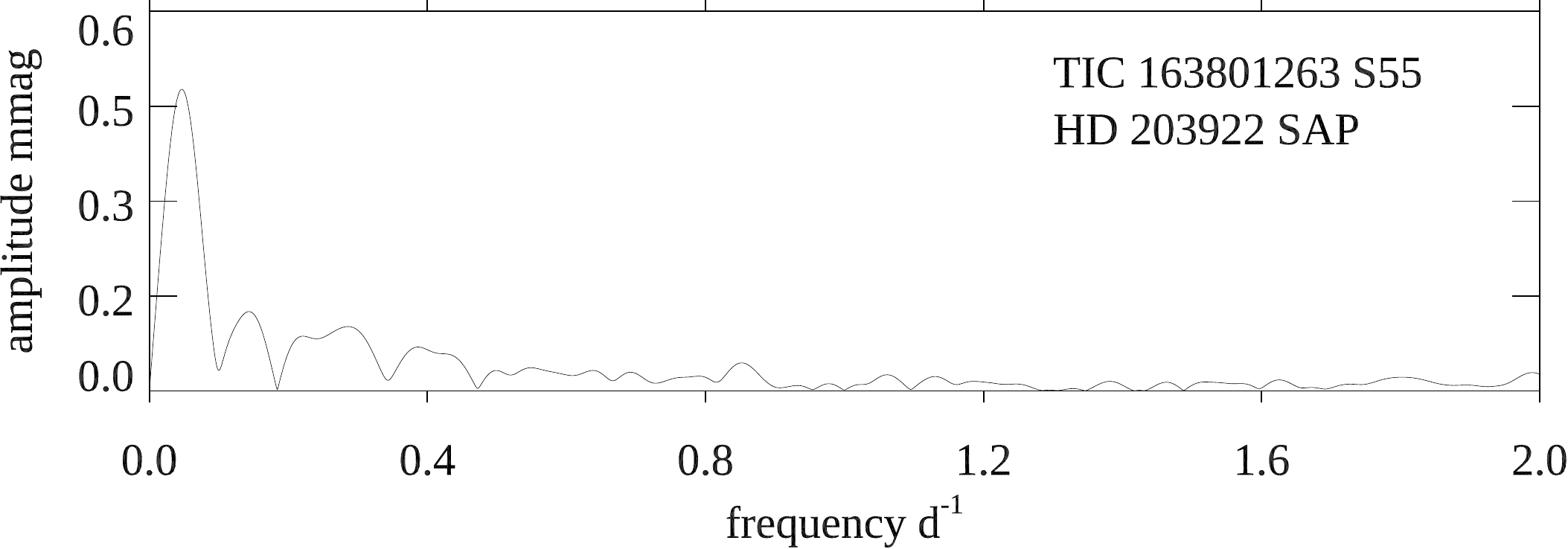}
 \caption{TIC~163801263 (HD~203922). Top: The light curve of the S55
   SAP data. Bottom: The amplitude spectrum of the S55 SAP data. The
   low frequency peak has a period of $\sim$22\,d. It could be
   the rotation period, or it could be instrumental.} 
\label{fig:163801263_lc_ftd}
\end{figure}

SAP data are available for Sectors 15 and
55. Figure~\ref{fig:163801263_lc_ftd} shows the light curve and
amplitude spectrum of the S55 SAP data. There is variability with an
amplitude of 0.5\,mmag on a timescale of $\sim$22\,d for the S55
data. This signal is neither evident nor ruled out for the S15 data,
and the two sectors are too far apart in time to meaningfully analyse
them together. Hence, it is possible that this star has a moderate
period of the order of 22\,d, although the signal could also be
instrumental.  The similarity to the light curves and amplitude
spectra for the next star, TIC~165446000, which was observed in the
same sectors with the same camera suggests that the variability for
both stars is instrumental. 

\subsection{TIC~165446000 (BD+39~4435)}
\label{sec:bd+39_4435}
The HARPS-N spectrum of BD+39~4435 was obtained 265 days after its
CAOS spectrum. The two components of the \Feline\ line are
magnetically resolved in both of them (see
Fig.~\ref{fig:spec6149}). The measurements of this line 
in the two spectra yield substantially different values of the mean
magnetic field modulus. While the $\Bm$ variability that they reflect
appears real, their uncertainties, which are large as the spectra are
noisy (especially the CAOS one), may be partly responsible for the
apparently large variation amplitude. This is all the more plausible
since the mean quadratic magnetic field, which could also be
determined at both epochs, suggests more moderate variability; however,
it is also difficult to constrain precisely due to the large formal
uncertainty of the $\Bq$ value derived from the CAOS spectrum.

In any event, the observed $\Bm$ and $\Bq$ variations do not rule out
super slow rotation, given the long time interval on which they
occurred. Moreover, 
no formally significant rotational broadening could be
detected as part of the $\Bq$ analysis of either spectrum. Thus, it is
highly probable that BD+39~4435 is a ssrAp star.

\begin{figure}
\centering  
\includegraphics[width=1.0\linewidth,angle=0]{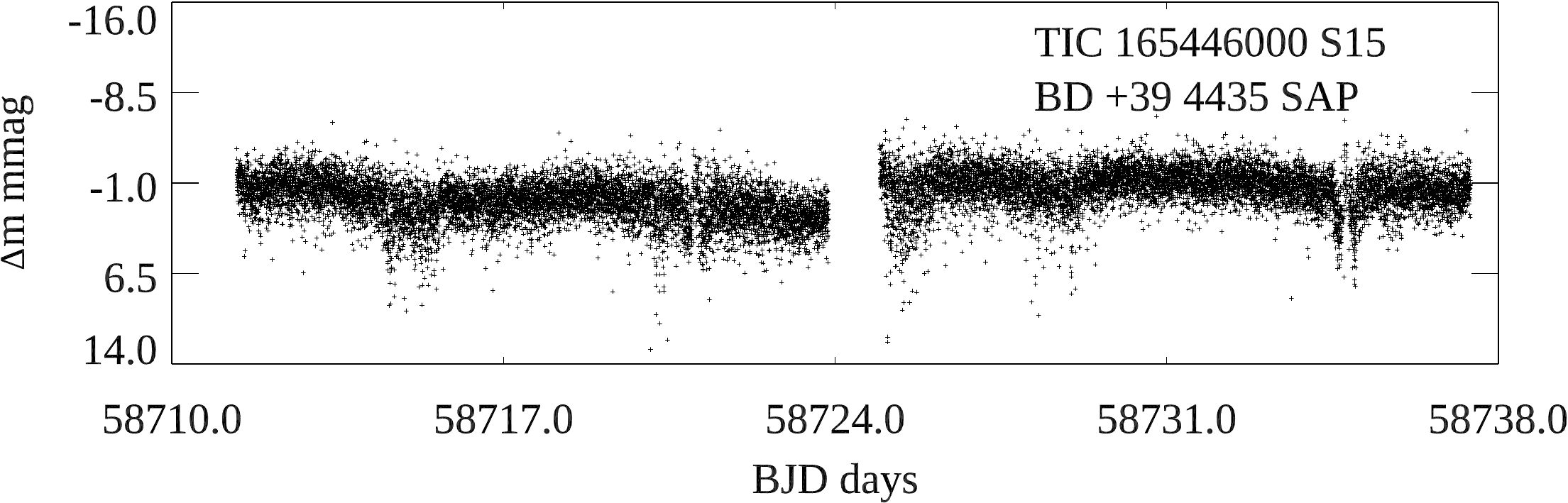}
\includegraphics[width=1.0\linewidth,angle=0]{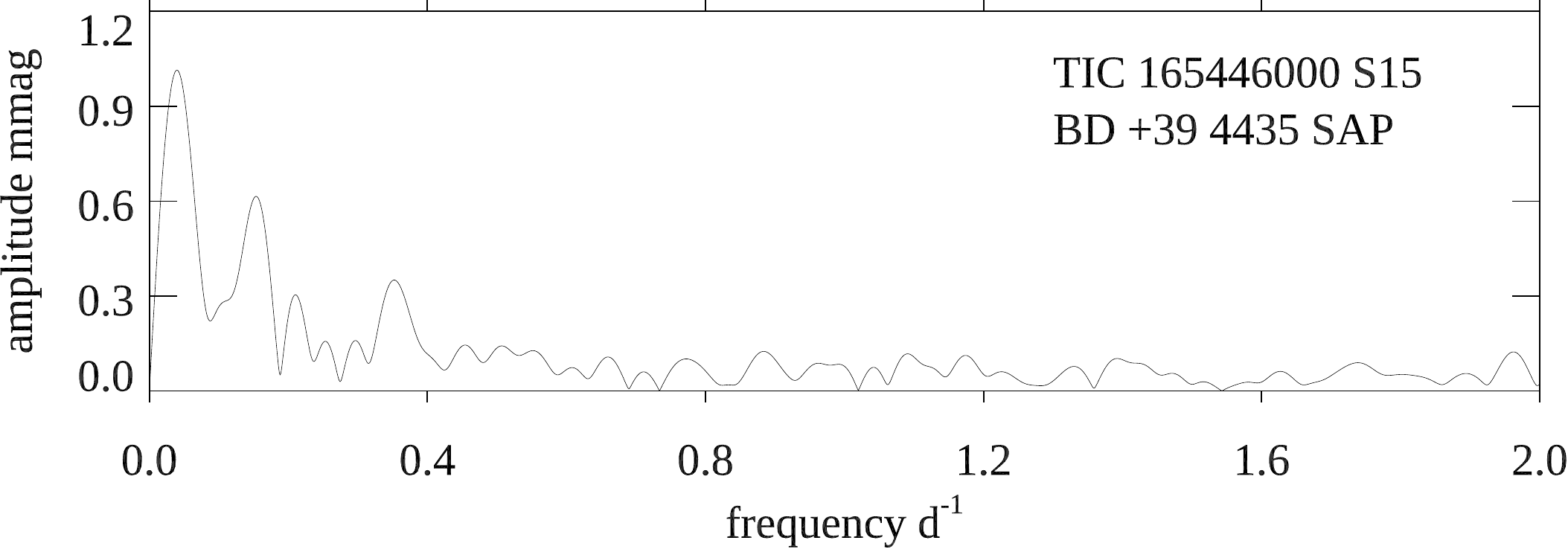}
\caption{TIC~165446000 (BD+39~4435). Top: The light curve of the S15
  SAP data. Bottom: The amplitude spectrum of the S15 SAP data. The
  low frequency peak has a period of $\sim$23\,d. This could be
  rotation period, or it could be instrumental. Its similarity to that
  of TIC~163801263 above suggests an instrumental origin.} 
\label{fig:165446000_lc_ftd}
\end{figure}

SAP data are available for Sectors 15 and
55. Figure~\ref{fig:165446000_lc_ftd} shows the light curve and
amplitude spectrum of the S15 SAP data. There is variability with an
amplitude of 1.0\,mmag on a timescale of $\sim$23\,d for the S15
data. The S55 data show a larger instrumental drift. Hence, it is
possible that this star has a moderate period of the order of 23\,d,
although the signal could also be instrumental.  The similarity to the
light curves and amplitude spectra for the previous star,
TIC~163801263, which was observed in the same sectors with the same
camera suggests that the variability for both stars is instrumental.  

\subsection{TIC~167695608 (TYC~8912-1407-1)}
\label{sec:j0651}
Five good SALT-HRS spectra of the roAp star TYC~8912-1407-1
\citep{2014MNRAS.439.2078H} were
obtained. The first one was recorded 146 days before
the second one; the last four span a time interval of only 12 nights.

All five spectra show the \Feline\ line marginally resolved into its two
magnetically split components. As in many Ap star spectra, the
\Feline\ line is blended on the blue side by the line of an
unidentified ion probably pertaining to a rare earth element (see
Fig.~\ref{fig:spec6149}). In TYC~8912-1407-1, the strength of this
blend tends to be above average, but it 
is separated enough from the blue component of \Feline\ to allow it to
be untangled reasonably well. We estimate the resulting uncertainty
affecting the derived values of the mean magnetic field modulus to be
of the order of 100~G. Within this uncertainty, no definite variation
of $\Bm$ is detected.

As described in Sect.~\ref{sec:Bqmeas}, we took advantage of the
availability of five observations of TYC~8912-1407-1 to achieve better
precision in the derived values of the mean quadratic magnetic
field. One can indeed note in Table~\ref{tab:meas} that the formal
uncertainties of the $\Bq$ values for this star are lower than for any
other star observed with 
SALT-HRS. Even with these rather low uncertainties, no significant
variations of the mean quadratic magnetic field are detected between
the considered epochs.

The applied procedure also allows in principle better precision to be
achieved in the determination of the upper limit of \vsi. This limit,
$\vsi\lesssim 1.96$\,\kms, is very low, hence consistent with the
identification of TYC~8912-1407-1 as a ssrAp star.

SAP data are available for Sectors~1-13, 27-39, and 61-68, as
  TIC~167695608 is in the southern continuous viewing zone. There is
  no indication of any rotational variability, consistent with this
  being a ssrAp star.

\subsection{TIC~170419024 (HD~151860)}
\label{sec:hd_151860}
The \Feline\ line is marginally resolved into its magnetically split
components 
in our SALT-HRS spectrum of HD~151860. The unidentified blend
affecting its blue component (see Fig.~\ref{fig:spec6149}) somewhat
complicates the determination of 
the mean magnetic field modulus. The difference between its value,
derived from the fit of three Gaussians, $\Bm=3.4$\,kG, and the 2.5\,kG
value determined by \citet{2013MNRAS.431.2808K} through spectrum
synthesis analysis of an observation obtained $\sim$13 years earlier,
seems significant and representative of actual 
variability. The fact that the \Feline\ line does not show any hint of
magnetic resolution in the FEROS spectrum analysed here, which was
recorded $\sim$14.5 years before our SALT-HRS spectrum, lends further
support to this interpretation. 

The value of the mean quadratic magnetic field that we derive from
this FEROS spectrum, $\Bq=2.5$\,kG is fully consistent with the field
determination of \citet{2013MNRAS.431.2808K}. Both are much lower than
the value based on the SALT-HRS observation, $\Bq=5.4$\,kG, which
however is affected by a considerably higher uncertainty. The
$\Bq/\Bm$ value derived from this spectrum also lies further above the
linear $\Bq$ vs $\Bm$ relation illustrated in Fig.~\ref{fig:Bq_vs_B} than
for any other star of this study. However, this deviation remains
compatible with the error bars. Furthermore, the distortion shown by
the profile of the Fe~{\sc i}~$\lambda\,6336.8$\,\AA\ line is indicative
of the presence of a rather strong magnetic field. Thus, the
difference between the $\Bq$ values derived from the 2008 and 2023
spectra appears to reflect real variations of the magnetic field.

On the other hand, the value of the projected equatorial velocity
determined by \citet{2013MNRAS.431.2808K}, $\vsi=4.5$\,\kms, seems
definitely too large compared to the upper limits set from our
analysis. We obtained $\vsi\lesssim1.87$\,\kms\ from consideration of
the FEROS spectrum while magnetic broadening is below the detection
threshold in the SALT-HRS spectrum. These results, which hint at a
value of \vsi\ significantly lower than 4.5\,\kms, are further
supported by the appearance of the Fe~{\sc i}~$\lambda\,5434.5$\,\AA\ line
(see Fig.~\ref{fig:spec5434}), whose FWHD hardly exceeds the estimated
intrinsic line width. The absence of a
detectable rotation signature in the pulsation anaysis of HD~151860
\citet{2021MNRAS.506.1073H} also tends to favour a long period.

Moreover, HD~151860 appears to be a spectroscopic binary. Indeed, the
radial velocity has significantly changed between the epochs of the
FEROS ($v_{\rm r}=-0.82$\,\kms) and SALT-HRS ($v_{\rm r}=4.10$\,\kms)
observations. Thus, while no constraints can be set on the orbital
period, HD~151860 is most probably another case of a ssrAp star in a
binary system.  

\begin{figure}
\centering  
\includegraphics[width=1.0\linewidth,angle=0]{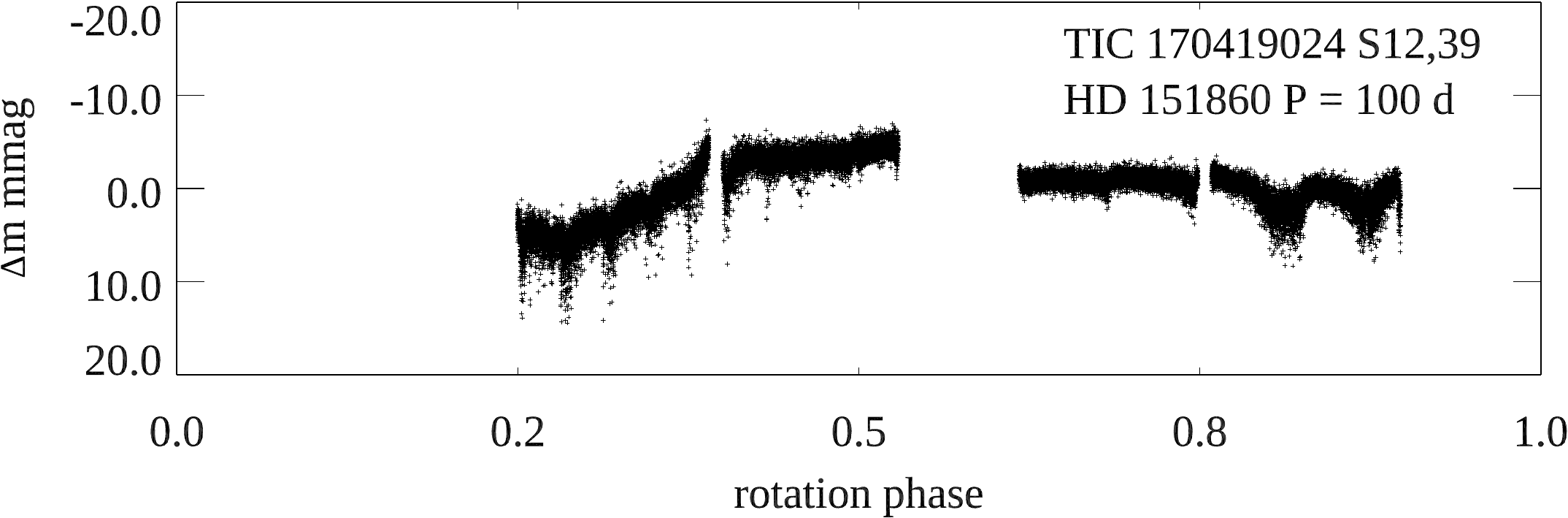}
 \caption{TIC~170419024 (HD~151860). The SAP light curves for Sectors
   12 and 
   39 are phased with $P_{\rm rot} = 100$\,d. This is simply to show what
   the problems are with the data and to indicate that if there is
   rotational variability, it is consistent with a ssrAp star. There
   is no claim that 100\,d may be the period, or even close to it.  } 
\label{fig:170419024_phases_ftd}
\end{figure}

SAP data are available for Sectors 12 and 39. The S12 data show a strong
slope; the S39 data have instrumental variations. Together a best fit
is found for $P_{\rm rot} = 40$\,d, but it is messy and
unconvincing. Figure~\ref{fig:170419024_phases_ftd} shows the two
sectors phased with $P_{\rm rot} = 100$\,d strictly for illustrative
purposes. The available SAP data are consistent with this being a
ssrAp star.

\subsection{TIC~202899762 (BD+46~570)}
\label{sec:bd+46_570}
The radial velocity of BD+46~570 shows a small but significant
variation between the CAOS and HARPS-N spectra of this star, which
were obtained 425 days apart. In the HARPS-N spectrum, the \Feline\
line is well resolved into its two components, and it is free from
significant blends (see Fig.~\ref{fig:spec6149}), so that the mean
magnetic field modulus can be 
determined with the best estimated precision of the present study. If
the field strength was of the same order ($\Bm=3.4$\,kG) at the time
of the CAOS observation, one could expect \Feline\ to be also
marginally resolved in that spectrum. This is not the case, but it may
be due to the rather low S/N rather than to magnetic field
variability.

The mean quadratic magnetic field is precisely determined through
analysis of the HARPS-N spectrum. The magnetic distortion of the
Fe~{\sc i}~$\lambda\,6336.8$\,\AA\ line in this spectrum is particularly
remarkable. A formally significant value of $\Bq$ is also
derived from the CAOS observation of BD+46~570, albeit with a large
uncertainty. Due to the latter, the possible variability of $\Bq$
cannot be constrained in a meaningful manner. Mean longitudinal
magnetic field measurements from \citet{2006MNRAS.372.1804K} and from
\citet{2020AstBu..75..294R} do not reveal any definite variations
either.

Rotational broadening is below the detection limit in the HARPS-N
spectra, consistent with the observed profile of the Fe~{\sc
  i}~$\lambda\,5434.5$\,\AA\ line (see Fig.~\ref{fig:spec5434}). The
high value of the upper limit of the projected equatorial velocity set
from analysis of the CAOS spectrum, $\vsi\lesssim8.36$\,\kms
definitely results from crosstalk between the $a_1$ and $a_3$ terms of
the linear regression. Indeed, the latter does not identify any
dependence of $\RI$ on $\ew^2\,\lambda_0^4/c^4$, which is physically
meaningless. As no
variations of any of the studied magnetic field moments 
were detected, BD+46~570 appears to be another example of a ssrAp star
in a binary system.

SAP data are available for Sector 18 only. There is no indication of
rotational variablity. 

\subsection{TIC~206461701 (HD~209364)}
\label{sec:hd_209364}
Despite the good quality of the SALT-HRS spectrum of HD~209364, its
mean quadratic magnetic field is below the detection 
limit. The \vsi\ upper limit that is derived for this star is higher
than for most other targets of this study. The rotational broadening
of the spectral lines is actually clearly seen even at the compressed
scale of Fig.~\ref{fig:spec6150_2}. No lines show any hint of
magnetic resolution. More strikingly, the absence of the Cr~{\sc
ii}~$\lambda\,6147.1$~\AA\ line from the spectrum of this star, together
with the weakness of the Nd~{\sc iii}~$\lambda\,6145.1$\,\AA\ line, suggests
that the classification of HD~209364 as ApSrEuCr
\citep{1988mcts.book.....H} may be 
mistaken.

SAP data are available for Sectors 1 and 68 only. There is no indication of
rotational variability in the S1 data. The S68 data have a slope that
may be instrumental. The best fit to the two sectors together is for
$P_{\rm rot} = 50$\,d, but that is plausibly instrumental. 

\subsection{TIC~207468665 (HD~148330)}
\label{sec:hd_148330}
The spectrum of HD~148330 does not show any of the lines typically
observed in the Ap stars of similar temperature (see
Fig.~\ref{fig:spec6150_4}). Even though \citet{2017MNRAS.471..926K}
found moderate, but not extreme, overabundances of a number of
elements, HD~148330 does not look like a typical Ap star. In
fact, \citet{1975A&AS...21...25F} had already noted inconsistencies
in the appearance of the lines used for peculiarity identification in
spectral classification studies and the possible variability of some
of them. Furthermore, not only no formally significant mean quadratic
magnetic field value was derived from the analysis of 12 ESPaDOnS
archive spectra (see Sect.~\ref{sec:Bqmeas}) nor of the CAOS spectrum
specifically obtained in the framework of this project, but also the
root mean square value of the mean longitudinal magnetic field,
$\Bzrms=(20\pm11)$\,G, determined from the same ESPaDOnS spectra, does
not significanly differ from zero. This is a stringent limit,
which shows that HD~148330 does not have the kind of magnetic
field that is found in typical Ap stars. The $\Bz$ values derived
by \citet{2017MNRAS.471..926K} from two of the
ESPaDOnS spectra from the series that we analysed, which have a much
higher uncertainty than our measurements, are not formally
significant. Neither are most of the individual values of this field
moment determined by \citet{1990BAICz..41..118Z}, whose uncertainties
are even larger; the argument given
by these authors that $\Bz$ phases with the photometric variations is
not valid since such variations are not detected from the much higher
precision TESS data \citepalias{2022A&A...660A..70M}. Finally, the
upper limit of 
the projected equatorial velocity derived from analysis of the
ESPaDOnS spectra, $\vsi\lesssim10.23$\,\kms, which is consistent with
the value reported by \citet{2017MNRAS.471..926K},
$\vsi=(9.8\pm1.0)$\,\kms, indicates that HD~148330 is not a super-slow
rotator. 

SAP data are available for many sectors: S16, 19, 23-25, 49-52, 56 and
76. Despite the non-negligible rotational line broadening, there is no
convincing rotational variability seen in these SAP 
data, consistent with the indications that this is not an Ap star. 

\subsection{TIC~233539061 (HD~174016-7)}
\label{sec:hd_174016}
The SB2 system HD~174016-7 has been studied in detail by
\citet{1999A&AS..140..279G}. The secondary component, HD~174017, is an
A0p SrCrEuSi star, which completes its orbit around the G6\,III primary,
HD~174016, in 3097.9\,d. A HARPS-N spectrum of this system was
obtained at orbital phase 0.10, at which the radial velocity
difference between the two components must be of the order of
3.5--4\,\kms\ \citep[see Fig.~2 of][]{1999A&AS..140..279G}. In our
spectrum, some Fe lines of the Ap component can indeed be
distinguished as a blue blend in the blue wing of the Fe lines
of the G giant (which is the brighter star). This is, for instance,
the case for the Fe~{\sc i}~$\lambda\,5434.5$\,\AA\ line, as can be seen
in Fig.~\ref{fig:spec5434}. The contributions of the
two components to the observed Fe lines cannot be
disentangled, so that there is no possibility to diagnose the mean
quadratic field modulus or the projected equatorial velocity for the
analysis of these lines, as was done in other stars of this
study. Nevertheless, the Fe lines of HD~174017 seem to be
sharp enough so that this may be a ssrAp star.

SAP data are available for many sectors: S14-17, 19-26, 40-41, 47, and
55-55. There is no convincing rotational variability seen in these SAP
data. 

\begin{figure}
\centering  
\includegraphics[width=1.0\linewidth,angle=0]{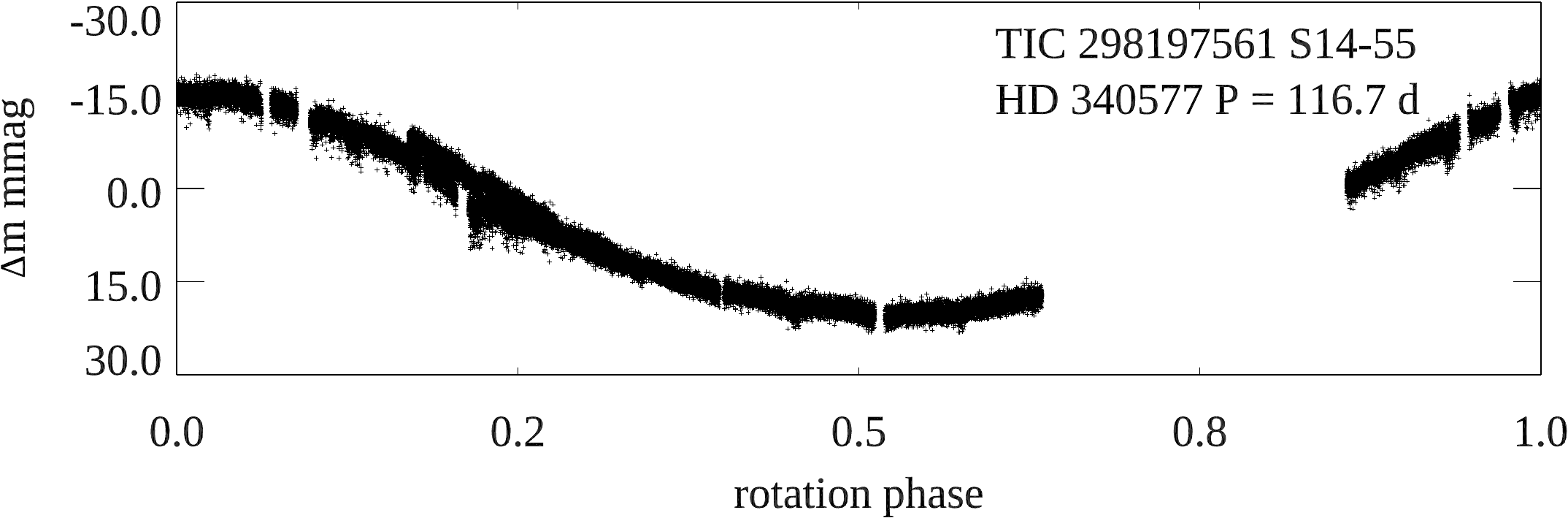}
 \caption{TIC~298197561 (HD~340577). There are four sectors of SAP
   data plotted here, S14, 15, 41, and 55. They have been phased
   with a rotation period of $P_{\rm 
     rot} = 116\fd7$ \citep{2016AJ....152..104H} after some zero-point
   adjustments up to 0.01 mag 
   have been made among the sectors. } 
\label{fig:298197561_phase}
\end{figure}

\subsection{TIC~286965228 (HD~127304)}
\label{sec:hd_127304}
The case of HD~127304 is, to some extent, similar to that of
HD~148330. Its spectrum also fails to show the characteristic lines
generally found in Ap stars of similar temperature (see
Fig.~\ref{fig:spec6150_4}). Here, too, the analysis of three ESPaDOnS
archive spectra and of one CAOS dedicated spectrum failed to detect
any formally significant mean quadratic magnetic field. A null value
of the root mean square longitudinal magnetic field is also determined
from the ESPaDOnS spectra, $\Bzrms=(9\pm7)$\,G.

However, HD~127304 definitely shows radial velocity variations: the
values that we measured range from $v_{\rm r}=-26.2$\,\kms\ to $v_{\rm
  r}=-2.1$\,\kms; \citet{2014A&A...562A..84R} determined $v_{\rm
  r}=-22.3$\,\kms; and \citet{2006AstL...32..759G} give $v_{\rm
  r}=-12.6$\,\kms. \citet{1989A&A...209..233R} suspect that this may
be an Am SB2. Nevertheless, we do not readily see evidence of the
secondary lines in the spectra that we analysed. The sets of
diagnostic lines that we measured in each spectrum are internally
consistent in terms of radial velocity, so that these lines must
pertain to a single component. The resulting upper limit of the
projected equatorial velocity, derived from the ESPaDOnS spectra,
$\vsi\lesssim7.34$\,\kms, does not support the view that this component,
whether it is an Ap star or not, rotates super slowly. 

SAP data are available for Sectors 23 and 50. There is no convincing
rotational variability seen in these SAP data, which supports the view
that HD~127304 is probably not an Ap star. 

\subsection{TIC~291561579 (HD~171420)}
\label{sec:hd_171420}
At the $2.2\sigma$ level, the value of the mean quadratic magnetic
field that is derived from the analysis our SALT-HRS spectrum of
HD~171420 is not formally significant . However, inspection of the
regression plots (Fig.~\ref{fig:bq90131}) suggests that it should be
readily measurable at 
somewhat higher resolution, with a probable value of the order of
1.5\,kG. It would be interesting to confirm this, to test whether
HD~171420 is a mild Ap star (rather than a star misclassified as Ap),
as conjectured in Sect.~\ref{sec:spectra}. However, it is definitely
not a slowly rotating Ap star, as indicated by the upper limit of the
projected equatorial velocity, $\vsi\lesssim11.51$\,\kms, which is
significant, and by the width of the Fe~{\sc i}~$\lambda\,5434.5$\,\AA\
line, as illustrated in Fig.~\ref{fig:spec5434}. 

SAP data are available for Sector 13. There is no convincing rotational
variability seen in these SAP data. 

\begin{figure}
\resizebox{\hsize}{!}{\includegraphics{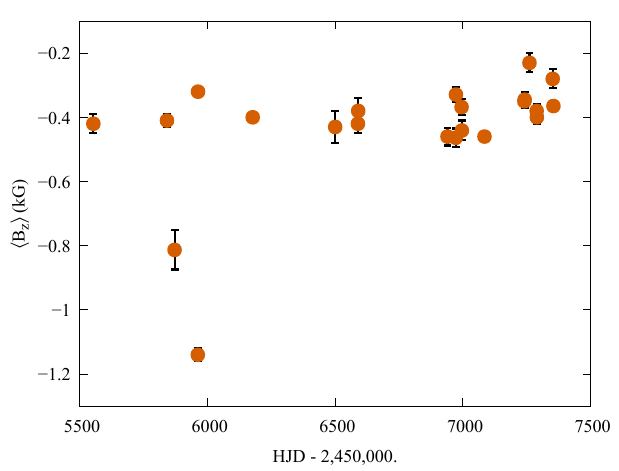}}
\caption{Mean longitudinal magnetic field measurements of HD~17330
  against date, from the Special Astrophysical Observatory team (see
  text for specific references). Except for two discrepant values,
  little variability is seen, except perhaps for a small long-term
  trend.}
\label{fig:bz17330}
\end{figure}

\subsection{TIC~298197561 (HD~340577)}
\label{sec:hd_340577}
Following the identification of HD~340577 as a ssrAp star candidate
that shows $\delta$~Sct pulsations, reported in
\citetalias{2022A&A...660A..70M}, 
\citet{2023MNRAS.526L..83H} detected the magnetic field of this star
through a single spectropolarimetric observation, from which they
measured a value $\Bz\sim-0.4$\,kG of its mean longitudinal
component. This rather low $\Bz$ value is consistent with the 
fact that the mean quadratic magnetic field is below the detection
threshold in the CAOS spectrum of HD~340577 that was analysed. The
rotational Doppler broadening is also below the formal detection
threshold, but the FWHD of the Fe~{\sc i}~$\lambda\,5434.5$\,\AA\ line
seems somewhat greater than could be accounted for by instrumental and
thermal broadening: the weakly magnetic Ap star
HD~340577 may either have a moderately long rotation period ($20\,{\rm
  d}\lesssim\Prot\lesssim50$\,d) or be a ssrAp star.

SAP data are available for Sectors 14, 15, 41 and
55. Figure~\ref{fig:298197561_phase} shows the phased light curve after
some problematic data have been excised and after some adjustments to
S14 and 55 by 0.01 mag were made. This star is definitely a ssrAp
  star, with a rotation period $\Prot=116\fd7$
  \citep{2016AJ....152..104H}.   

\subsection{TIC~301918605 (HD~17330)}
\label{sec:hd_17330}
The magnetic field of HD~17330 was detected in 2010 by
\citet{2017AstBu..72..391R}. In the following years, this group
obtained 22 additional observations of the star, which are shown in
Fig.~\ref{fig:bz17330} 
\citep{2018AstBu..73..178R,2020AstBu..75..294R,2022AstBu..77...94R,2022AstBu..77..271R,2023AstBu..78..567R}. Twenty
of the measurements yield values $\Bz\sim-400$\,G, with a hint of a
long-term trend towards less negative values at the more recent
epochs. Two early determinations look discrepant, with mean
longitudinal field values departing considerably from
$\Bz\sim-400$\,G. The origin of these deviations is unclear -- the
authors regard instrumental effects as unlikely -- but they do not look
at all like typical rotational modulation of Ap star magnetic
fields. We believe that the long-term near stability of the 21 other
$\Bz$ values, with a possible slow upward trend, is much more likely
to represent the true variability of this field moment. If so,
HD~17330 must have a rotation period $\Prot>5$\,yr; it is a ssrAp
star. 

Romanyuk and his colleagues (see references above) also detected
radial velocity variations, which they regard as significant, hence
which indicate that HD~17330 is a spectroscopic binary. Their
published measurements are insufficient to constrain the orbital
period. The single radial velocity value that we derived is well
within the range of those of the Romanyuk group. 

For determination of the mean quadratic magnetic field of HD~17330
from our CAOS spectrum, we used diagnostic
lines from Fe~{\sc ii}. Indeed this is one of the few stars in the
present study that have 
effective temperatures $\Teff>10^4$\,K, so that Fe~{\sc i} lines tend
to be 
too weak and too few to be used. The mean quadratic magnetic field is
below the detection threshold, but a significant upper limit of the
projected equatorial velocity is derived, $\vsi<8.15$\,\kms, which is
larger than one would expect if the rotation period is longer
than 5\,yr, as inferred from the mean longitudinal field
variations. However, this apparent inconsistency results from the
definite occurrence of significant crosstalk between the $a_1$ and
$a_2$ terms of the right-hand side of Eq.~(\ref{eq:Bq}), so that it
does not question the conclusion that HD~17330 is very probably a
ssrAp star in a binary system. 

SAP data are available for Sector 18. There is no convincing rotational
variability seen in these SAP data. 

\subsection{TIC~301946105 (HD~7410)}
\label{sec:hd_7410}
\begin{figure}
\centering  
\includegraphics[width=1.0\linewidth,angle=0]{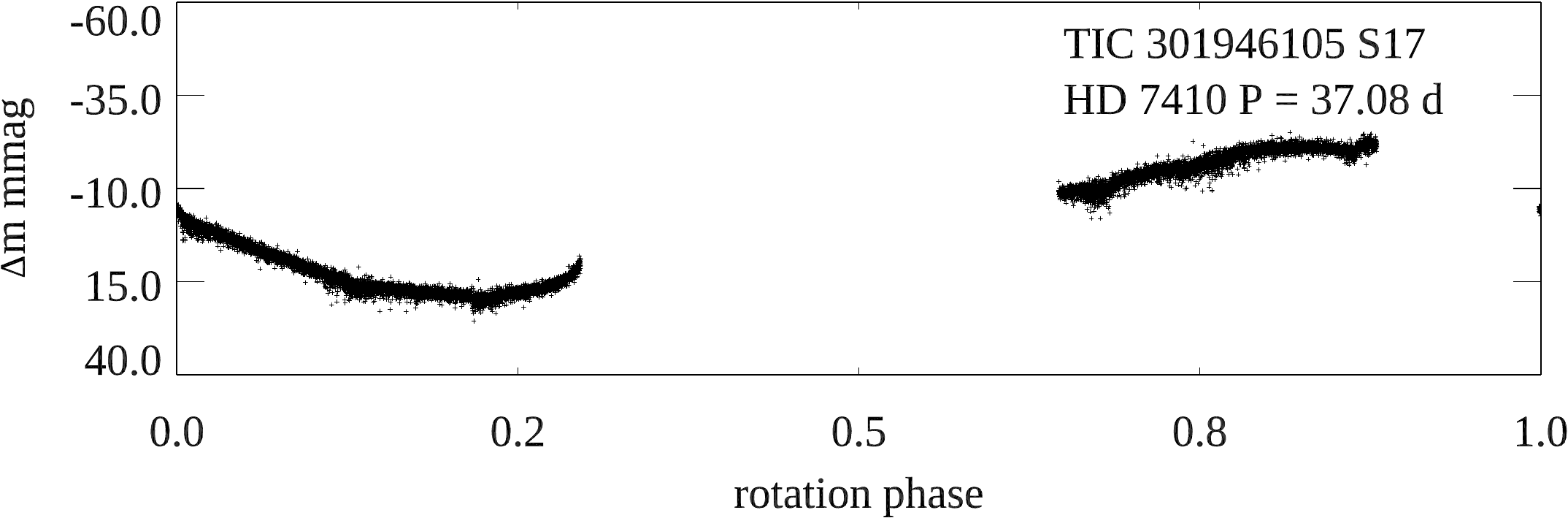}
 \caption{TIC~301946105 (HD~7410). This shows the light curve of the
   S17 SAP data phased with the rotation period found by
   \citet{2020MNRAS.493.3293B} of $\Prot=37\fd08$. The SAP data do not
   confirm this period, but they are consistent with it. } 
\label{fig:301946105_phase}
\end{figure}

\citet{2020MNRAS.493.3293B} reported the occurrence of photometric
variations in HD~7410 with a period $\Prot=37\fd08$. This was
overlooked in \citetalias{2022A&A...660A..70M}. This variability is
unduly removed by the reduction process in the PDCSAP data, but it is
clearly seen in the SAP data. The star was observed with TESS only in
Sector~17, over a time base too short to allow the rotation period to
be clearly determined. But  the variations are consistent with the
published \Prot\ value, as can be seen in
Fig.\,\ref{fig:301946105_phase}. Hence, HD~7410 is not a ssrAp star,
but 
an Ap star with a moderately long rotation period
\citepalias[see Sect.~2.4 of][]{2024A&A...683A.227M}.

We have obtained a CAOS spectrum and a HARPS-N spectrum of HD~7410,
392 days apart. The radial velocity shows a small but formally
significant change between the two epochs. This is probably a
spectroscopic binary, possibly one with a long orbital period. 

The analysis of the HARPS-N spectrum allowed us to derive a precise
value of the mean quadratic magnetic field, $\Bq=2086$\,G. This value
is below the detection limit with CAOS. Meaningful upper limits of
\vsi\ were determined from from both spectra: $\vsi\lesssim7.16$\,\kms
(CAOS) and $\vsi\lesssim4.82$\,\kms (HARPS-N). 

\subsection{TIC~334505323 (HD~106322)}
\label{sec:hd_106322}
Spectral lines typical of Ap stars, such as those of Cr, Si, and Nd,
are not seen in the spectrum of HD~106322 (see
Fig.~\ref{fig:spec6150_3}): although it has been assigned the MK
type Ap EuCr \citep{1978mcts.book.....H}, this star looks more like a
normal A star. In particular, 
the spectral lines of Fe~{\sc i} in the red-arm SALT-HRS spectrum of
HD~106322 are weak and their density is low, so that they cannot be
used to diagnose the 
mean quadratic magnetic field. The Fe~{\sc i} lines of the blue-arm
spectrum were used instead. Of course, because the Zeeman effect
depends quadratically on the wavelength, the magnetic sensitivity of
the lines is lower in the blue arm than in the red arm.
In fact, no definite detection of the mean quadratic
magnetic field was achieved, but a meaningful upper limit of the
projected equatorial velocity, $\vsi\lesssim14.77$\,\kms, was
derived. Thus, HD~106322 is not rotating extremely slowly, as is also 
apparent from consideration of the profile of the Fe~{\sc
  i}~$\lambda\,5434.5$\,\AA\ line in Fig.~\ref{fig:spec5434}. 

SAP data are available for Sectors 10 and 37. There is no convincing
rotational variability seen in these SAP data, lending plausibility to
the suspicion that HD~106322 is not an Ap star. 

\subsection{TIC~347202840 (HD~236298)}
\label{sec:hd_236298}
As for some other stars, a precise determination of the mean quadratic
magnetic field of HD~236298, $\Bq=2218$\,G, could be obtained through
analysis of a HARPS-N spectrum, but this field moment remained below
the detection threshold of a CAOS spectrum. As the star is one of the
hottest ones of the present sample ($\Teff=10,700$\,K), Fe~{\sc ii}
lines were used to diagnose the magnetic field. A meaningful upper
limit of \vsi\ is obtained from both spectra: $\vsi\lesssim5.50$\,\kms\
(CAOS) and $\vsi\lesssim3.07$\,\kms\ (HARPS). The CAOS value is
definitely an overestimate, because of crosstalk between the $a_2$ and
$a_3$ terms of the regression analysis (the former is negative, which
is physically meaningless). Crosstalk in this case appears to result
primarily from the weakness of the spectral lines and the very narrow
range of equivalent widths that they span, so that $a_3$ is poorly
constrained. The \vsi\ upper limit determined from the
HARPS-N spectrum is consistent with a long rotation period.

The radial velocity does not appear to have changed at all over the
435 days that separate our two observations. All elements of
information available until now indicate that HD~236298 must either be
be a single magnetic ssrAp star, or a single Ap star with a moderately
long rotation period. 

\begin{figure}
\centering  
\includegraphics[width=1.0\linewidth,angle=0]{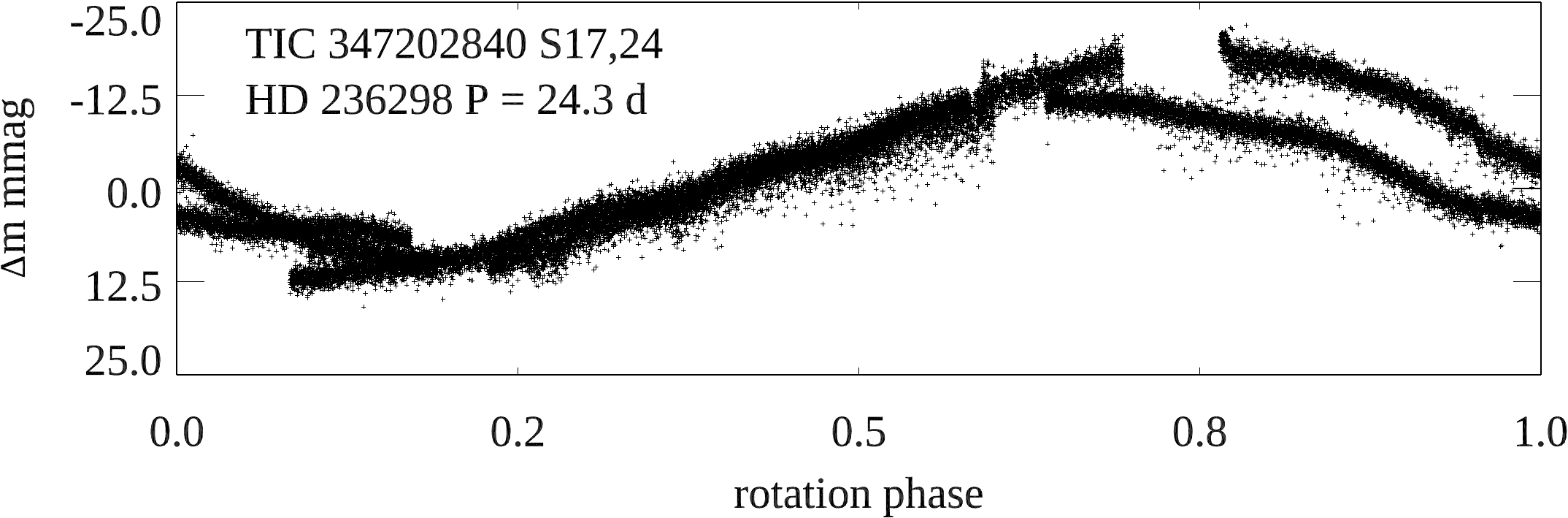}
 \caption{TIC~347202840 (HD~236298). Sectors 17 and 24 SAP data. Those have
   been phased with a rotation period of $P_{\rm rot} = 24.3$\,d after
   some zero-point adjustments have been made. It is clear that there
   are further instrumental zero point shifts.} 
\label{fig:347202840_phase}
\end{figure}

SAP data are available for Sectors 17 and
24. Fig.\,\ref{fig:347202840_phase} shows the phased light curve after
some zero-point shifts were made. It is clear that further zero-point
shifts exist within the SAP data, but the rotational variation  is
clear with a period near $P_{\rm rot} = 24.3$\,d, or twice this
value. (As pointed out in \citetalias{2024A&A...683A.227M}, double-wave
light curves are not infrequent in Ap stars, which cannot be easily
distinguished using TESS observations when the one-wave period value
is close to the sector length.) This star is an Ap
star with a moderately long rotation period. 

\subsection{TIC~352787151 (BD+35~5094)}
\label{sec:bd+35_5094}
The fact that no Cr~{\sc ii} lines are visible in the portion of the
spectrum of BD+35~5094 illustrated in Fig.~\ref{fig:spec6150_1}
probably results primarily from a temperature effect. Indeed, this is
one of the coolest stars of the present sample, and the lines of
Si~{\sc i} and Nd~{\sc iii} are similar to those of the other Ap stars
in the same temperature range. The Ap classification of BD+35~5094
also received recent confirmation from the analysis of LAMOST spectra
by \citet{2022ApJS..259...63S}. 

The mean quadratic magnetic field of BD+35~5094 is below the detection
threshold of the CAOS spectrum that we obtained for this
star. However, a meaningful upper limit of the projected equatorial
velocity, $\vsi\lesssim5.92$\,\kms, was obtained as a by-product of
our analysis, but there appears to be some crosstalk between the $a_1$
and $a_2$ terms of the regression analysis. Indeed, the latter is
negative, which is not physically meaningful. This implies that the
value of $a_1$ is overestimated. More likely, the rotational
broadening of the spectral lines is at most marginally above the
detection threshold, consistent with the  appearance of the Fe~{\sc
  i}~$\lambda\,5434.5$\,\AA\ line profile in Fig.~\ref{fig:spec5434}.
Accordingly, BD+35~5094 is probably a weakly magnetic ssrAp
star, but a moderately long rotation period cannot be ruled
out. 

\begin{figure}
\centering  
\includegraphics[width=1.0\linewidth,angle=0]{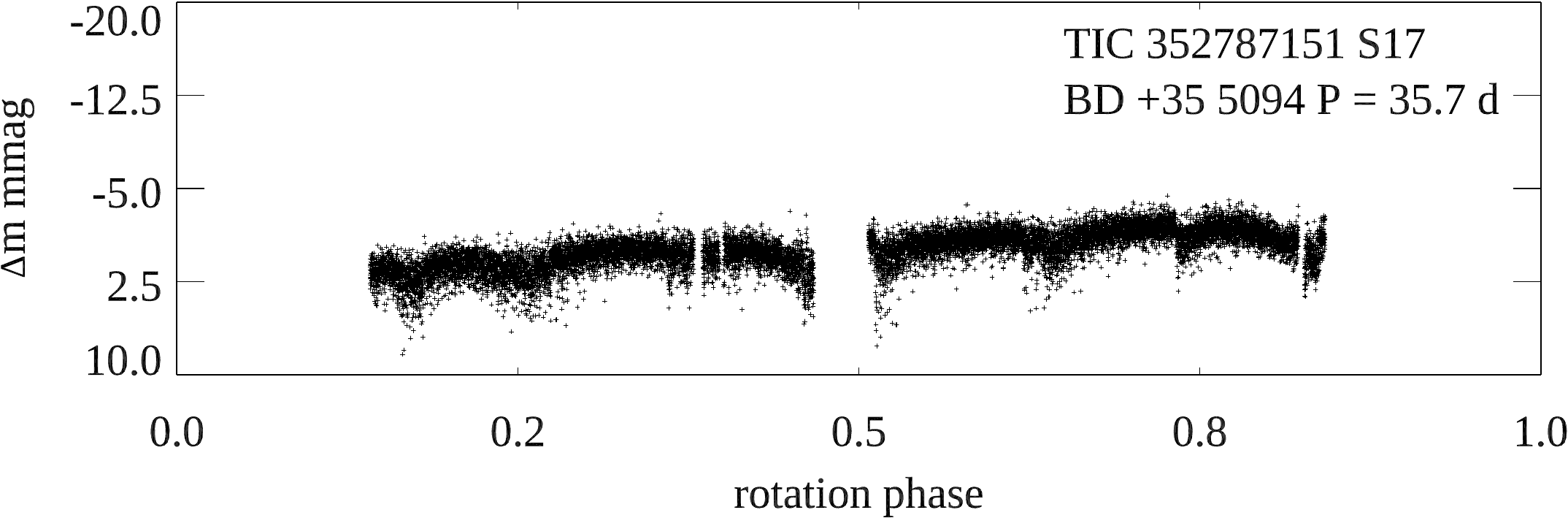}
 \caption{TIC~352787151 (BD +35 5094). S17 SAP data. Those have been
   phased with a rotation period of $P_{\rm rot}=35\fd7$. There is
   no convincing rotational variability seen in these SAP data. The
   small amplitude is plausibly instrumental.} 
\label{fig:352787151_phase}
\end{figure}

SAP data are available for Sector 17. Figure~\ref{fig:352787151_phase}
shows the phased light curve with a period of $P_{\rm
  rot}=35\fd7$. The variations are plausibly instrumental.    

\subsection{TIC~403625657 (HD~11187)}
\label{sec:hd_11187}
\citet{1958ApJS....3..141B} detected the magnetic field of HD~11187
and obtained seven measurements of its mean longitudinal
component. The latter shows apparently significant variability over a
couple of days, raising the suspicion that the rotation period may be
short. Re-analysing the TESS data, short-term photometric variations
were indeed detected, which suggest a value of the rotation period of
the order of $5\fd27$. However, this value is not unambiguously
constrained. A series of 15 determinations of the mean longitudinal
magnetic field of HD~11187 obtained at the Special Astrophysical
Observatory (Romanyuk, private communication) also indicates that the
rotation period must be of the order of days to weeks. In particular,
sign reversals of $\Bz$ are observed over a few days, but these
measurements are not sufficient yet for unambiguous determination of
the rotation period. 

When fitting the observed values of $\RI$ in our CAOS spectrum of
HD~11187 by a function of the form
given in Eq.~(\ref{eq:Bq}), the rotational broadening term ($a_1$) is
by far the dominant one. The derived upper limit of the projected
equatorial velocity, $\vsi\lesssim16.96$\,\kms, is compatible with a
period of a few days. The mean quadratic magnetic field is below the
detection threshold. In summary, HD~11187 is definitely a magnetic Ap
star that does not rotate particularly slowly. 

\begin{figure}
\centering  
\includegraphics[width=1.0\linewidth,angle=0]{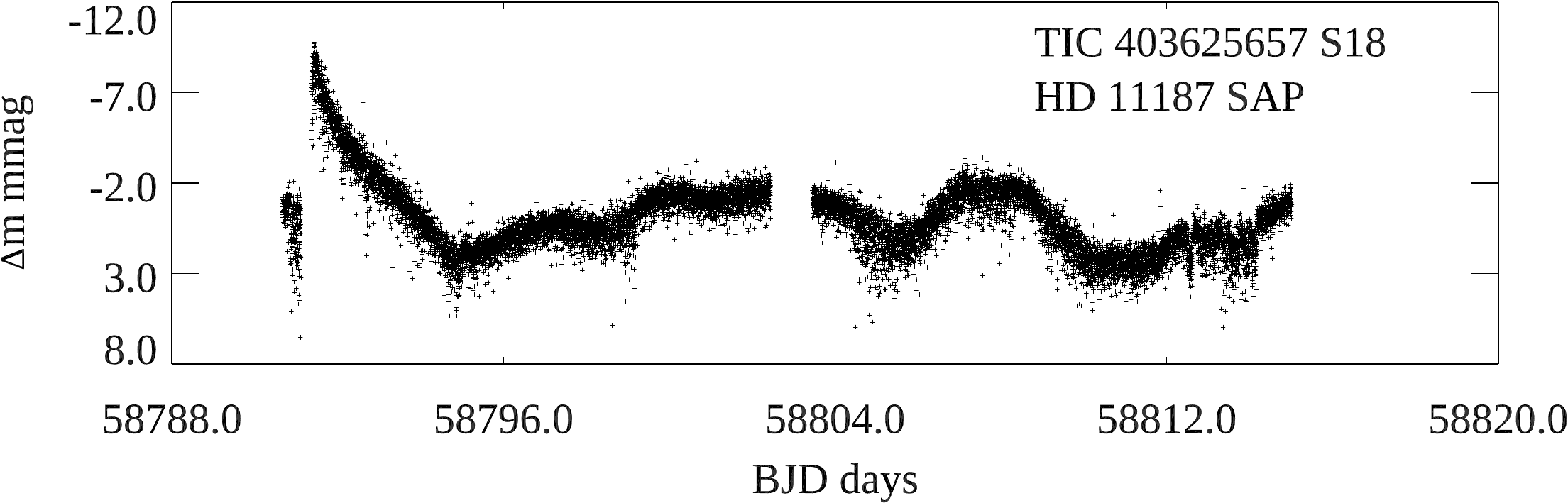}
 \caption{TIC~403625657 (HD 11187). The SAP light curve for S18. The
   variations are  not periodic and do not look rotational. They are
   plausibly instrumental in origin. } 
\label{fig: 403625657_lc_ftd}
\end{figure}

SAP data are available for Sector 18. Over the sector the light curve
  shows variations up to 0.015 mag that do not look periodic (see
  Fig.~\ref{fig: 403625657_lc_ftd}). Hence 
  these may be instrumental. No clear rotational variation is seen. We
  do not know how to reconcile this with the apparently significant
  variability of $\Bz$ over a timescale of days and the clear
  rotational broadening that affects the spectral lines.

\subsection{TIC~444094235 (HD~85284)}
\label{sec:hd_85284}
The effective temperature listed in the TIC for HD~85284,
$\Teff=13,640$\,K, is definitely much too high. The magnetic field of
this typical Ap star can be best constrained by analysing its Fe~{\sc
  i} lines. Analysing a FEROS spectrum of this star, we could
determine the mean quadratic magnetic field at the $5\sigma$ level,
but the rotational broadening of the spectral lines is below the
detection threshold. Indeed, the contribution of the $a_1$ term of the
regression appears to be somewhat below the expected intrinsic line
broadening. However, the latter is overestimated because of the too
large value of the effective temperature. If it was of the
order of $\Teff\sim10,700$\,K, as seems more plausible from the
appearance of the spectrum (see Sect.~\ref{sec:spectra}), the
observed line width would no longer be significantly less than the
intrinsic line width estimate. Even so, no rotational broadening would
be detected. 

At the resolution of our SALT-HRS spectrum, the mean
quadratic magnetic field is below the detection threshold. The upper
limit of the projected equatorial velocity, $\vsi\lesssim4.49$\,\kms,
is formally significant, but this results from a considerable amount
of crosstalk between the $a_1$ and $a_2$ terms of the regression
analysis ($a_2$ is negative, which is unphysical). There is little
doubt that the conclusion drawn from consideration of the FEROS
spectrum, which implies that HD~85284 is most likely a ssrAp star,
is correct. It can also be noted that a number of lines show
significant intensity differences between the two available spectra,
which are indicative of (probably) long-term spectroscopic
variability. Detailed characterisation of these variations is beyond
the scope of the present study.

Almost 13 years elapsed between the FEROS and SALT-HRS observations of
HD~85284. The values of the radial velocity differ considerably
between the two epochs: $v_{\rm r}=0.62$\,\kms\ and $v_{\rm
  r}=9.47$\,\kms, respectively. Both also significantly differ from
the value published by \citet{1987A&AS...70..373D}, $v_{\rm
  r}=16$\,\kms. Thus, HD~85284 appears to be another ssrAp star in a
spectroscopic binary. The existing data are too sparse to set any
constraint on the orbital period. 

SAP data are available for Sectors 9, 10, 36, and 37. They show
no obvious rotational variations.  

\subsection{TIC~461161123 (HD~95811)}
\label{hd_95811}
Although \citet{2009A&A...498..961R} note that HD~95811 may be a
$\delta$~Del Am star rather than an Ap star, inspection of our
SALT-HRS spectrum of this star confirms that the latter classification is
correct. The mean quadratic magnetic field is below the detection
threshold. The derived upper limit of the
projected equatorial velocity, $\vsi\lesssim10.89$\,\kms, indicates
that the lines are subject to some rotational broadening, which is
also visible in the profile of the Fe~{\sc i}~$\lambda\,5434.5$\,\AA\
line in Fig. \ref{fig:spec5434}. Thus, HD~95811 is
not a ssrAp star. It may be an Ap star with a moderately long
rotation period, even though no photometric variability signal was
detected in the TESS SAP data, which are available for Sectors 9 and
36. 

\subsection{TIC~468507699 (HD~206977)}
\label{sec:hd_206977}
The analysis of our CAOS spectrum of HD~206977 yields a $3.1\sigma$
determination of the mean quadratic magnetic field. While it is at
the limit of formal significance, this detection is likely real, as the
fit appears well defined. At $\Bq=1.65$\,kG, this is the
lowest mean quadratic field value that could be measured in this study
with either CAOS or SALT-HRS; weaker fields could only be measured
with HARPS-N. This illustrates the importance of recording spectra at
the highest possible resolving power for characterisation of the low
end of the field strength distribution in the ssrAp stars. Indeed, the
upper limit of the projected equatorial velocity that was derived for
HD~206977, $\vsi\lesssim1.75$\,\kms, is consistent with it being a
ssrAp star, or at least an Ap star with a moderately long rotation
period ($20\,{\rm d}\lesssim\Prot\lesssim50\,{\rm d}$). 

SAP data are available for Sectors 9 and 36. There is no clear
rotational variation.

\end{document}